\documentclass[12pt,preprint]{aastex}
\usepackage[T1]{fontenc}
\newcommand{\nraoblurb}{The National Radio Astronomy Observatory is
a facility of the National Science Foundation operated under cooperative
agreement by Associated Universities, Inc.}

\newcommand{\degree}{\ensuremath{\,^\circ}}

\newcommand{\yr}{\ensuremath{\,{\rm yr}}}
\newcommand{\myr}{\ensuremath{\,{\rm Myr}}}
\newcommand{\gyr}{\ensuremath{\,{\rm Gyr}}}
\newcommand{\khz}{\ensuremath{\,{\rm kHz}}}
\newcommand{\mhz}{\ensuremath{\,{\rm MHz}}}
\newcommand{\ghz}{\ensuremath{\,{\rm GHz}}}

\newcommand{\K}{\ensuremath{\,{\rm K}}}

\newcommand{\cm}{\ensuremath{\,{\rm cm}}}
\newcommand{\persec}{\ensuremath{\,{\rm s^{-1}}}}
\newcommand{\percc}{\ensuremath{\,{\rm cm^{-3}}}}
\newcommand{\kpc}{\ensuremath{\,{\rm kpc}}}
\newcommand{\pc}{\ensuremath{\,{\rm pc}}}
\newcommand{\kms}{\ensuremath{\,{\rm km\, s^{-1}}}}
\newcommand{\msun}{\ensuremath\,M_\odot}
\newcommand{\s}{\,s}

\newcommand{\dexkpc}{\ensuremath{\rm \,dex\,kpc^{-1}}}
\newcommand{\Kkpc}{\ensuremath{\rm \,K\,kpc^{-1}}}
\newcommand{\te}{\ensuremath{T_{\rm e}}}
\newcommand{\teff}{\ensuremath{T_{\rm eff}}}
\newcommand{\Ne}{\ensuremath{n_{\rm e}}}
\newcommand{\NL}{\ensuremath{N_{\rm L}}}
\newcommand{\rgal}{\ensuremath{{R_{\rm gal}}}}
\newcommand{\dsun}{\ensuremath{{d_{\rm Sun}}}}
\newcommand{\hy}[2]{${\rm H}#1#2\alpha$}
\newcommand{\expo}[1]{${10^{#1}}$}

\newcommand{\hi}{H~{\sc i}}

\newcommand{\hii}{H~{\sc ii}}

\newcommand{\oii}{O~{\sc ii}}
\newcommand{\oiii}{O~{\sc iii}}

\newcommand\urltilda{\kern -.15em\lower .7ex\hbox{\~{}}\kern .04em}
\begin{document}


\title{H~{\textbf{\textsc{ii}}} Region Metallicity Distribution in the Milky Way Disk}

\author{Dana S. Balser\altaffilmark{1}, RobertT. Rood\altaffilmark{2},
T. M. Bania\altaffilmark{3}, \& L. D. Anderson\altaffilmark{3,4}}

\altaffiltext{1}{National Radio Astronomy Observatory, 520 Edgemont Rd., 
Charlottesville, VA 22903, USA.}
\altaffiltext{2}{Astronomy Department, University of Virginia, 
P.O. Box 400325, Charlottesville VA 22904-4325, USA.}
\altaffiltext{3}{Institute for Astrophysical Research, Department of Astronomy,
Boston University, 725 Commonwealth Avenue, Boston MA 02215, USA.}
\altaffiltext{4}{Current address: Laboratoire d'Astrophysique de Marseille 
(UMR 6110 CNRS \& Universit\'{e} de Provence), 38 rue F. Joliot-Curie, 
13388 Marseille Cedex 13, France.}

\begin{abstract}

The distribution of metals in the Galaxy provides important
information about galaxy formation and evolution.  \hii\ regions are
the most luminous objects in the Milky Way at mid-infrared to radio
wavelengths and can be seen across the entire Galactic disk.  We used
the NRAO Green Bank Telescope (GBT) to measure radio recombination
line and continuum emission in 81 Galactic \hii\ regions.  We
calculated LTE electron temperatures using these data.  In thermal
equilibrium metal abundances are expected to set the nebular
electron temperature with high abundances producing low temperatures.
Our \hii\ region distribution covers a large range of Galactocentric
radius (5 to 22\kpc) and samples the Galactic azimuth range
$330$\degree\ to $60$\degree.  Using our highest quality data (72
objects) we derived an O/H Galactocentric radial gradient of $-0.0383
\pm 0.0074$\dexkpc.  Combining these data with a similar survey made
with the NRAO 140 Foot telescope we get a radial gradient of $-0.0446
\pm 0.0049$\dexkpc\ for this larger sample of 133 nebulae.  The data
are well fit by a linear model and no discontinuities are detected.
Dividing our sample into three Galactic azimuth regions produced
significantly different radial gradients that range from $-0.03$ to
$-0.07$\dexkpc.  These inhomogeneities suggest that metals are not
well mixed at a given radius.  We stress the importance of homogeneous
samples to reduce the confusion of comparing data sets with different
systematics.  Galactic chemical evolution models typically derive
chemical evolution along only the radial dimension with time.  Future
models should consider azimuthal evolution as well.

\end{abstract}


\section{Introduction}\label{sec:intro}

Galactic chemical evolution (GCE) models are important for
understanding how galaxies form and evolve.  A key observational
constraint for these models is the spatial and temporal distribution
of abundances in the Galaxy \cite[][and references within]{pagel97,
  chiappini97, tosi00, prantzos03, carigi05, colavitti08, fu09,
  schonrich09}.  The isotopes of hydrogen, helium, and lithium were
produced in the Big Bang \citep[e.g.,][]{steigman07}.  Further
processing, primarily in low-mass ($< 2 \msun$) stars, altered these
primordial abundances through many generations of star formation and
evolution in the Galaxy, therefore light element abundance
measurements yield important constraints for models of Big Bang
nucleosynthesis and low-mass stellar evolution \citep[][and references
within]{boesgaard85, wilson94, steigman07}.

In contrast, heavier elements or metals are produced in higher mass
stars.  The two most common abundance tracers of metallicity are iron
and oxygen since they have bright spectral lines and are a measure of
heavy element production from hydrogen \citep{henry99}.  Iron
absorption lines are typically observed in stars, while oxygen
emission lines are observed in the interstellar medium (ISM).  All
optical diagnostics within the disk are restricted to the local Solar
neighborhood because of extinction from dust.  While far infrared
tracers penetrate the dust, they are significantly weaker than their
optical counterparts.

Early studies of radio recombination line (RRL) and continuum emission
toward \hii\ regions revealed positive Galactic radial gradients of
the derived electron temperatures \citep{churchwell75, churchwell78,
  wink83, shaver83}.  The electron temperature radial gradient was
interpreted as a metallicity gradient since the balance of heating and
cooling within \hii\ regions is sensitive to the abundance of metal
coolants, such as carbon and oxygen. \citet{shaver83} calibrated the
electron temperature-metallicity relation by comparing oxygen
abundances from collisionally excited lines with derived electron
temperatures from RRLs.  Over the last decade, significant
improvements in telescope instrumentation have increased the
sensitivity of RRL observations \citep[e.g.,][]{balser06,
  vonprochazka10} and increased the Galactic \hii\ region sample size
\citep{bania10, anderson11}.  Here we discuss new RRL and continuum
measurements made with the National Radio Astronomy Observatory
(NRAO)\footnote{\nraoblurb} Green Bank Telescope (GBT).

\section{H~{\textbf{\textsc{ii}}} Region Target Sample}\label{sec:sample}

We selected \hii\ regions from the following catalogs: \citet{felli72,
  felli78, shaver83, digel94, lockman89, lockman96, rudolph96,
  rudolph97, brand07, anderson11}.  The \hii\ region sample was
chosen, in part, to compliment the NRAO 140 Foot telescope survey made
by \citet{quireza06a}; hereafter called the ``140 Foot Sample''.  We
specifically selected \hii\ regions at large Galactocentric radius to
study the radial metallicity gradient discontinuity at 10\kpc\
suggested by several studies \citep[e.g.,][]{twarog97}.
\citet{quireza06b} identified possible structure in the azimuthal
distribution of nebular electron temperature so we chose sources that
produced a more uniform spatial distribution between Galactic azimuth
$330$\degree\ and $60$\degree.  Only objects that were well isolated,
so that a continuum zero-level could be measured, and bright enough to
accurately measure the RRL parameters were chosen.  We re-observed 28
objects from \citet{quireza06b} for cross-calibration.

Table~\ref{tab:prop} summarizes the properties of our GBT \hii\ region
sample; hereafter called the ``GBT Sample''.  Listed are the source
name, the Galactic and Equatorial coordinates, the Galactocentric
azimuth, $Az$, and radius, \rgal, the distance from the Sun, \dsun,
the helium abundance ratio by number, $y$, and the LTE electron
temperature, \te.  Details about the derivation of helium abundances
and electron temperatures are given in \S{\ref{sec:te}}.
Figure~\ref{fig:AzRgal} shows both \hii\ regions samples, hereafter
called the ``Green Bank Sample'', projected onto the Galactic plane.
Plotted are the GBT Sample (green points) and the 140 Foot Sample
(blue crosses) as a function of Galactocentric position ($Az$, \rgal).
The GBT Sample includes 44 \hii\ regions at $R_{\rm gal} > 10$\kpc\
compared with only 26 sources for the 140 Foot Sample.

We can determine kinematic distances for all but two of the \hii\
regions in our sample.  Kinematic distances are derived using the
measured source velocity and a rotation curve model for the Galaxy.
For all kinematic distances we use our measured hydrogen RRL
velocities (see Table~\ref{tab:line}) and the \citet{brand86} rotation
curve model.  In the inner Galaxy, rotation curve models have two
possible distances for each measured velocity, a ``near'' and a
``far'' distance.  This problem is known as the ``kinematic distance
ambiguity,'' or KDA, and to resolve it one must use auxiliary data.
For \hii\ regions, \hi\ along the line of sight, which emits at 21\cm,
will absorb against the broadband \hii\ region continuum.  \hi\ is
nearly ubiquitous in our Galaxy and emits at all allowed velocities.
The maximum velocity of detected absorption in an \hi\ spectrum in the
direction of an \hii\ region can distinguish between the near and the
far distance \citep[see][]{kuchar94, fish03, kolpak03, anderson09}.
Using the \hi\ absorption spectrum to resolve the KDA is known as the
\hi\ Emission/Absorption (\hi\,E/A) method.

Our sample contains 22 inner Galaxy \hii\ regions affected by the KDA.
For the seven sources in common with \citet{anderson09}, we use their
KDA resolution.  These sources are: G32.797+0.19 (far), G46.495$-$0.25
(near), G52.75+0.3 (far), K47 (far), NRAO584 (near), W51 (near), and
W43 (far).  There are four inner Galaxy sources in our sample that are
not in \citet{anderson09}.  For these we make an \hi\,E/A measurement
using the 21\cm\ \hi\ VLA Galactic Plane Survey \citep{stil06} and
apply the same methodology as \citet{anderson09}.  For S87, S88, and
S93, we conclude that the near distance is more consistent with the
\hi\ absorption spectrum, whereas for S90 the \hi\ absorption spectrum
supports the far distance.  For two additional inner Galaxy sources
not included in \citet{anderson09}, S76 and W48, there is no public
\hi\ data at sufficient angular resolution to perform the \hi\,E/A
method.  For these two sources, we assume the KDA resolution given in
\citet{quireza06b} and place both sources at the near kinematic
distance.  One additional source, G39.728$-$0.396, is part of the GBT
\hii\ Region Discovery Survey \citep{bania10, anderson11}, and we use
the far kinematic distance from L.\,Anderson et al. (2011, in
preparation).  There are seven \hii\ regions whose RRL velocity is
within 10\kms\ of the tangent point velocity, which is the maximum
velocity along the line of sight: G78.03+0.6, G79.42+2.4, G80.88+0.4,
G81.25+1.1, S106, S108, and S112.  The KDA resolutions for such
sources are unreliable and we therefore locate them at the tangent
point distance.  The remaining 62 sources are in the outer Galaxy and
are thus not affected by the KDA.

There are two sources with kinematic distances that are clearly
incorrect.  For Orion~A (``OriA'' in
Tables~\ref{tab:prop}-\ref{tab:cont}), the measured RRL velocity of
$-2.25$\kms\ is not defined for its Galactic location in the
\citet{brand86} rotation curve.  We instead use a distance of 0.4\kpc\
\citep{menten07}, which was derived using VLBI maser parallax
measurements.  S235 is located near the Galactic anti-center and the
measured velocity of $-25.61$\kms\ produces an unreliable kinematic
distance using the \citet{brand86} rotation curve.  We instead use a
distance of 1.6\kpc\ for this source \citep{fich89, blitz82}, as in
\citet{bania97}.

We choose to use kinematic distances here since we are able to derive
kinematic distances to almost all of our sources, thus providing a
homogeneous sample for our electron temperature analysis.  The radial
gradients, moreover, do not suffer from the KDA since they only depend
on \rgal.  Reliably detecting azimuthal structure, however, does
require \dsun\ and that we resolve the KDA.  In future work we shall
make further analyzes of the Galactic distribution of nebular \te\ by
including spectroscopic and trigonometric parallaxes, especially those
made using VLBI measurements of \hii\ region masers.  It is the case,
however, that the $\sim 20$\% errors typically found for kinematic
distances are much smaller than the separation between a source's near
and far kinematic distance.  We should thus have sufficient accuracy
in the distances that we use here to resolve azimuthal structure in
the \te\ distribution on scales of approximately a few kpc.

\section{Observations}\label{sec:obs}

Here we discuss RRL and continuum observations at X-band (8 to 10\ghz)
toward 81 Galactic \hii\ regions with the GBT.  The spectral line data
consist of total power, position switched spectra in which a reference
position (Off) was observed offset $\sim 6$ minutes in right ascension
from the source and then the target position (On) was observed.  The
reference and target positions were observed for 6 minutes each for a
total of 12 minutes.  The GBT auto-correlation spectrometer (ACS) was
configured with 8 spectral windows, hereafter sub-bands, and two
orthogonal, circular polarizations for a total of 16 independent
spectra.  Each sub-band had a bandwidth of 50\mhz\ and 4096 channels,
yielding a frequency resolution of 12.2\khz.  These spectra included 7
Hn$\alpha$ RRLs (\hy87\ to \hy93) that span the range of the X-band
receiver with half-power beam-widths (HPBWs) from 73 to 90
arcsec.\footnote{Our configuration is identical to \citet{balser06}
  except we replaced the sub-band centered at the \hy86\ line, which
  is confused by higher order RRL transitions, with a sub-band
  centered at 8665.3\mhz.  This sub-band includes the H114$\beta$ and
  H130$\gamma$ RRLs that are used to monitor system performance.}  The
center rest frequencies were: 8045.60495992323, 8300.0,
8584.82315062037, 8665.3, 8877.0, 9183.0, 9505.0, and 9812.0 MHz.

We measured the \hii\ region continuum emission with the GBT Digital
Continuum Receiver (DCR) by scanning in R.A. and Decl. through the
center of the source at a frequency of 8665\mhz\ and a bandwidth of
320\mhz.  The catalog target coordinates were updated based on GBT
pointing observations in R.A. and Decl.---that is, we peaked up on
each \hii\ region.  Each continuum observation consisted of 4 scans,
whereby the telescope was driven at a rate of 80 arcmin per minute in
both the forward and backward directions in R.A. and then in Decl.
Data were simultaneously taken at two orthogonal, circular
polarizations and were sampled every 0.1\s\ over an angular path 40
arcmin in length.

Details of the data reduction and calibration are discussed in
Appendix~\ref{sec:app}.  To increase the signal-to-noise ratio the 7
Hn$\alpha$ RRLs were averaged.  In the final averaged RRL spectra we
excluded the H90$\alpha$ RRL because nearby, higher order RRLs made it
difficult to measure a good spectral baseline.  The averaging process
required three steps: (1) the velocity scale of the \hy88\ to \hy93\ RRLs
were re-gridded to the velocity scale of the \hy87\ RRL; (2) the RRL
spectra were shifted to match the Doppler tracked \hy89\ RRL; and (3)
the intensity scale was adjusted to correct for the different HPBWs by
assuming Gaussian source brightness distributions and HPBWs.  The
primary goal here is to calculate the nebular electron temperature
which is a function of the line-to-continuum ratio.  Since the
electron temperature is derived from this ratio of two intensities,
many observing errors cancel and therefore a relative flux scale
calibration is sufficient.  We used all target sources with a
continuum antenna temperature $> 5$\K\ to derive an average relative
calibration accurate to within 5\%.

The data were analyzed using TMBIDL, an IDL single-dish software
package.\footnote{See
  http://www.bu.edu/iar/files/script-files/research/dapsdr/index.html.}
The line and continuum parameters were determined after the data
reduction and calibration.  For each spectral average we typically
modeled the baseline with a third-order polynomial function that was
subtracted from the data.  The hydrogen, helium, and carbon RRL
profiles were fit with a Gaussian function using a Levenberg-Marquardt
least-squares method \citep{markwardt09} to derive the peak intensity,
the full-width half-maximum (FWHM) line width, and the LSR velocity.
The continuum data were analyzed in a similar way.  A polynomial
function was used to model the background signal and to measure the
zero level.  The background emission consists of the non-thermal
Galactic background, the atmosphere, and receiver noise.  The R.A. and
Decl. scans were independently modeled with Gaussian functions to
derive the peak continuum intensity and the angular FWHM source size
in each direction.

\section{Results}\label{sec:results}

Table~\ref{tab:line} lists the Gaussian parameters of the H, He, and C
RRLs together with their associated errors.  These include the peak
intensity, $T_{\rm L}$, the FWHM line width, $\Delta{V}$, and the
LSR\footnote{ The RRL velocities here are in the kinematic local
  standard of rest (LSR) frame using the radio definition of the
  Doppler shift. The kinematic LSR is defined by a solar motion of
  20.0\kms\ toward ($\alpha$, $\delta$) = ($18^{\rm h}$, $+30$\degree)
  [1900.0] \citep{gordon76}.}  velocity, $V_{\rm LSR}$.  Also shown
are the total integration time, $t_{\rm intg}$, and the
root-mean-square noise of the line free region, $rms$.  The
integration time listed in Table~\ref{tab:line} is derived from the
spectral average of both polarizations of the H87$\alpha$,
H88$\alpha$, H89$\alpha$, H91$\alpha$, H92$\alpha$, and H93$\alpha$
transitions.  Helium and carbon RRLs were detected in 66\% and 33\% of
the \hii\ region targets, respectively.  Sample RRL spectra that range
from our best to worse quality data are shown in
Figure~\ref{fig:spectra}.  The antenna temperature is plotted as a
function of LSR velocity.  The velocity scale is referenced with
respect to the H89$\alpha$ RRL.

Table~\ref{tab:cont} lists the radio continuum Gaussian parameters for
the peak intensity, $T_{\rm C}$, and FWHM angular width, $\Theta$, for
the R.A. scans and the Decl. scans, together with the average source
properties.  We use the arithmetic mean for the average peak intensity
and the geometric mean for the average FWHM width.  Sample continuum
scans are shown in Figure~\ref{fig:cont}.  The antenna temperature is
plotted as a function of the offset sky position relative to the
target coordinates listed in Table~\ref{tab:prop}.  Both the R.A. and
Decl. scans are displayed for the Figure~\ref{fig:spectra} \hii\
regions.

Following \citet{quireza06a} we assign a letter grade, called the
quality factor or QF, to the line and continuum data for each source.
The QF is a measure of data quality taking into account the
signal-to-noise ratio, $snr$, the baselines, confusion, and the
Gaussian fit.  The errors in the Gaussian fit and the $rms$ baseline
noise yield a quantitative measure of the data quality, but these
metrics do not assess systematic uncertainties.  For example, a poor
fit of the baseline can result in significant errors in the peak
intensity that are not reflected by the formal errors.  For the
continuum data the QF was determined by visual inspection of the data,
including the baseline and Gaussian fits as is described in
\citet{quireza06a}.  The QF scale ranges from A (excellent) to D (poor).

Visual inspection of the spectral line data revealed little variation
in the quality of the spectral baselines---the spectral baselines were
excellent.  Unlike the 140 Foot telescope spectra, the unblocked
aperture of the GBT yields significantly improved spectral baselines.
Moreover, we selected \hii\ region targets with a single component H
RRL so there was no line confusion.  For these reasons we derived the
QF for the H RRLs quantitatively using the $snr$ and the percent error
in the Gaussian fit of the line area, $\epsilon_{\rm a} = 100\,(\sigma
T_{\rm L}\,\sigma \Delta V)/(T_{\rm L}\,\Delta V)$.  We assigned the
spectral line QF in the following way: A ($snr \ge 50$, $\epsilon_{\rm
  a} \le\ 1.0$); B ($50 > snr \ge\ 20$, $1.0 < \epsilon_{\rm a} \le\
2.0$); C ($20 > snr \ge\ 7.5$, $2.0 < \epsilon_{\rm a} \le\ 5.0$); or
D otherwise. The QF's are listed in the last column of
Tables~\ref{tab:line} and \ref{tab:cont}.  The QF distribution for the
RRLs is: 22 A, 25 B, 31 C, 3 D.  For the continuum it is: 19 A, 38 B,
18 C, and 6 D.  Thus 96\% of the RRLs have QFs of C or better whereas
93\% of the continuum scans have QFs of C or better.

\subsection{Electron Temperature}\label{sec:te}

Following \citet{quireza06b} we derived the LTE electron temperature using
the equation:
\begin{equation}\label{eq:te}
{\left(T_{\rm e} \over K\right)} = \left[7103.3 {\left(\nu_{\rm L}
\over {\rm GHz}\right)}^{1.1} 
\left({T_{\rm C} \over\strut T_{\rm L} ({\rm H{^+}})} \right)
{\left(\Delta V ({\rm H{^+}}) \over \kms \right)}^{-1} 
\left( 1 + {y}\right)^{-1} \right]^{0.87}
\end{equation}
where
\begin{equation}\label{eq:y}
y \equiv {n({^4}{\rm He}{^+}) \over n({\rm H}{^+})} = {{T_L ({\rm
{^4}He{^+}})\, \Delta V ({\rm {^4}He{^+}})}\over{T_L ({\rm H{^+}})\,
\Delta V ({\rm H{^+}})}},
\end{equation}
and $\nu_{\rm L}$ is the H RRL rest frequency, $T_{\rm C}$ is the
continuum peak intensity, $T_{\rm L}$ is the H or He RRL intensity,
and $\Delta V$ is the H or He RRL FWHM line width.  We used $\nu_{\rm
  L} = 9$\ghz, the average H RRL frequency.  The values of $T_{\rm e}$
and $y$ are listed in Table~\ref{tab:prop} for each \hii\ region.  The
uncertainties were calculated by propagating the Gaussian fitting
errors through Equation 1 for the line and continuum observations.  As
discussed by \citet{quireza06b}, the LTE electron temperature should
be a very good approximation to the real average nebular electron
temperature.  Non-LTE effects, such as stimulated emission and
pressure broadening from electron impacts should be small.

Helium RRL emission was not detected in 27 of the 81 \hii\ regions.
In these cases we estimated an upper limit for $y$ by assuming $T_{\rm
  L}({\rm {^4}He{^+}}) = 3*rms$ and $\Delta V ({\rm {^4}He{^+}}) =
0.75\, \Delta V ({\rm H{^+}})$.  This is consistent with gas at \te\ =
8000\K\ and a hydrogen FWHM line width of 25\kms.  In all cases the
upper limit is larger than the canonical value of the He/H abundance
ratio \citep[0.08; see][]{quireza06b}, indicating that we lack the
sensitivity to detect helium emission in these objects and not that
these \hii\ regions have low excitation.  For sources with no detected
helium emission we therefore assume $y = 0.08$ and list them with zero
errors in Table~\ref{tab:prop}.  Nonetheless, the electron temperature
is not very sensitive to $y$.  Using Equation~\ref{eq:te} and
parameter values typical of our \hii\ region sample, a 10\% change in
$y$ results in only a 0.6\% change in $T_{\rm e}$.

Figure~\ref{fig:cross-cal} compares the electron temperatures derived
for nebulae that are in both the GBT and 140 Foot samples.  We only
plot sources that have a QF of C or better for both line and continuum
data.  For these 16 objects the mean value of the electron temperature
is $8813 \pm 1114$\K\ and $8844 \pm 1084$\K\ for the GBT Sample and
the 140 Foot Sample, respectively.  While there is no systematic
variation in electron temperature there are differences in $T_{\rm e}$
that are significantly larger than the formal error.  The error bars
shown in Figure~\ref{fig:cross-cal} correspond to the measurement
errors and do not include calibration errors.  We estimate that errors
in $T_{\rm e}$ due to calibration are less than 5\% for the GBT (see
Appendix B) and less than 10\% for the 140 Foot telescope (see \S{5.1}
in \citet{quireza06b}).  Since the telescope HPBW's vary by a factor
of 2.4, moreover, we are sampling different volumes of nebular gas and
so cannot rule out true fluctuations in $T_{\rm e}$ within each nebula
that can vary by as much as 10\% \citep[e.g.,][]{roelfsema92}.

\subsection{Electron Temperature Radial Gradient}\label{sec:tegrad}

\begin{figure}
\includegraphics[angle=0,scale=0.8]{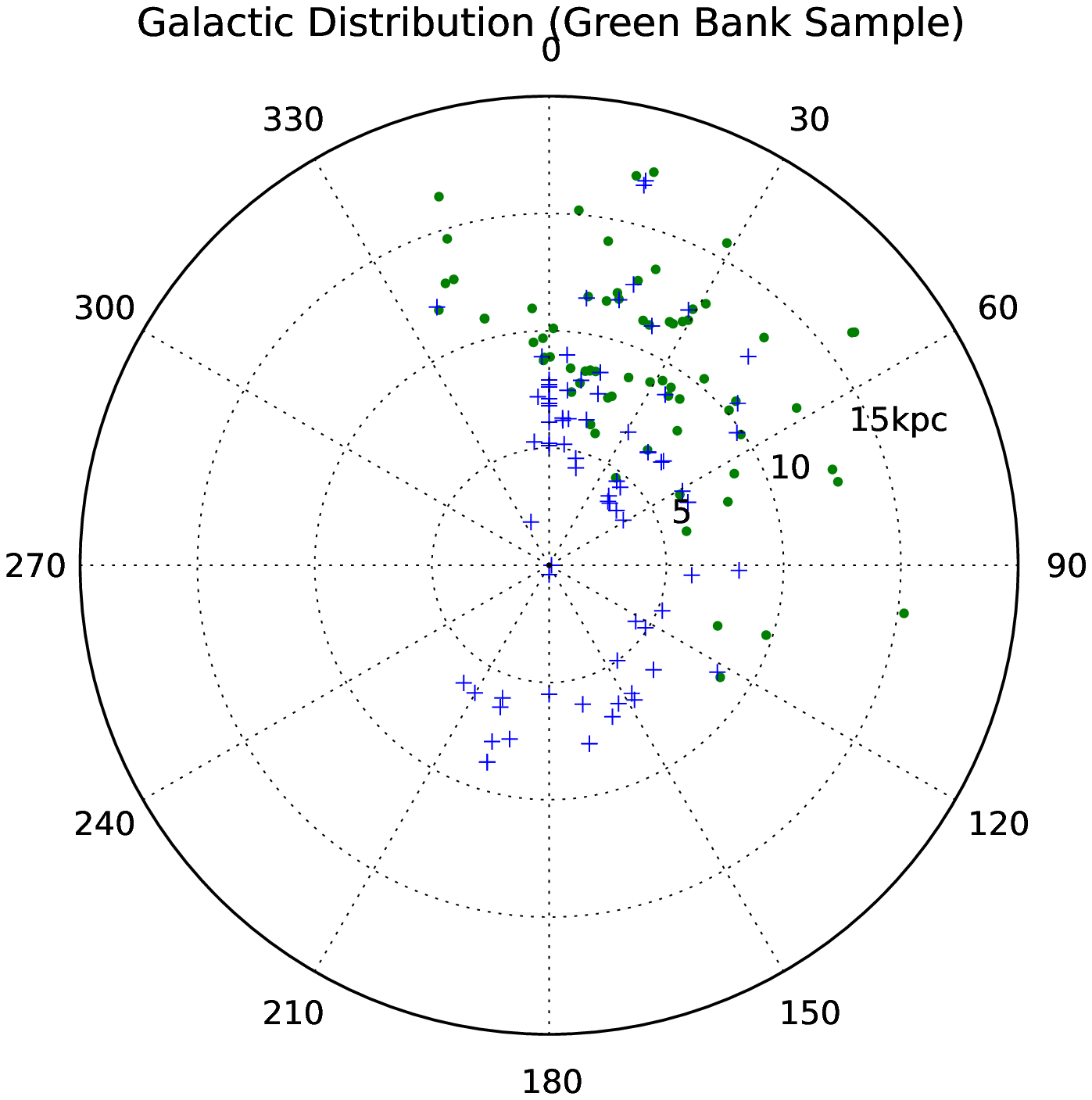} 
\caption{\hii\ region survey Galactic distribution ($Az$, \rgal).
  Shown are the \hii\ regions from the GBT (green points) and 140 Foot
  telescope (blue crosses) surveys.  Only quality factor values of C
  and better are included (see \S{\ref{sec:results}}).}
\label{fig:AzRgal}
\end{figure}

\begin{figure}
\includegraphics[angle=90,scale=0.30]{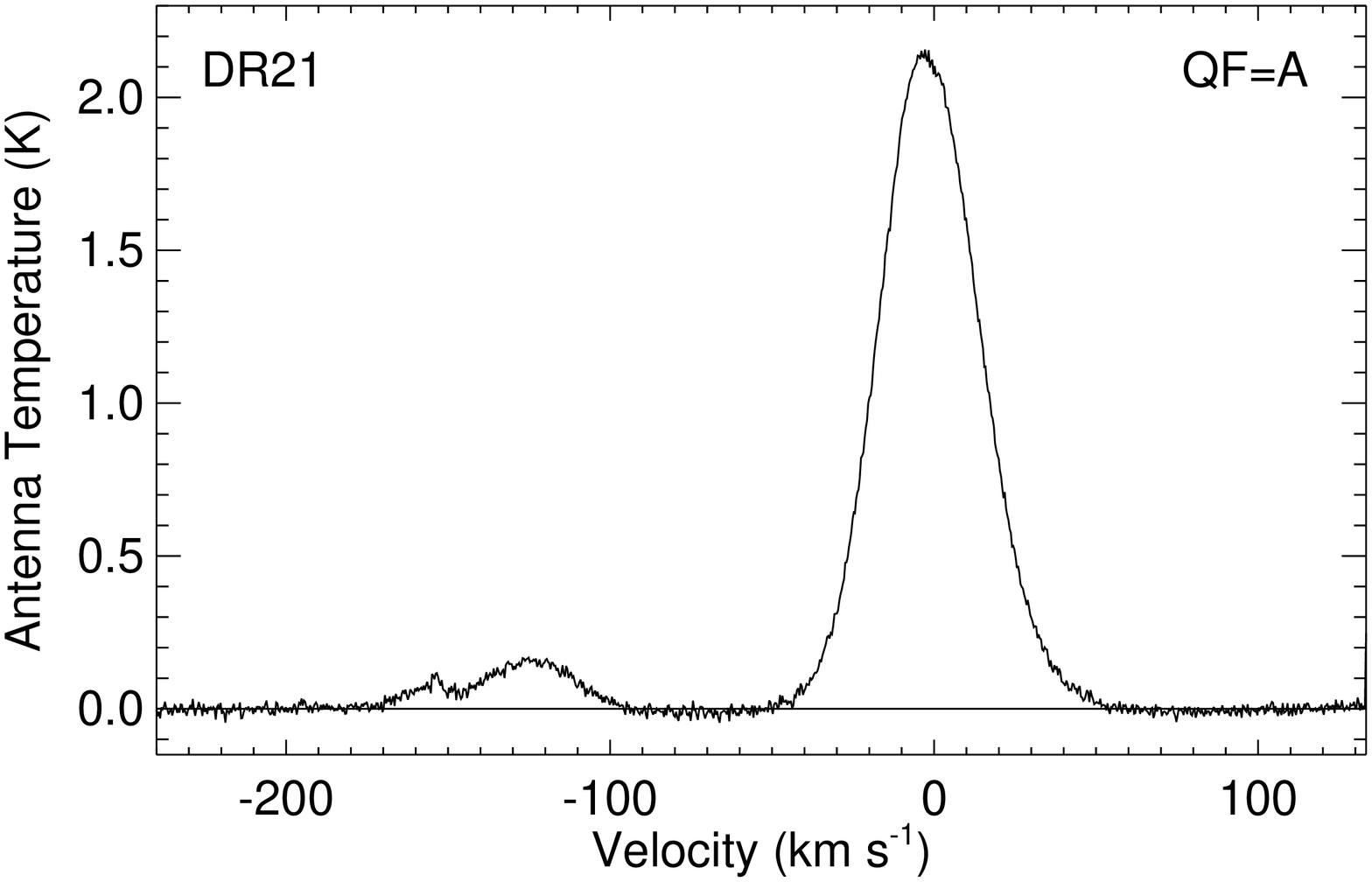} 
\includegraphics[angle=90,scale=0.30]{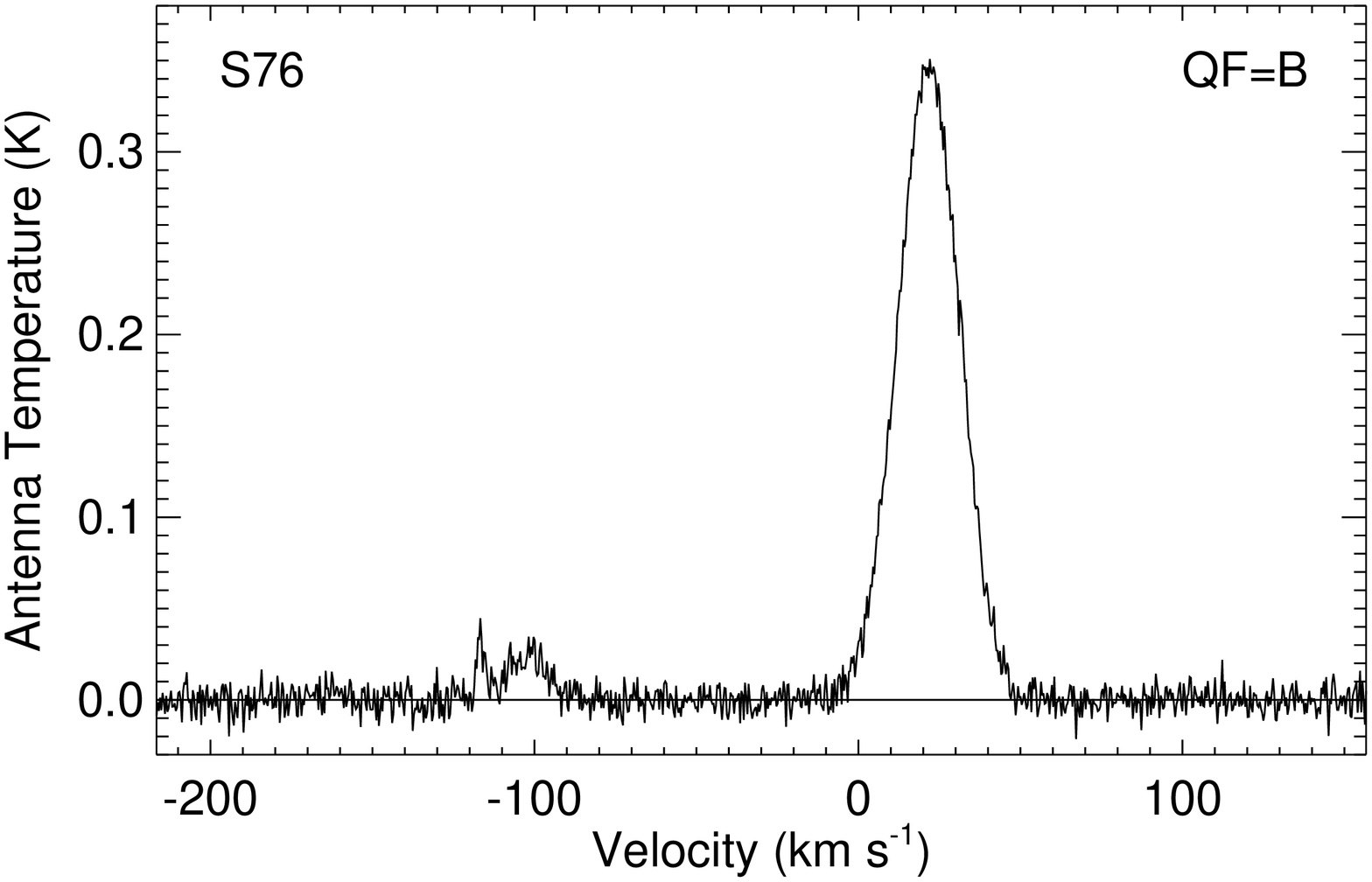}
\includegraphics[angle=90,scale=0.30]{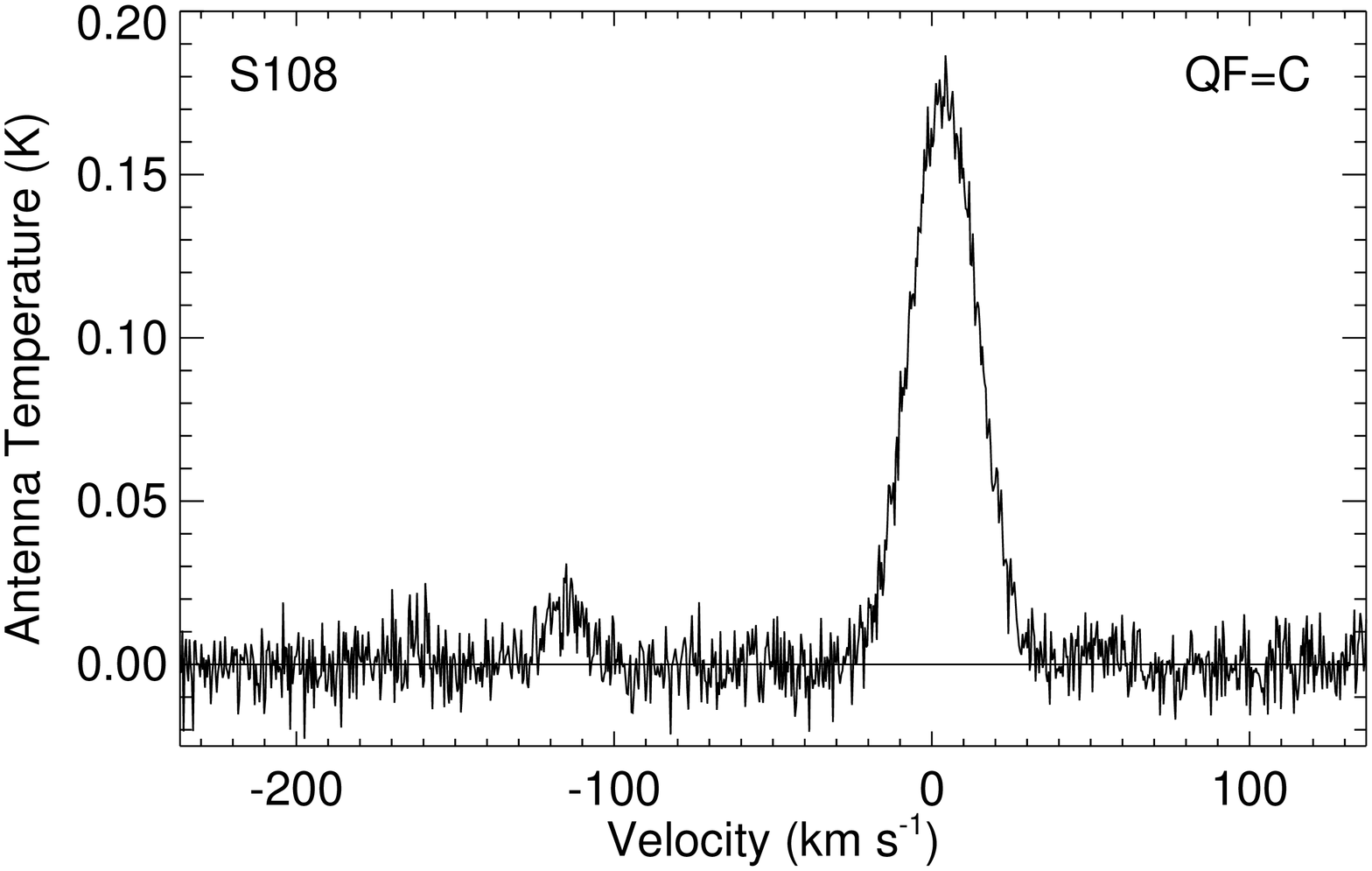}
\includegraphics[angle=90,scale=0.30]{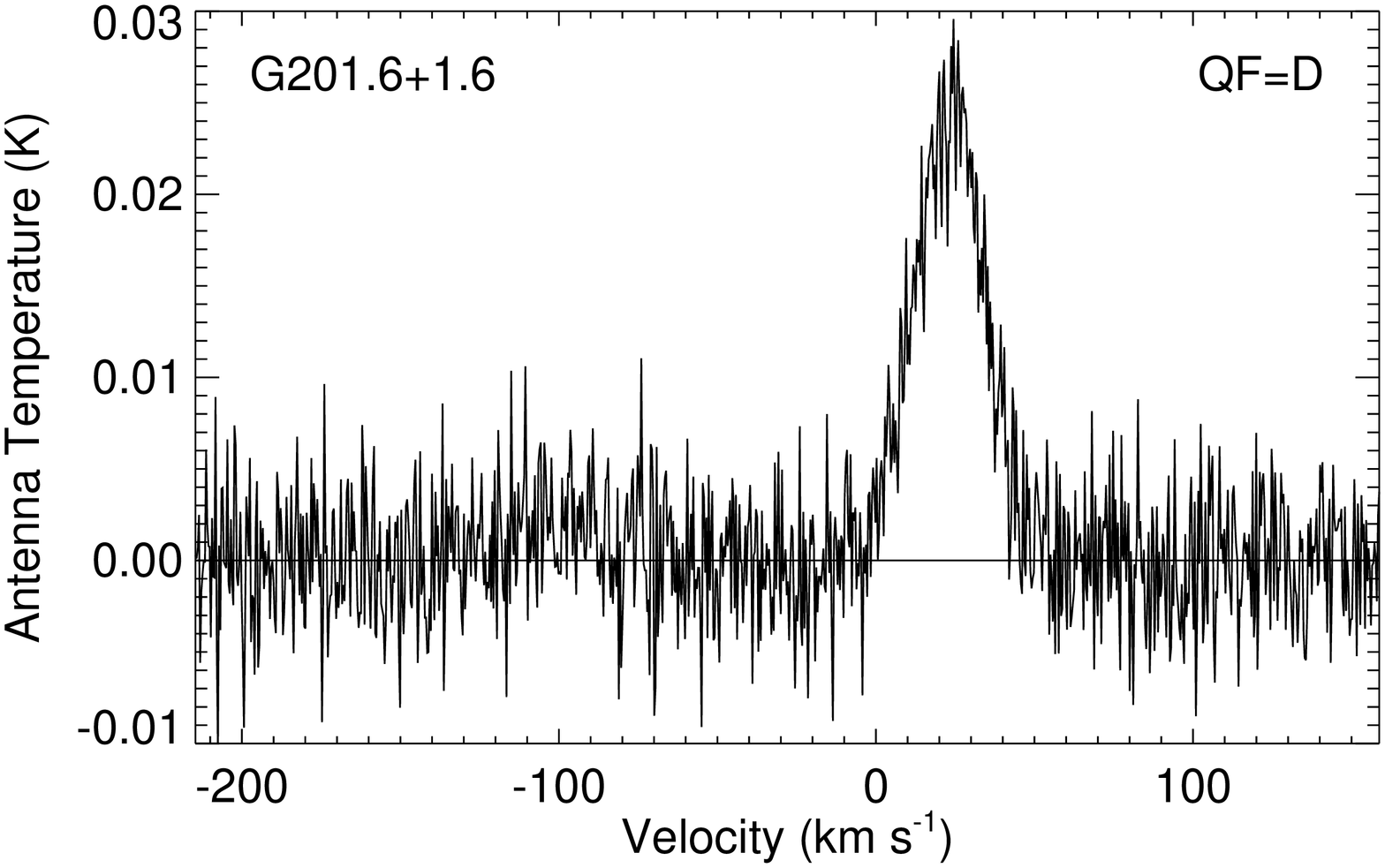}
\caption{Sample \hii\ region RRL spectra.  The antenna temperature is
  plotted as a function of the LSR velocity.  The LSR velocity is
  referenced with respect to the H89$\alpha$ RRL.  For some sources
  the He and C RRLs, located about $-125$\kms\ from the H line, were
  detected.  A third or forth order polynomial function was fitted to
  the line free regions and removed from the data.  The quality factor
  is shown at the top right-hand corner of each plot, where QF=A is
  excellent and QF=D is poor.}
\label{fig:spectra}
\end{figure}

\begin{figure}
\includegraphics[angle=0,scale=0.70]{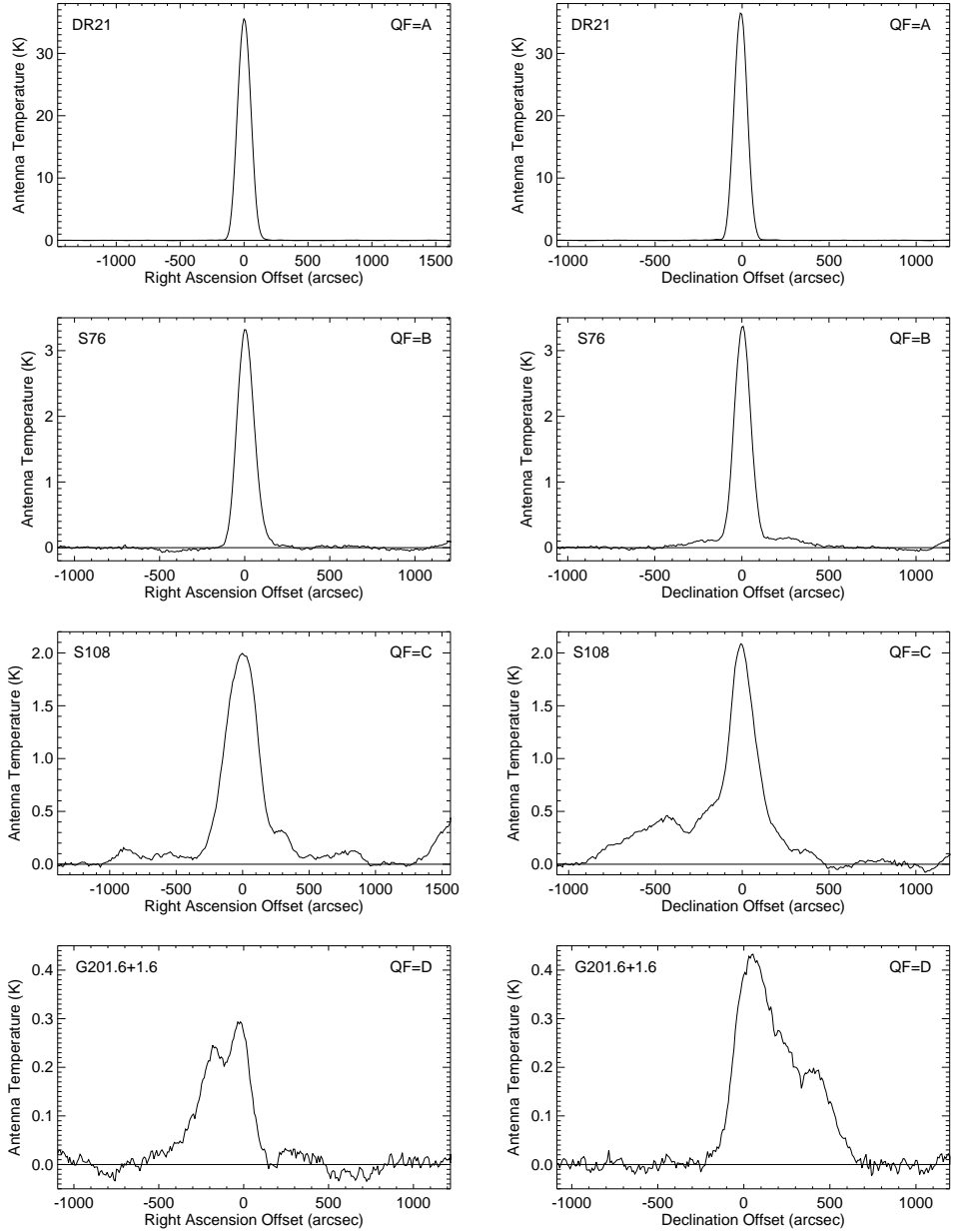} 
\caption{Continuum scans for the Figure~\ref{fig:spectra} \hii\
  regions.  The antenna temperature is plotted as a function of the
  offset position relative to the target coordinates listed in
  Table~\ref{tab:prop}.  For each source the R.A. and Decl. scans are
  shown.  A third or fourth order polynomial function was fitted to
  the baseline and removed from the data.  The quality factor is shown
  at the top right-hand corner of each plot, where QF=A is excellent
  and QF=D is poor.}
\label{fig:cont}
\end{figure}

\begin{figure}
\includegraphics[angle=0,scale=0.8]{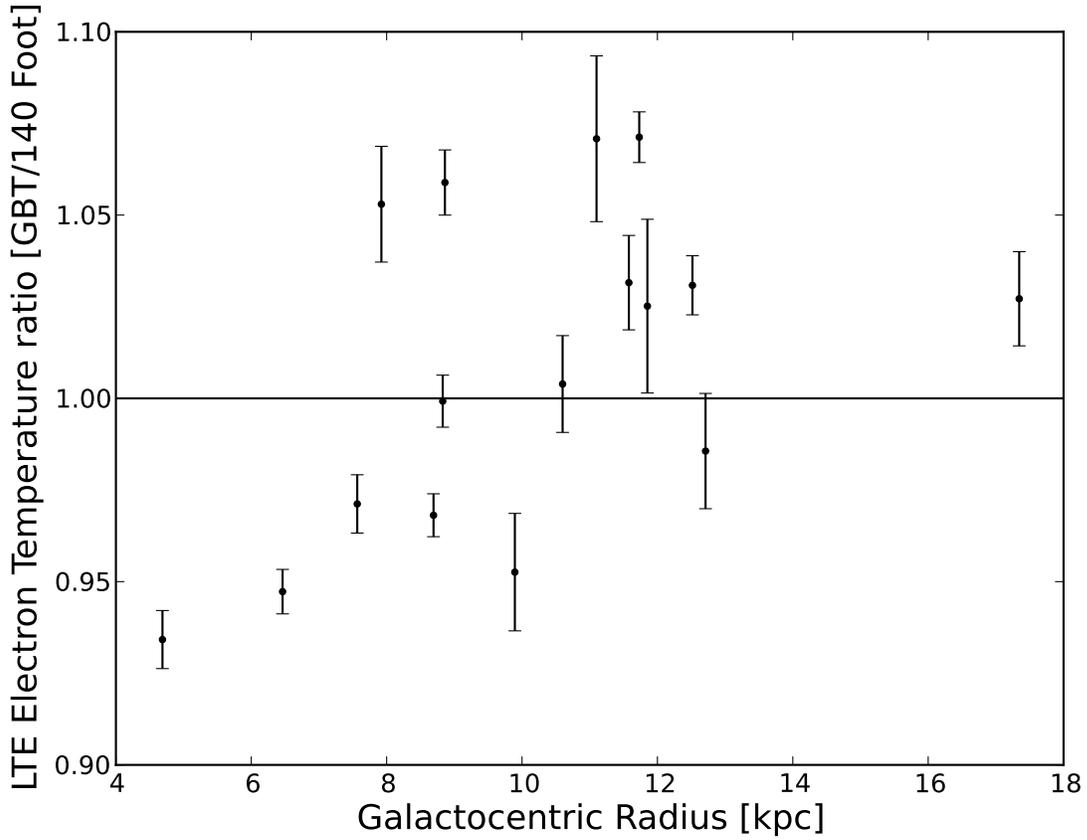} 
\caption{LTE electron temperature ratio (GBT/140 Foot) for sources in
  common between the GBT and 140 Foot telescope samples.  Only quality
  factor values of C and better for both line and continuum data are
  included.  A least-squares fit to the data ($y = ax + b$) yields: $a
  = 0.932 \pm\ 0.045$ and $b = 0.0075 \pm\ 0.0043$.}
\label{fig:cross-cal}
\end{figure}

\begin{figure}
\includegraphics[angle=0,scale=0.60]{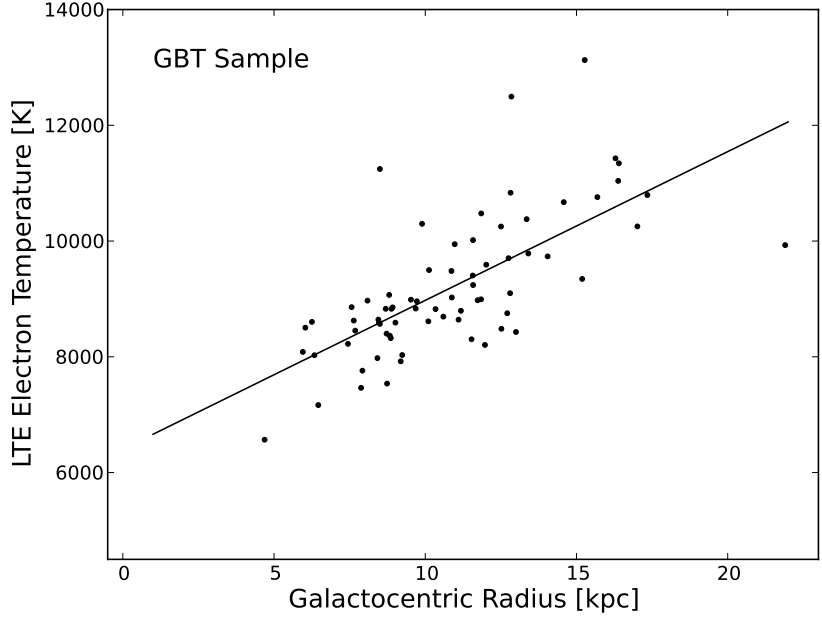} 
\includegraphics[angle=0,scale=0.60]{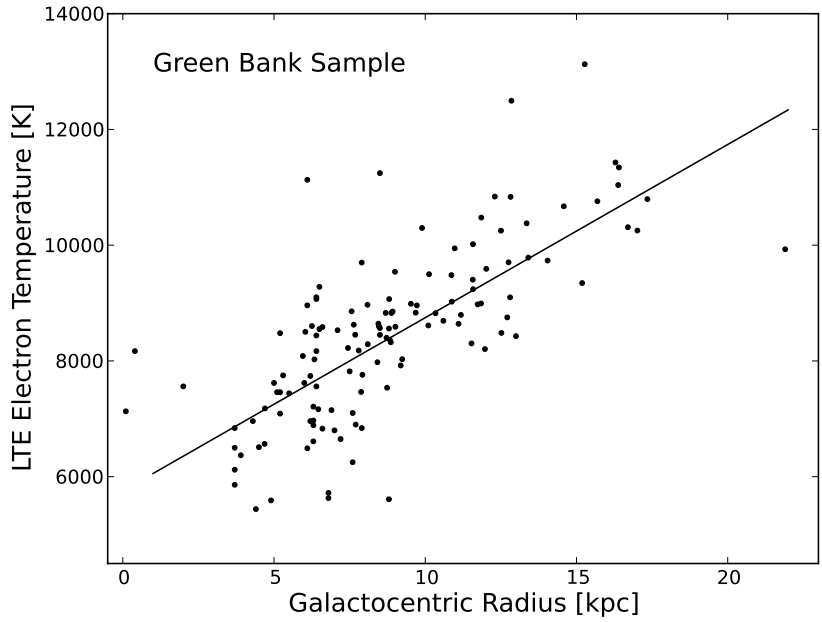} 
\caption{Electron temperature radial gradient for the GBT Sample (top)
  and the Green Bank Sample (bottom).  Only quality factor values of C
  and better for both line and continuum data are included.  The solid
  line is a linear least-squares fit to the data.}
\label{fig:TeRgal}
\end{figure}

The electron temperature radial gradient is plotted in
Figure~\ref{fig:TeRgal} for the GBT Sample (top panel) and the Green
Bank Sample (bottom panel).  Only sources with QFs C or better for
both the line and continuum data are included.  For clarity we do not
display the electron temperature errors which are typically less than
5\%.  As was also found by \citet{quireza06b}, these errors are
significantly smaller than the dispersion of the data points.  For
example, using the GBT Sample the $rms$ of the measured electron
temperatures relative to the fitted value is 854\K, or 9\% of the mean
electron temperature.  The dispersion for the Green Bank Sample is
larger, especially for $R_{\rm gal} < 10$\kpc.  This can be explained
by the presence of azimuthal structure because the azimuth range
sampled by our \hii\ regions with $R_{\rm gal} < 10$\kpc\ is larger
(see Figure~\ref{fig:AzRgal} and below).  The solid lines are
least-squares linear fits to the data.  The program
SLOPES\footnote{See
  http://www.astro.psu.edu/users/edf/research/stat.html.} was used to
perform an ordinary least-squares regression using asymptotic error
formulae \citep{isobe90, feigelson92}.  These uncertainties are valid
for $N > 50$ but underestimate the true error for a smaller sample.
We used the jackknife resampling procedure in SLOPES to derive more
accurate uncertainties for smaller samples, although the slopes are
unchanged.  The electron temperature radial gradient fits are shown in
Tables~\ref{tab:GBTTeFits} and \ref{tab:GBT140TeFits} for the GBT
Sample and the Green Bank Sample, respectively.  Listed are the
azimuth range of the nebulae included in each fit, the fitted
coefficients $a$ and $b$ (where $T_{\rm e} = a + b\,R_{\rm gal}$), the
correlation coefficient, $r$, the number of \hii\ regions in each fit
sample, $N$, and the Galactocentric radius range of the sample
nebulae.  The uncertainties for the correlation coefficient were
derived using the jacknife method discussed by \citet{efron79}.  The
fit for the GBT Sample is consistent with the gradient measured for
Sample B by \citet{quireza06b} using the 140 Foot telescope.  Sample B
includes all \hii\ regions with QF values of C and better.  {\it
  \citet{quireza06b} made an error in their determination of the
  continuum temperature of S209N.}  Here we derive $T_{\rm e} = 10310
\pm 360$\K\ for S209N.  A linear fit to their Sample B now gives: $a =
5856 \pm 340$\K\ and $b = 254 \pm 41\,$K$\,$kpc$^{-1}$.  This
corresponds to a 20\% change in \te\ for S209N and a 10\% change in
the slope.  Since $R_{\rm gal} = 16.7$\kpc\ for S209N it has
significant leverage in the fit for $b$.

\begin{figure}
\includegraphics[angle=0,scale=0.60]{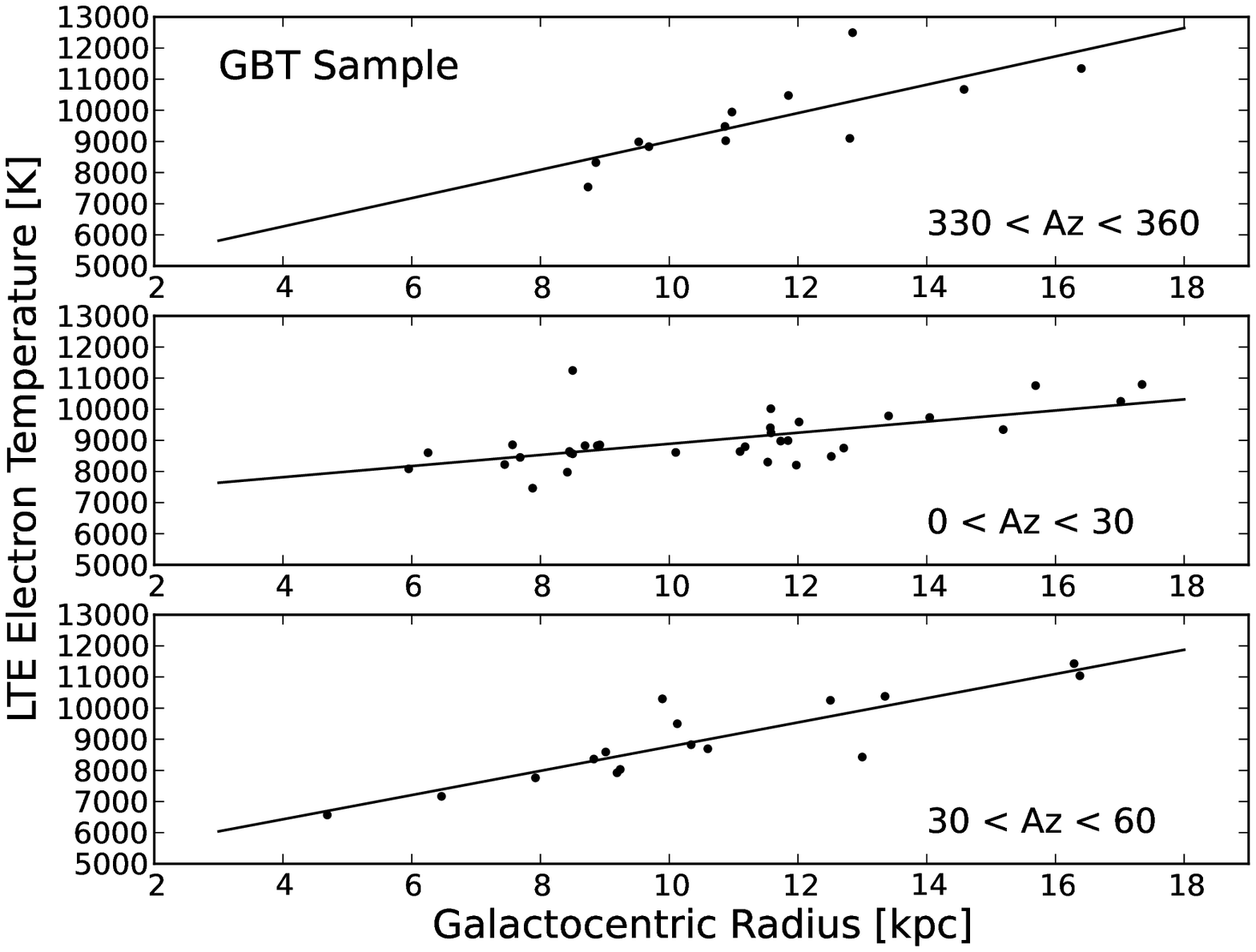} 
\includegraphics[angle=0,scale=0.60]{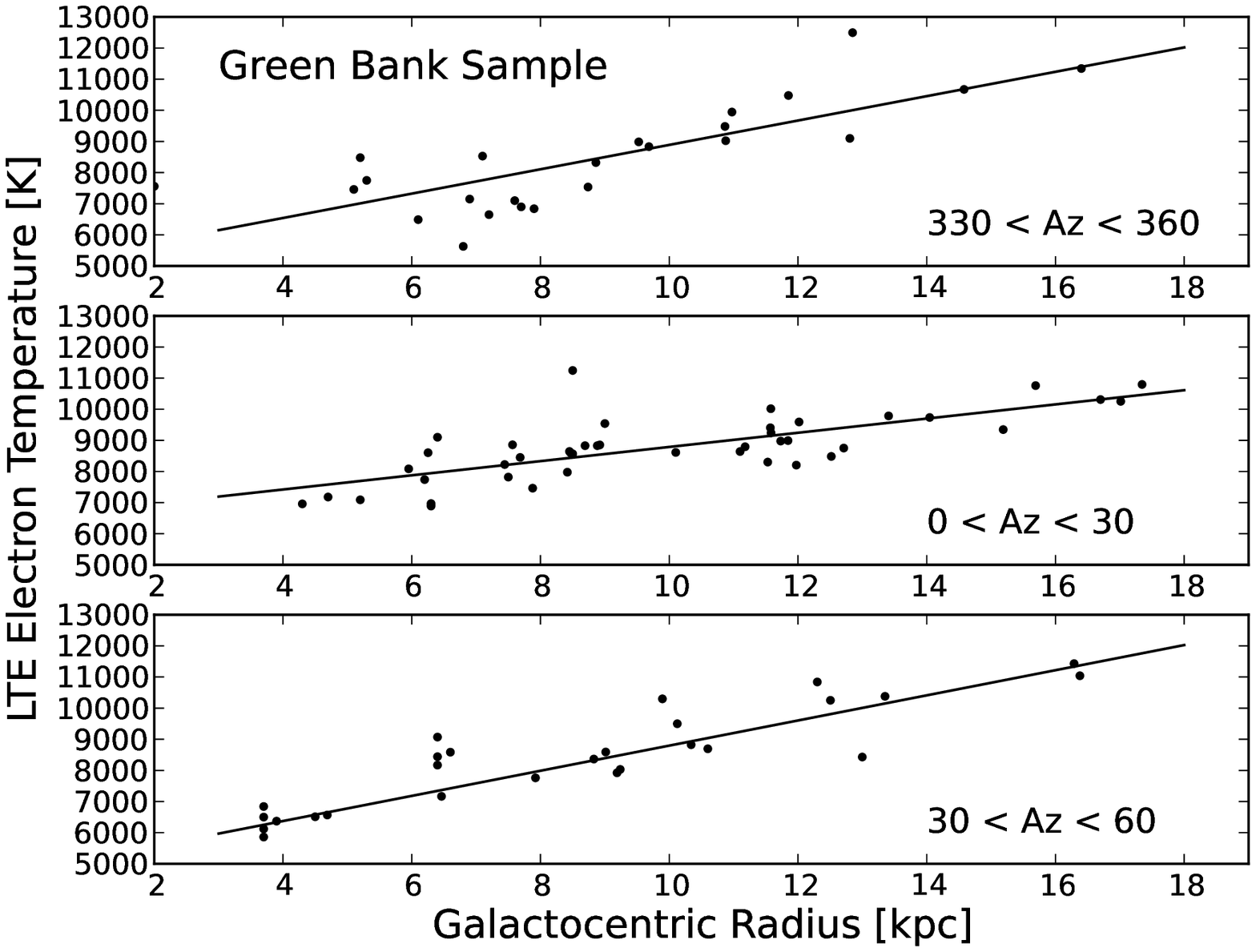}
\caption{Electron temperature radial gradients for the GBT Sample (top)
  and the Green Bank Sample (bottom). Only quality factor values
  of C and better for both line and continuum data are included.  The solid
  lines are linear least-squares fits to the data.  Top panel:
  $330\,^\circ\ < Az < 360\,^\circ$.  Middle panel: $0\,^\circ\ < Az <
  30\,^\circ$. Bottom panel: $30\,^\circ\ < Az < 60\,^\circ$.}
\label{fig:TeRgalC}
\end{figure}

We explored azimuthal structure in the electron temperature gradient
by dividing the data into three azimuth ranges: $330\,^\circ < Az <
360\,^\circ$; $0\,^\circ < Az < 30\,^\circ$; and $30\,^\circ < Az <
60\,^\circ$.  Figure~\ref{fig:TeRgalC} plots the electron temperature
radial gradient for each azimuth range, and Tables~\ref{tab:GBTTeFits}
and \ref{tab:GBT140TeFits} list the linear least-squares fit
parameters.  The data within each azimuth range are well modeled by a
linear fit with correlation coefficients between $0.62-0.88$.  The
gradient varies significantly within each azimuth range with a smaller
slope between azimuth $0-30$\degree.

\begin{figure}
\includegraphics[angle=0,scale=1.3]{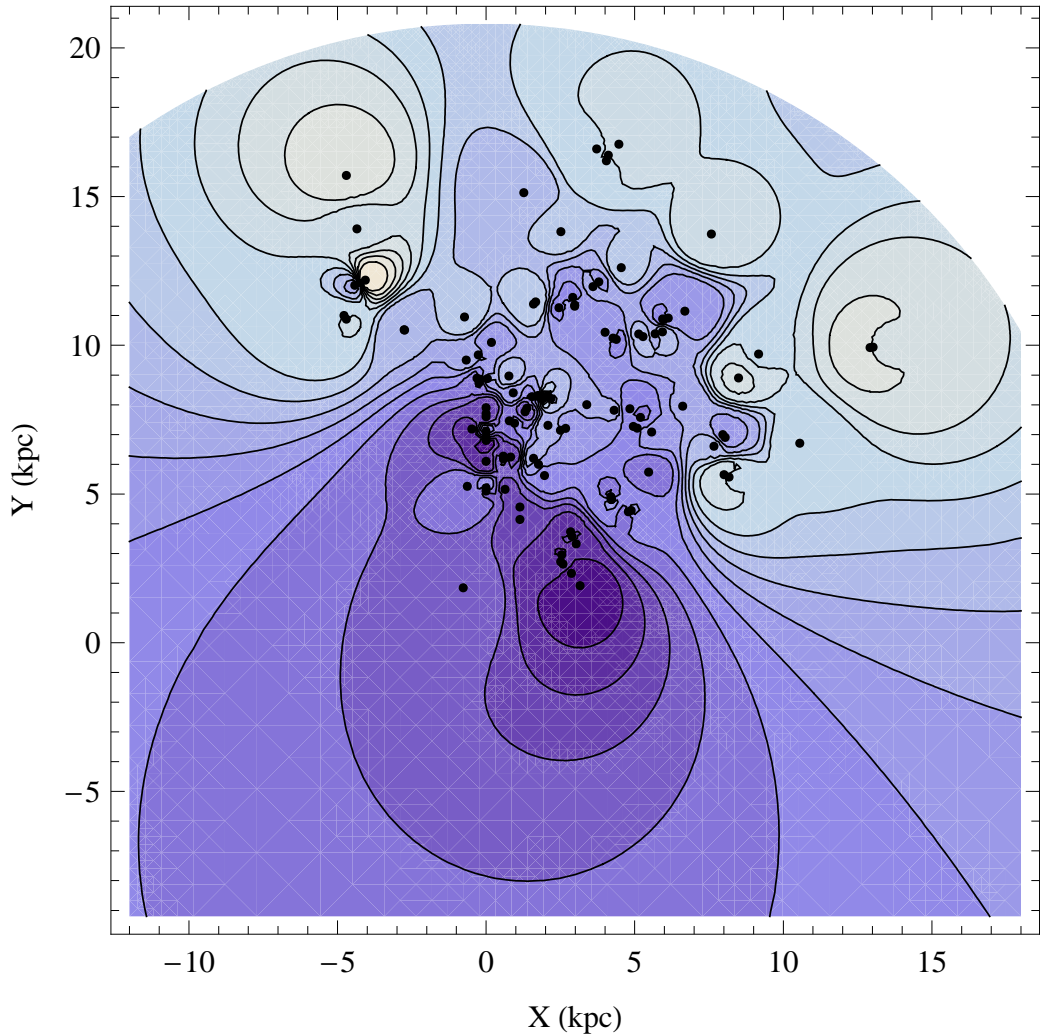}
\caption{Image of the Galactic distribution of nebular electron
  temperatures produced from the discrete \hii\ regions located
  between Galactic azimuth 330\degree\ and 60\degree\ for the Green
  Bank Sample (110 sources).  The image was generated by using
  Shepard's method with $\alpha = 5$ (see text).  The contours range
  between 6400 and 11200\K\ at intervals of 400\K.  The darker shades
  are lower temperatures.  The orientation is the same as in
  Figure~\ref{fig:AzRgal} with the Galactic Center located at ($x = 0,
  y = 0$) and the Sun at 8.5\kpc\ above the Galactic Center at zero
  azimuth.  The points indicate the location of the discrete \hii\
  regions.}
\label{fig:shepard}
\end{figure}

To visualize structure in the electron temperature distribution in the
Galactic plane we employed Shepard's method to interpolate the
electron temperature values over the discrete \hii\ region locations
\citep{shepard68, gordon78}.  \citet{shepard68} developed an algorithm
to interpolate over arbitrarily spaced, discrete data.
\citet{gordon78} introduced a free parameter, $\alpha$, that when
increased flattens the resulting image near the discrete data
locations.  Figure~\ref{fig:shepard} plots the results of this
interpolation when applied to the Green Bank Sample for $\alpha = 5$.
The black points denote the location of the \hii\ regions used in the
interpolation (cf.  Figure~\ref{fig:AzRgal}).  The interpolated image
is represented by the contour map.  The contour levels range between
6400 and 11200\K\ at intervals of 400\K.  The darker shades are lower
temperatures.  We only included data spanning an azimuth range from
330\degree\ to 60\degree.  The radial \te\ gradients are visible at
all Galactocentric azimuths with higher electron temperatures at
larger \rgal.  At a constant Galactocentric radius the
electron temperature is not constant, revealing azimuthal structure
and that the gas is not well mixed in the Galaxy.  Positions in the
Galaxy that are located far from any of our \hii\ regions have an
interpolated electron temperature equal to the average \te\ value in
our sample.

\subsection{O/H Abundance Ratio}\label{sec:o2h}

The metallicity is the dominant factor that controls the \hii\ region
equilibrium electron temperature.  There are at least four physical
properties that effect \hii\ region electron temperatures: the
effective temperature of the ionizing star, the electron density of
the surrounding medium, dust-particle interactions, and the
metallicity.  The stellar effective temperature, \teff, sets the
hardness of the radiation field that excites and heats the gas.
\citet{rubin85} predicts an increase in \te\ of 1300\K\ for a change
in \teff\ from 33,000 to 45,000\K\ (B0 to O5 spectral type).  The
electron density, \Ne, alters the rate of collisional de-excitation.
High values of \Ne\ will inhibit cooling and thus increase \te.
\citet{rubin85} estimates an increase in \te\ of 2900\K\ for a change
in \Ne\ from 100 to \expo{5}\percc.  Dust effects the electron
temperature in complex ways: photo-electric heating occurs as
electrons are ejected from dust grains and collide with atoms, whereas
cooling of the gas can result when there are collisions of fast
particles with dust grains \citep{mathis86, baldwin91, shields95}.
\citet{oliveira86} estimate that the net effect of dust is on the
order of 500\K.  Heavy elements within the ionized gas will increase
the cooling primarily through collisionally excited lines and lower
\te\ \citep{garay83}.  \citet{rubin85} predicts changes in \te\ of
7,000\K\ for a factor of 10 change in metal abundance.  

Given the typical range of values for these physical properties in
Galactic \hii\ regions, the metallicity should thus be the dominant
factor in producing variations in the electron temperature.  We
explored, for example, the effects of density and excitation on the
derived electron temperatures for the GBT Sample.  Assuming that each
\hii\ region is a spherical, homogeneous, optically thin nebula we
derived the electron density using Equation G7 by \citet{balser95}.
(See \citet{quireza06b} for a similar analysis.)  To assess the effect
of excitation we calculated the number of H-ionizing photons emitted
per second, \NL, as a proxy for \teff\ using Equation G9 by
\citet{balser95}. This provides an upper limit to \teff\ since more
than one early-type star may be ionizing the nebula.  We found no
significant correlation of the electron temperature with the electron
density ($r=0.10$) or \NL\ ($r=0.03$), where $r$ is the correlation
coefficient.  This suggests that metallicity is the most important
factor in determining the electron temperature for our sample of \hii\
regions.

Given the homogeneity of our sample, the derived electron temperatures
should provide an excellent proxy for metallicity.  Following
\citet{shaver83}, we divided our \hii\ regions into two groups having
either high or low values of both \NL\ or \Ne.  The threshold values
were ${\rm Log}(\NL) = 49.5$\persec\ and \Ne\ = 150\percc.  Similar to
\citet{wink83} and \citet{quireza06b} we find no systematic trend in
the electron temperature for these two samples.  \citet{shaver83}
found that the high (\Ne, \NL) sample produced higher values of \te\
for a given Galactocentric radius.  But this trend is weak since the
correlation is strongest at small \rgal, where there are
systematically low \te\ values \citep{quireza06b}.  The physical
effect of \Ne\ on the electron temperature has clearly been observed,
however, by comparing electron temperatures derived from single-dish
telescopes with those from interferometers.  \citet{afflerbach96}
observed RRLs with the VLA toward ultracompact \hii\ regions and found
electron temperatures about 1000\K\ hotter than the \citet{shaver83}
single-dish values.  Interferometers are sensitive to compact, high
density gas, whereas single-dish telescopes probe the more extended,
low density gas.

We calculated O/H abundance ratios from our \hii\ region electron
temperatures using the relationship derived by \citet{shaver83}.  This
is based on electron temperatures from RRL and continuum emission and
collisionally excited lines (CELs) of oxygen.  They estimate an $rms$
uncertainty of 0.1 dex in Log(O/H).  \citet{pilyugin03} claim that the
oxygen abundances calculated by \citet{shaver83} are systematically
overestimated by $0.2-0.3$ dex.  They determined O/H abundance ratios
using the P-method, where an excitation parameter is employed to
derive the physical conditions in the \hii\ regions.  Regardless, any
systematic O/H abundance calibration error should not effect gradient
measurements.  The \citet{shaver83} relationship between Log(O/H) and
\te\ is:
\begin{equation}\label{eq:cel}
12 + {\rm Log(O/H)} = (9.82 \pm 0.02) - (1.49 \pm 0.11)\,T_{\rm e}/10^{4}. 
\end{equation}

Here we only use CELs to convert electron temperatures to O/H
abundance ratios.  Some authors have advocated the use of optical
recombination lines (ORLs) since, unlike CELs, deriving the O/H
abundance ratio using ORLs is not a strong function of the electron
temperature or density structure \citep[e.g.,][]{esteban05}.  O/H
abundance ratios derived from ORLs are typically larger than those
determined from CELs.  This trend is observed with several abundance
tracers in both \hii\ regions and planetary nebulae and is called the
``abundance discrepancy problem'' \citep{garcia-rojas07a}.
\citet{garcia-rojas07b} suggest that the problem stems from the
presence of temperature fluctuations, whereas \citet{liu00} argue that
it comes from hydrogen deficient clumps in the nebula.  Recently,
\citet{nahar10} have suggested that the abundance discrepancy problem
is caused by an underestimate of the low-temperature
(\expo{2}$-$\expo{4}\K) dielectronic recombination rates.  Based on
theoretical calculations they argue that resonant features in the
low-energy photoionization cross section of \oii\ must be included.
They are in the process of calculating recombination rate coefficients
into all recombined levels of \oii\ up to n = 10.

Regardless of how accurately we can measure the O/H abundance, this
ratio corresponds to the gas phase abundance and does not account for
depletion of oxygen onto dust grains.  \cite{peimbert10} estimate an
oxygen depletion that varies from about 0.08 dex for the most metal
poor \hii\ regions to 0.12 dex for the most metal rich \hii\ regions
with an uncertainty of about 0.03 dex.  Given the small changes in the
O/H abundance ratio and the large uncertainty we assume oxygen
depletion onto dust grains is negligible.

\begin{figure}
\includegraphics[angle=0,scale=0.60]{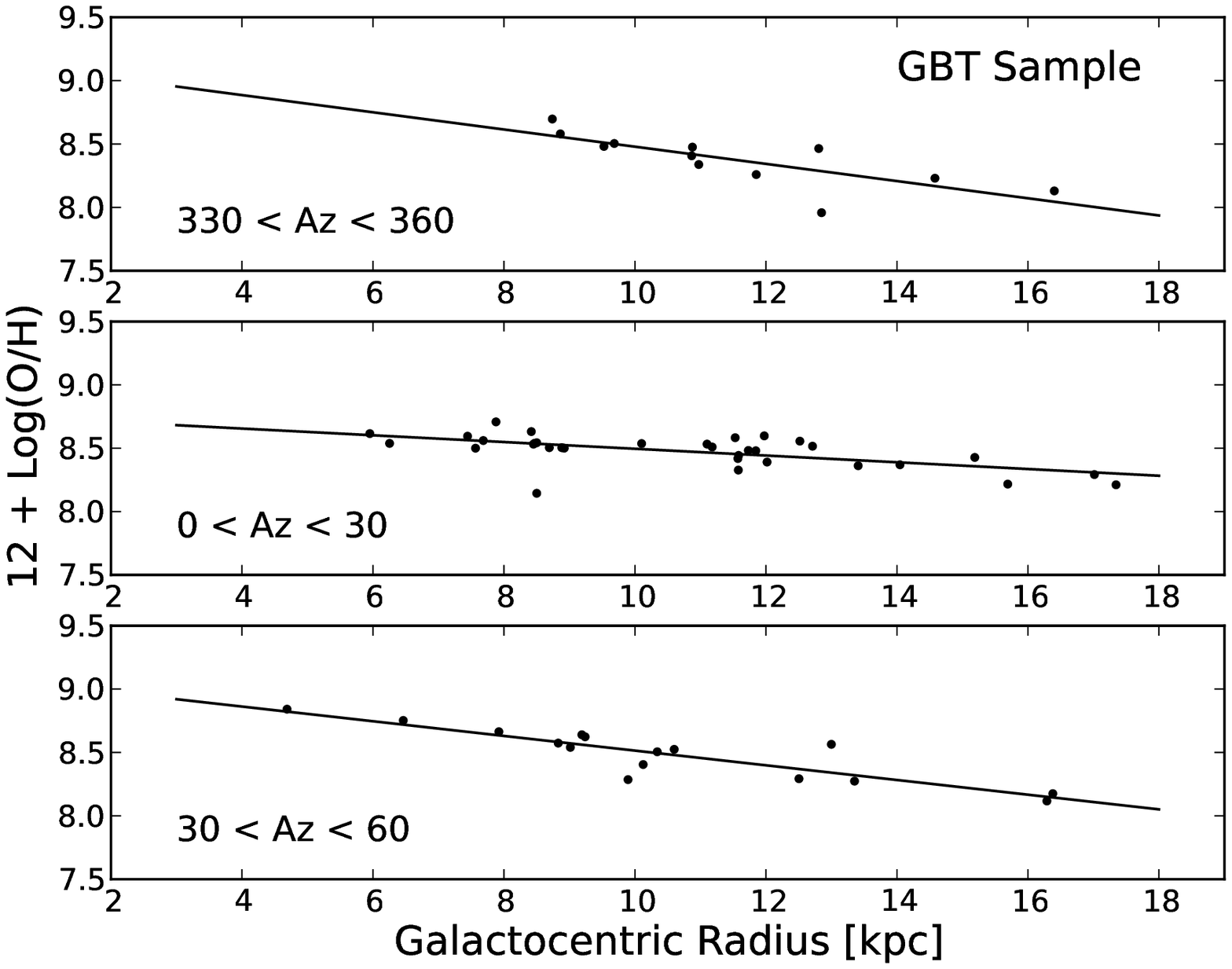} 
\includegraphics[angle=0,scale=0.60]{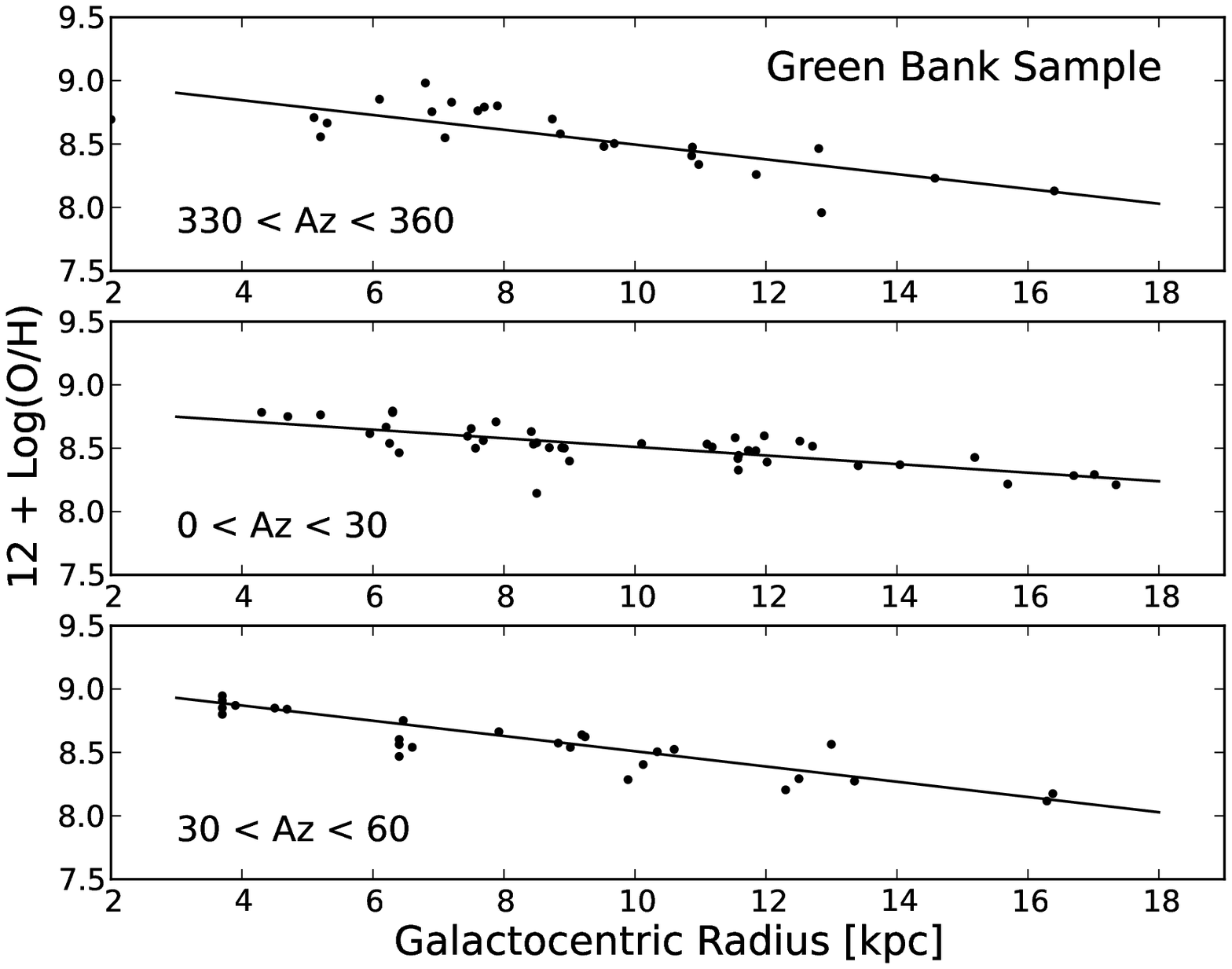} 
\caption{O/H abundance ratio radial gradient for the GBT Sample (top)
  and the Green Bank Sample (bottom). Only quality factor values of C
  and better for both line and continuum data are included.  The solid
  lines are linear least-squares fits to the data.  Top panel:
  $330\,^\circ\ < Az < 360\,^\circ$. Middle panel: $0\,^\circ\ < Az <
  30\,^\circ$. Bottom panel: $30\,^\circ\ < Az < 60\,^\circ$.}
\label{fig:O2HRgalC}
\end{figure}

Our O/H radial gradient fits are summarized in Table~\ref{tab:O2HFits}
and Figure~\ref{fig:O2HRgalC} which plots the O/H abundance ratio
derived from Equation~\ref{eq:cel} as a function of \rgal\ for the GBT
Sample (top) and the Green Bank Sample (bottom).  The data are
divided into the same azimuth ranges as shown in
Figure~\ref{fig:TeRgalC}.  The O/H gradient varies from about $-0.03$
to $-0.07$\dexkpc\ depending on the Galactic azimuth.

\section{Discussion}\label{sec:dis}

Studies of chemical evolution are important because they connect to
many other fields of astronomy, such as cosmology, stellar
nucleosynthesis, and the origin and evolution of galaxies
\citep{king71, vandenbergh75, peimbert75, trimble75, audouze76,
  pagel81, henry99}.  Measuring the spatial and temporal distribution
of abundances is a key constraint to understanding chemical evolution
in galaxies.  Radial gradients were first detected in nearby galaxies
from optical observations of CELs of oxygen, nitrogen, and sulfur in
\hii\ regions \citep{searle71, rubin72, smith75}.  Similar
measurements in the Milky Way disk are significantly more sensitive
but they are hampered by extinction from dust and the difficulty of
determining relative positions within the disk. RRLs provide a unique
probe of metallicity since \hii\ regions can be detected at radio
frequencies throughout the Galactic disk \citep{bania10, anderson11},
and significant progress has been made in deriving accurate distances
\citep{anderson09, reid09}.  Here we summarize previous abundance
measurements in the Galactic disk, discuss our GBT RRL results, and
explore how these data can constrain chemical evolution models.

\subsection{Previous Studies}\label{subsec:previous}

The primary tracers of metals in the Galactic disk are open clusters,
Cepheids, OB stars, planetary nebulae, and \hii\ regions.  Open
clusters cover a wide range of metallicity and age (30\myr\ to
11\gyr); hence they directly probe metallicity with time
\citep{friel95}.  Early studies of abundances in open clusters were
based on photometry with radial metallicity gradients of $-0.05 \pm
0.01$\dexkpc\ \citep{janes79}.  More recent work has focused on
spectroscopic measurements that provide an abundance and a velocity
that can be used to determine cluster membership.  \citet{friel02}
measured a radial gradient of $0.059 \pm 0.010$\dexkpc\ from a sample
of 39 open clusters.  They found evidence for a flattening of the
radial gradient with time: $-0.072 \pm 0.016$\dexkpc\ and $-0.046
\pm 0.012$\dexkpc\ for open clusters older and younger than 3\gyr,
respectively \citep[also see][]{chen03}.  \citet{twarog97} analyzed 76
open clusters using photometry and spectroscopy and suggested that
there is a discontinuity in the radial abundance distribution near
10\kpc.  They associated this break in the radial abundance with the
edge of the initial Galactic disk as defined by the disk globular
clusters or the thick disk.  More recent observations of open clusters
in the outer Galaxy support their findings with a break in the radial
gradient near 10\kpc\ and a flattening in the outer Galaxy
\citep{yong05, sestito08, andreuzzi10}.  The shallow slope in the
outer Galaxy does not appear to be caused by accretion of metal poor
extragalactic material (e.g, dwarf galaxies).  Measurements of the
location, kinematics, and chemistry of outer Galaxy open clusters are
consistent with these objects being part of the Galactic disk
\citep{carraro07}.

Cepheids are bright, evolved stars that are detected at large
distances and are relatively young ($< 200$\myr).  They are easy to
identify and the photometric distances are reasonably accurate.  Many
studies using Cepheids support non-uniform radial gradients
\citep{caputo01, andrievsky02, luck03, andrievsky04, pedicelli09}.
Furthermore, there is evidence of chemical inhomogeneities across the
Galactic quadrants \citep{luck06, lemasle08, pedicelli09}.  Using a
sample of 24 Cepheids spanning $R_{\rm gal}$'s between 12 and
17.2\kpc, \citet{yong06} measured a radial distribution that flattens
at 14\kpc, in contrast to the older population open clusters that
flatten at 10\kpc.  Since the Cepheid population is younger than the
open cluster population this suggests that the Galactic disk has grown
in radius by a few kpc over the past several billion years via
accretion.  Young et al. find a bimodal [Fe/H] distribution in their
outer Galaxy sample with one population, ``Galactic Cepheids'', that
agrees with the extrapolation of the gradient measured at smaller
$R_{\rm gal}$, and another population, ``Merger Cepheids'', with
significantly lower values of [Fe/H].

OB-type stars are young ($< 10\,$Myr) and their abundances trace the
metallicity near their current location.  This assumes that mixing in
the outer layers of OB-type stars is negligible \citep{sofia01,
  daflon04}.  Early studies using OB-type stars found little or no
radial metallicity gradient \citep{fitzsimmons90, kaufer94,
  kilian-montenbruck94}, whereas more recent work shows radial
gradients between $-0.03$ and $-0.07$\dexkpc\ \citep{gummersbach98,
  smartt97, rolleston00, daflon04}.  Using 69 OB stars in 25 clusters
or associations with Galactocentric distances between 4.7 and
13.2\kpc, \citet{daflon04b} claim a metallicity discontinuity
consistent with that found for open clusters and Cepheids.
\citet{rolleston00}, however, see no measurable discontinuity in
their sample of about 80 B-type stars from 19 open clusters with
Galactocentric radii between 6 and 18\kpc.

Planetary nebula (PN) trace metallicity in the Galactic disk, bulge,
and halo.  Their ages span 1 to 8\gyr\ and thus PNe can in principle
probe metallicity with time \citep{maciel03}.  The difficulty is in
separating the different populations and determining accurate
distances.  One approach is to use Type II PNe \citep{peimbert78}
which are disk objects with lower masses that are less likely to be
contaminated by nucleosynthetic products from the progenitor star.
Radial gradients of the O/H abundance for this population range
between $-0.02$ and $-0.06$\dexkpc\ \citep{maciel99, henry04,
  perinotto06}.  \citet{stanghellini06} selected PNe with either
elliptical or bipolar core morphologies since they are uniformly
distributed in the disk and have more reliable distances.  They
derived essentially flat gradients.  \citet{pottasch06} observed both
IR and optical transitions to better calculate the ionization
correction and therefore derive more accurate abundances.  They found
a higher radial gradient of $-0.085$\dexkpc, although their sample
size is only 26 PNe with limited Galactocentric range (3.5 to 9.8\kpc).
\citet{maciel03} developed a method to determine the age of the PNe
central star and reported a flattening of the O/H radial gradient from
$-0.11$\dexkpc\ to $-0.06$\dexkpc\ during the last 9\gyr.
\citet{stanghellini10}, however, found the opposite trend: the radial
abundance gradients are steepening with time. They used the most
up-to-date abundances and distances for 147 PNe, and the Peimbert PNe
classification to discriminate age.  \citet{henry10} analyzed 124 PNe
and derived a radial O/H gradient of $-0.058 \pm 0.006$\dexkpc\
between Galactocentric radii 0.9 to 21\kpc\ with no variation in time.
They concluded ``From our complete exercise we consider it very likely
that the true slope is within the range $-0.04$ to $-0.06$\dexkpc, but
we cannot refine the number beyond that point.''

Galactic \hii\ regions are the formation sites of massive OB stars and
they reveal the locations of current Galactic star formation.  Their
chemical abundances indicate the present state of the interstellar
medium and reveal the elemental enrichment caused by the nuclear
processing of many stellar generations.  Collisionally excited lines
are bright in \hii\ regions and provide a sensitive measure of
metallicity at optical wavelengths \citep{pagel97}.  Early studies of
O/H radial gradients from optical CELs measured radial gradients from
$-0.04$ to $-0.13$\dexkpc\ using \hii\ regions with Galactocentric
radii of 8 to 14\kpc\ \citep{hawley78, peimbert78, talent79}.  Using
both radio and optical spectroscopy \citet{shaver83} significantly
expanded the \hii\ region sample and derived an O/H radial gradient of
$-0.07 \pm 0.015$\dexkpc.  More recent work has focused on the outer
Galaxy.  \citet{vilchez96} extended the \hii\ region sample out to
$R_{\rm gal} = 18$\kpc\ and found a flatter gradient in the outer
Galaxy \citep[also see][]{fich91}.  \citet{deharveng00} detected no
significant flattening, but measured an overall lower gradient of
$-0.0395 \pm 0.0049$\dexkpc.

Observations of CELs in the far infrared (FIR) are less sensitive to
extinction by dust and can probe further into the Galactic disk.
Unlike their optical counterparts they are not very sensitive to
temperature fluctuations; the FIR CELs, however, are weaker and they
cannot be used to determine the hydrogen column density
\citep{rudolph06}.  O/H radial gradients in \hii\ regions derived
using FIR lines range from $-0.06$ to $-0.08$\dexkpc\
\citep{simpson95, afflerbach97, rudolph97}.  \citet{rudolph06}
reanalyzed data from 8 studies in a self consistent way and found a
vertical shift of 0.25 dex in the O/H ratio between the optical and
FIR data that they cannot explain.  Moreover, they found no evidence
for a flattening of the gradient or a discontinuity in the outer
Galaxy.

Large optical telescopes have allowed the detection of optical
recombination lines (ORLs) from Galactic \hii\ regions.  Unlike CELs,
abundance ratios derived from ORLs are not a strong function of
temperature fluctuations within the \hii\ region (See
\S{\ref{sec:o2h}}).  Using ORLs, \citet{esteban05} calculated an O/H
radial gradient of $-0.044 \pm 0.010$\dexkpc\ with a small sample of 8
\hii\ regions with a Galactocentric radius range of 6 to 10\kpc.

Only a decade after their discovery RRLs were used to derive electron
temperatures and to indirectly probe metallicity structure in the
Galaxy \citep{churchwell75, churchwell78, mezger79, lichten79,
  wilson79}.  Converting \te\ values to O/H abundances produced radial
gradients between $-0.04$ and $-0.06$\dexkpc, consistent with both
optical and IR observations of the O/H abundance ratio \citep{wink83,
  afflerbach96, quireza06b}.

What do we conclude from these studies?  Radial metallicity gradients
have been measured to be between $0$ and $-0.1$\dexkpc; to be
increasing with time, decreasing with time, or to have no time
dependence; to be linear with Galactocentric radius or to have
discontinuities in the radial profile.  The bulk of the data do
indicate that the Milky Way disk has a negative radial metallicity
gradient, but not much else is very certain.

\subsection{This Work}\label{subsec:thisWork}

A Galactic map of \hii\ region metallicities can inform both Galactic
chemical evolution (GCE) models and the merger history of the Milky
Way, especially in the outer disk.  Using RRLs to derive Galactic
metallicity structure has several advantages.  RRLs are detected
throughout the Galactic disk.  Unlike many stellar probes of
metallicity, the ISM gas is less likely to suffer from radial mixing on
a Galactic scale.  It thus preserves the local enrichment history in
the disk.  \hii\ regions are confined to the thin disk with a scale
height of 100\pc, and thus cannot be confused with other Galactic
components.  Large volumes of gas are sampled and therefore
metallicities derived here should not be contaminated by nearby stars.

Using our highest quality data we calculated an O/H radial gradient of
$-0.0446 \pm 0.0049$\dexkpc. The observed dispersion in these fits
appears to be real and not due to measurement error \citep[also
see][]{shaver83, quireza06b}.  Filtering the sample by Galactic
azimuth increases the correlation coefficient and produces gradients
between $-0.03$ and $-0.07$\dexkpc, consistent with the range of
values for \hii\ regions in the literature (see
\S{\ref{subsec:previous}}).  There is no evidence for non-linear
structure or discontinuities in the metallicity distribution with
Galactocentric radius when the data are azimuthally averaged.

\begin{figure}
\includegraphics[angle=0,scale=0.42]{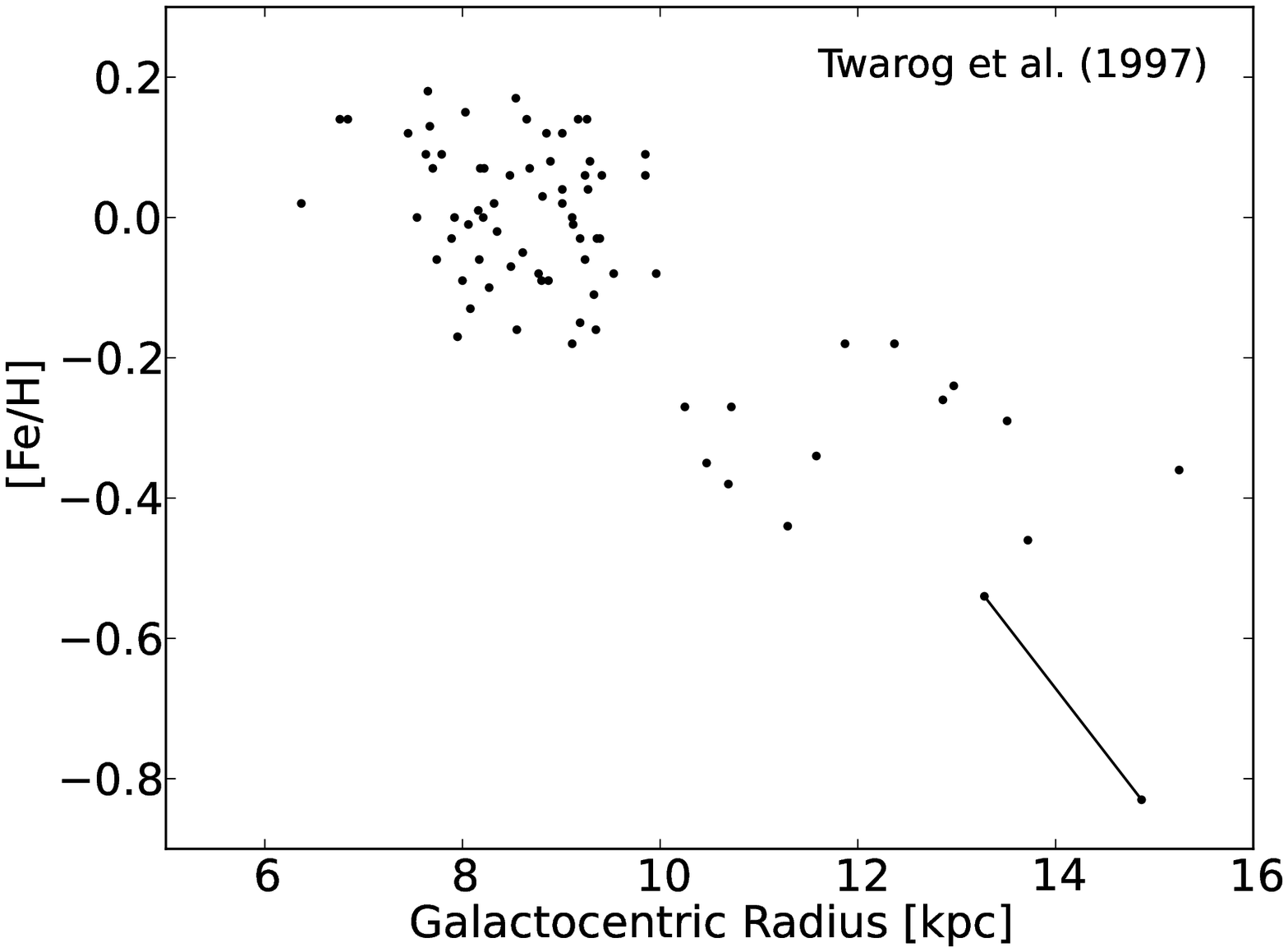}
\includegraphics[angle=0,scale=0.42]{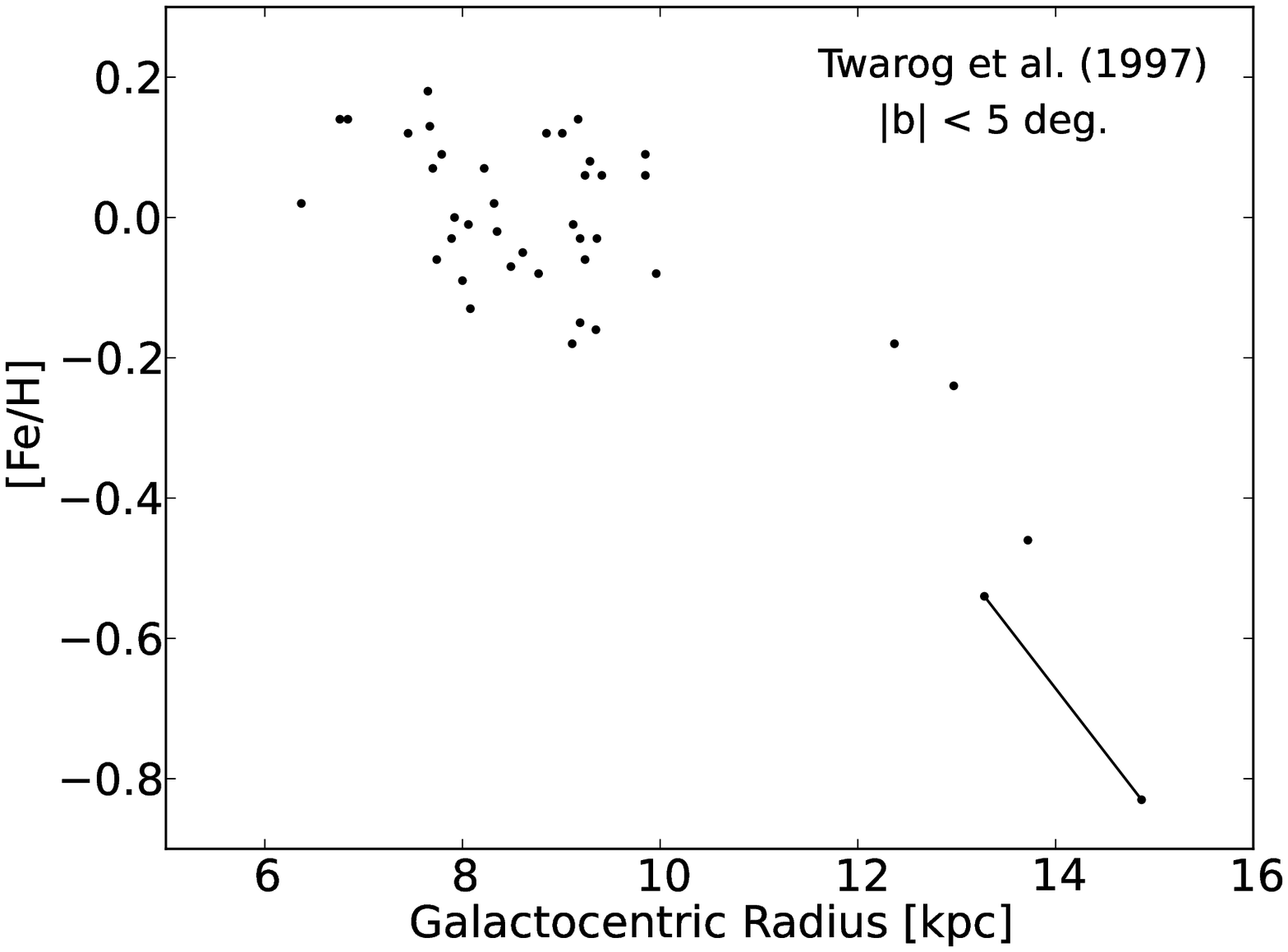}
\includegraphics[angle=0,scale=0.42]{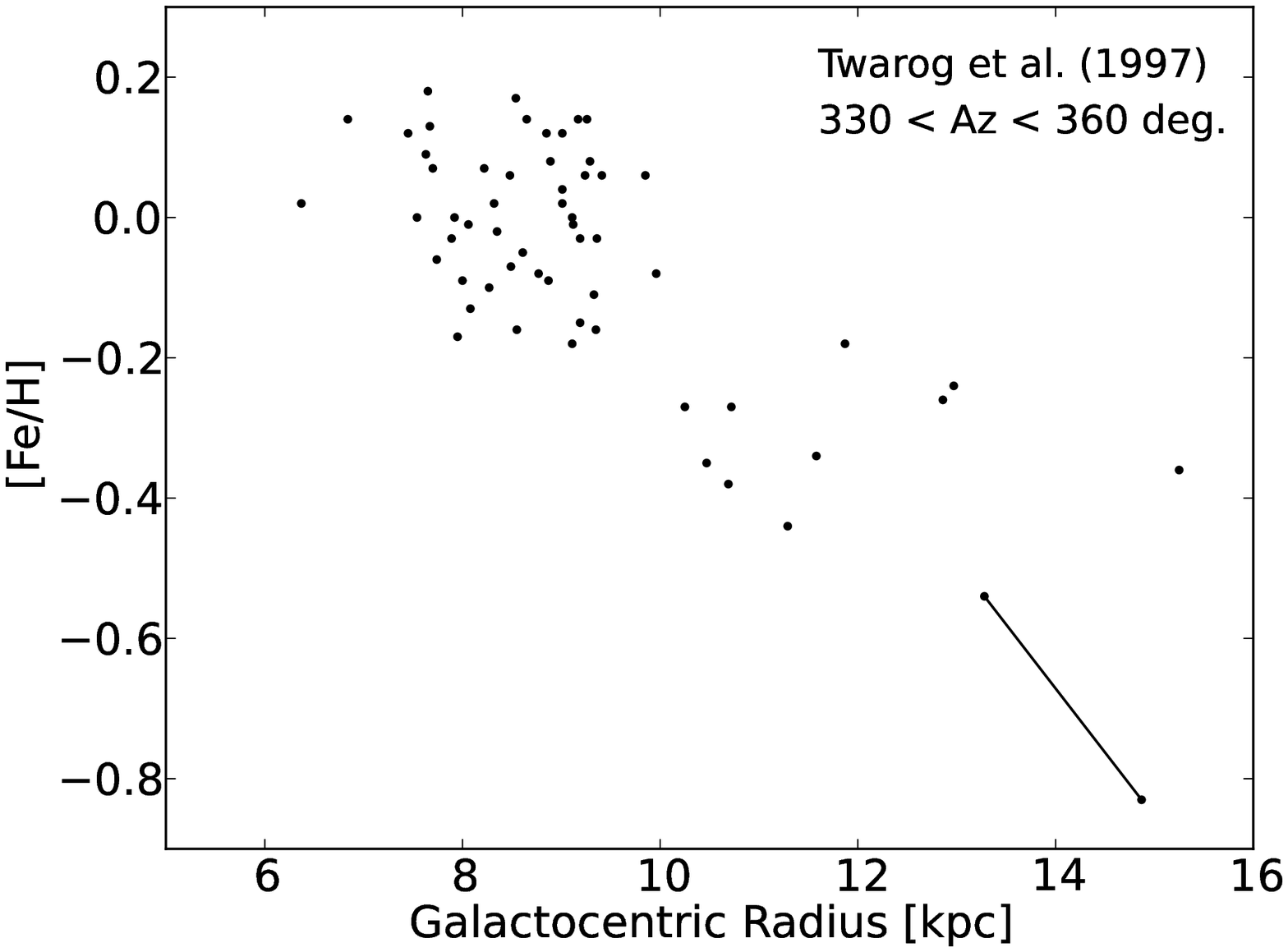}
\includegraphics[angle=0,scale=0.42]{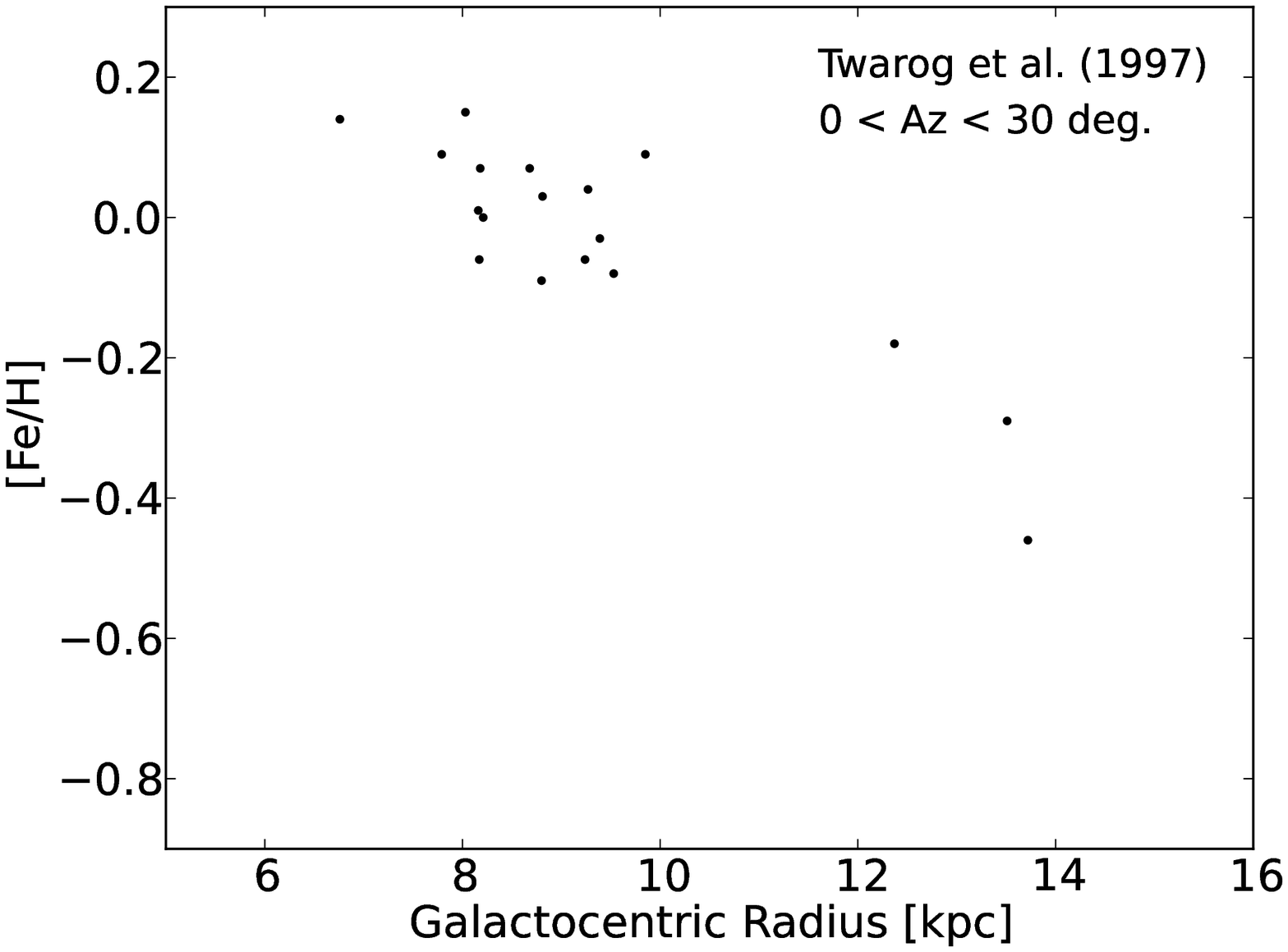}
\caption{[Fe/H] abundance plotted as a function of Galactocentric
  radius from the open cluster data of \citet{twarog97}.  Top left
  panel: all data in their sample (cf. Figure~3 of Twarog et al.).
  Top right panel: only data within 5\degree\ of the Galactic plane
  are included.  Bottom left panel: only data within the Galactic
  azimuth range $330-360$\degree.  Bottom right panel: only data
  within the Galactic azimuth range $0-30$\degree.  The solid line
  connects the two results for open cluster BE21.}
\label{fig:twarog}
\end{figure}

\citet{twarog97} were the first to notice a discontinuity in the
radial gradient from observations of iron in open clusters.  In
Figure~\ref{fig:twarog} we plot [Fe/H] versus \rgal\ from their open
cluster sample (cf.  Figure~3 of Twarog et al.).  We explored whether
azimuthal metallicity structure could be responsible for the observed
discontinuity by filtering their sample into two azimuth ranges:
$330-360$\degree\ and $0-30$\degree\ (bottom panels).  The metallicity
jump near $R_{\rm gal} = 10$\kpc\ is still present for ${\rm Az} =
330-360$\degree.  Given the large range of ages for open clusters,
however, any azimuthal structure that existed when the open clusters
were formed would confuse the interpretation of the radial gradient.
We also filtered their data by Galactic latitude.  When open clusters
at high Galactic latitudes, $\mid b \mid > 5$\degree, are excluded
there is no indication of a step in the radial gradient.  The open
clusters that provided the evidence for such a discontinuity are
located between $10 < R_{\rm gal} < 12$\kpc\ and have larger Galactic
latitudes.  

Observations of Cepheids also reveal azimuthal metallicity structure.
The [Fe/H] abundance decreases for sources near $R_{\rm gal} \approx
10$\kpc\ while scanning in Galactic azimuth from Galactic quadrant II
to III \citep{luck06, lemasle08, pedicelli09}.  There is a similar
trend in our RRL electron temperature data (cf.,
Figure~\ref{fig:shepard} and recall that the metallicity is inversely
proportional to the electron temperature).  But one should be careful
when comparing metallicities in Cepheids with \hii\ regions since they
have different ages and may be from a different population.  Even
though Cepheids are relatively young ($< 200$\myr), it only takes the
Sun $\sim\ 230$\myr\ to complete one orbit, whereas \hii\ region
abundances probe metallicities at their observed locations.  Cepheids
are also more likely to be influenced by radial mixing and therefore
their current radius may be different than the radius of their
birthplace.  \citet{pedicelli09} claim, however, that the detected
metallicity structure from Cepheids was not affected by age or height
above the Galactic Plane.

There are reasons to expect that the Galaxy is well-mixed at a given
radius.  Galactic differential rotation, cloud-cloud collisions, and
turbulent diffusion should mix the ISM on time scales of
\expo{6}-\expo{9}\yr\ depending on the mixing mechanism and spatial
scale \citep{roy95, scalo04}.  Local pollution and infalling material,
however, can modify these homogeneous abundances.  Our \hii\ region
RRL results indicate that the gas is not well-mixed in azimuth.  We
calculate a dispersion of $0.18\,$dex relative to our linear, radial
fit using the Green Bank Sample with QFs of C or better.  Observations
of the O/H abundance in spiral galaxies measure a dispersion of
$0.1-0.2\,$dex at a given radius using the metallicity indicator
$R_{23}$ \citep{kennicutt96, vanzee98}\footnote {$R_{23} =
  ($[\oii]$\lambda{3727} + $[\oiii]$\lambda{4959,5007})/{\rm H}\beta$
  \citep{pagel79}.}  or the [\oiii]$\lambda{4363}$ auroral line method
\citep[e.g.,][]{rosolowsky08}.  But it has been shown that this
dispersion is consistent with measurement errors \citep{kennicutt96,
  bresolin11}.  More recent, high quality abundance measurements in
spiral galaxies find no evidence of azimuthal structure with a $rms$
scatter in the derived abundance gradients on the order of $0.06\,$dex
\citep{bresolin09, bresolin11}.

Many GCE models consider chemical evolution along the radial dimension
as a function of time and allow for infalling gas
\citep[e.g.,][]{matteucci89, chiappini97, tosi98, boissier99}.  A key
constraint is the radial metallicity gradient as a function of time.
Depending on the star formation rate with radius and the composition
of infalling material, GCE models can predict a steepening of the
metallicity gradient with time \citep[e.g.,][]{tosi88, chiappini97}, a
flattening with time \citep[e.g.,][]{molla97, hou00, fu09}, or a
constant gradient with time \citep[e.g.,][]{magrini09}.  To our
knowledge all such models are axisymmetric and therefore cannot make
predictions about metallicity structure along the azimuthal direction.

Chemodynamical models consider both the chemical and dynamical
evolution of the Galaxy \citep[e.g.,][]{samland97}.  Some models
predict radial metallicity gradients that steepen with time with
present-day values of $\approx -0.07$\dexkpc\ \citep{samland97,
  roskar08b, schonrich09}.  There are several interesting dynamical
effects that are important for chemical evolution.  \citet{sellwood02}
show that radial mixing will result from resonance scattering by
spiral arms \citep[also see][]{roskar08a}.  \citet{schonrich09}
developed chemical evolution models that incorporated these radial
flows for both stars and gas.  These models predict the coevolution of
the thick and thin discs without requiring accreted material from
outside the Galaxy.  \citet{minchev10} proposed a new mechanism of
radial migration involving the resonance overlap of the bar and spiral
structure that produced shallower radial metallicity gradients.
\citet{samland03} model the formation and evolution of a disk galaxy
within a growing dark halo using cosmological simulations of structure
formation.  A bar naturally forms and reduces the radial metallicity
gradient in the thin disk to $-0.02$\dexkpc.  This is consistent with
iron abundances in external galaxies where metallicity gradients are
found to be shallower in galaxies with bars
\citep[e.g.,][]{vila-costas92}.  But to our knowledge all of these
models do not consider or show results of the chemical evolution in
the azimuthal direction.

Many investigators that propose either a discontinuity in the radial
gradient or multiple components often use heterogeneous data sets
\citep{vilchez96, yong05, luck06, carraro07, lemasle08, sestito08,
  pedicelli09, andreuzzi10}.  In contrast, studies that use a
homogeneous sample and process the data in a self-consistent way
typically find linear gradients with no significant discontinuities
\citep{rolleston00, deharveng00, quireza06b, rudolph06, henry10}.
Although here we have combined data from two different telescopes, the
GBT and the 140 Foot telescope, the observing technique and data
reduction process were identical.  Moreover, we get the same results
even when using only GBT data.  Future efforts should focus on
homogeneous samples of a given tracer to reduce systematic errors; for
example, the Bologna Open Cluster Chemical Evolution (BOCCE) project
\citep{bragaglia06}.

\section{Summary}\label{sec:summary}

We are discovering \hii\ regions over a wide range of Galactic azimuth
in the outer Galaxy \citep{bania10, anderson11}.  Most metallicity
tracers are at optical wavelengths and are restricted to probing the
Galactic disk in just the Solar neighborhood near Az = 0\degree.
Deriving metallicities for this new sample of \hii\ regions will
compliment the Apache Point Observatory Galactic Evolution Experiment
(APOGEE) that will measure the metallicity of red giant stars at
infrared wavelengths over transgalactic distances
\citep{eisenstein11}.

Here we have calculated the metallicity of 81 Galactic \hii\ regions
using RRLs measured with the GBT.  Most of these objects are located
between Galactocentric azimuth 330\degree\ and 60\degree.  We derived
radial metallicity gradients in the range $-0.03$ to $-0.07$\dexkpc.
There is no evidence for any breaks or discontinuities in the radial
gradient as probed here by our RRLs observations.  We do, however,
find evidence for azimuthal structure in the spatial pattern of \hii\
region metallicities, especially in the outer Galactic disk.

\acknowledgements

D.S.B. thanks Dick Henry for discussions about planetary nebulae and
chemical evolution; Anil Pradhan for pointing out recent studies
concerning the photoionization cross section of oxygen; and Fred
Schwab for providing information about linear regression and metric
interpolation.  T.M.B. was partially supported by NSF award AST
0707853.  LDA was partially supported by SNF and by the NSF through
GSSP awards 08-0030 and 09-005 from the NRAO.  This research has made
use of NASA's Astrophysics Data System.

\appendix

\section{Data Reduction and Calibration}\label{sec:app}

\subsection{Averaging Radio Recombination Lines}

For hydrogenic recombination line transitions with high principal
quantum numbers ($n > 50$), the line parameters of adjacent
transitions should be similar.  For example, the classical oscillator
strength for the \hy51\ RRL differs from the \hy50\ RRL by only 2\%
indicating that these lines should have very similar intensities
\citep{menzel68}.  We averaged the \hy87\ to \hy93\ RRLs in order to
improve the signal-to-noise ratio of our measurements.  Below we
discuss the three steps required to average GBT RRL data in this way.

\subsubsection{Velocity Scale}

Because adjacent RRLs are at different frequencies they have different
velocity resolutions.  Since the \hy87\ sub-band has the lowest spectral
resolution, the spectra are re-gridded to the same velocity scale as
the \hy87\ sub-band.  We assume the velocity resolution is constant across
the 50 MHz bandwidth.  We use a Sin(x)/x interpolation to re-grid the
spectra of each RRL sub-band onto the \hy87\ sub-band (e.g., see Balser 2006).

\subsubsection{Velocity Offset}

The current GBT system only properly Doppler tracks one RRL sub-band.
We used the \hy89\ sub-band as the Doppler tracked sub-band since it
is located near the center of the X-band receiver system.  The other
RRL sub-bands are Doppler tracked at the sky frequency of the \hy89\
sub-band.  Since Doppler tracking is a function of sky frequency the
other RRL sub-bands will be offset, typically by a few channels,
relative to the \hy89\ sub-band.  Moreover, since the sky frequency is
a function of time the offset will vary slowly in time.  The variation
in the offset is negligible assuming the data are averaged over
time-scales of days and not months.

The relativistic Doppler frequency is given by:
\begin{equation}
{ \nu_{\rm sky} = \nu_{\rm rest} - \nu_{\rm rest}\,\frac{1 -
V/c}{\sqrt{1 - (V/c)^2}} }\label{eqn:doppler}
\end{equation}
where $\nu_{\rm sky}$ is the sky frequency, $\nu_{\rm rest}$ is the
rest frequency, $V$ is the radial velocity, and $c$ is the speed of
light.  The offset is just given by the difference in the derived sky
frequency between the RRL sub-band in question and the H89$\alpha$
sub-band.

\subsubsection{Intensity Scale}  

Since the telescope's HPBW varies with RRL sub-band, the line
intensities may vary depending on the convolution of the source
structure with the telescope's beam.  Assuming a Gaussian telescope
beam pattern and a Gaussian source brightness distribution, the
relationship between the brightness temperature, $T_{\rm B}$, and the
antenna temperature, $T_{\rm A}$, is given by
\begin{equation}
{ T_{\rm B} = \frac{T_{\rm A}}{\eta_{\rm b}}\biggl(\frac{\theta_{\rm s}^{2} + 
\theta_{\rm b}^{2}}{\theta_{\rm s}^{2}}\biggr) }\label{eqn:gauss}
\end{equation}
where $\eta_{\rm b}$ is the beam efficiency, $\theta_{\rm b}$ is the
FWHM beam size, and $\theta_{\rm s}$ is the angular source size.
Observations of a source with a brightness temperature $T_{\rm
B}$ and size $\theta_{\rm s}$ with two different Gaussian beams
$\theta_{\rm b}(1)$ and $\theta_{\rm b}(2)$ will produce antenna
temperatures $T_{\rm A}(1)$ and $T_{\rm A}(2)$ that are related by
\begin{equation}
{ T_{\rm A}(1) = \frac{ T_{\rm A}(2)\,[\theta_{\rm obs}(1)^2 - \theta_{\rm b}(1)^2 + \theta_{\rm b}(2)^2] }
{ \theta_{\rm obs}(1)^2 } }\label{eqn:scale}
\end{equation}
where $\theta_{\rm obs}(1)$ is the observed (convolved) source size
for beam 1 $\theta_{\rm obs}(1)^2 = \theta_{\rm s}^2 + \theta_{\rm
  b}(1)^2$.  If the source is larger than the beam size, $\theta_{\rm
  obs}(1) >> \theta_{\rm b}(1)$, then $T_{\rm A}(1) = T_{\rm A}(2)$,
whereas if the source size is much smaller than the beam size,
$\theta_{\rm obs}(1) \approx \theta_{\rm b}(1)$, then $T_{\rm A}(1) =
T_{\rm A}(2) (\theta_{\rm b}(2) / \theta_{\rm b}(1) )^2$.

The continuum data were observed centered at 8665\ghz\ with a
bandwidth of 320 MHz.  The \hy91\ transition lies within this observed
band.  We therefore scaled the line intensity of the other RRL
transitions to the \hy91\ transition using Equation~\ref{eqn:scale}.
In principle any flux density calibration should cancel when
calculating the line-to-continuum ratio but since we only have a
continuum measurement near one RRL this is not true when the RRLs are
averaged.  Any atmospheric or gain fluctuations should be
approximately constant across the 8 to 10\ghz\ range of our
measurements; they should not cause any significant error in
determining the line-to-continuum ratio.

\subsection{Flux Density Calibration}\label{sec:cal}

During both the line and continuum observations a noise diode was used to
inject a calibration signal with an intensity of about $5-10$\%
of the nominal total system temperature.  The noise diodes have been
calibrated in the laboratory with 50\mhz\ resolution to about 10\%
accuracy.  For the X-band receiver system the noise diodes are $T_{\rm
cal} \approx 2\,$K.

Astronomical calibration sources can be used to validate or revise the
$T_{\rm cal}$ values, whereby a correction factor, CF, the ratio of
the expected intensity to the measured intensity, is derived.  Below
we explore several different calibration methods.  We assume a
telescope gain of $2\,$K$\,$Jy$^{-1}$ \citep{ghigo01} and the flux
densities from \cite{peng00}.  We neglect corrections due to opacity
and elevation gain because the former are typically a few percent
under most conditions and the latter are less than 5\% at 10\ghz.

\subsubsection{Flux continuum calibrator with the Digital Continuum
Receiver (DCR)}

DCR measurements of the flux density calibrator 3C147 were used to
derive correction factors.  The DCR was configured to center an
80\mhz\ bandwidth at each RRL sub-band frequency, and a 320\mhz\
bandwidth at the continuum band frequency of 8665\mhz.  The spectral
line data had bandwidths of 50\mhz, but the nearest available filter
to this bandwidth was 80\mhz.  We assume that the true calibration
does not significantly vary across the 50\mhz\ RRL sub-bands.

Observations of 3C147 were taken near transit on 6 January 2008, 20
February 2008, and 22 February 2008.  The weather was not optimal for
these observations although the conditions were best on 6 January 2008
(see Table~\ref{tab:cal}).  For this epoch the correction factors are
typically less than 10\%, consistent with the expected accuracy of the
laboratory $T_{\rm cal}$ measurements.  The intensity of 3C147 varies
by about $1-2$\% between the 80 MHz and 320 MHz bandwidth observations
centered at similar frequencies.

Given the marginal weather conditions we did not apply these CF's
to the data.  The primary goal here, moreover, is to calculate electron
temperatures which are a function of the line-to-continuum ratio (c.f.,
Equation~\ref{eq:te}).  Since many errors cancel when taking the ratio
of these intensities only a relative calibration is required.

\subsubsection{Flux continuum calibrator with the Auto-correlation
  Spectrometer (ACS)}

The flux density calibrator 3C147 was also observed using the ACS in
the same configuration as for the RRL measurements.  From these data
$T_{\rm cal}$ values were calculated for each spectral channel with
the same spectral resolution.  The $T_{\rm cal}$ vector may be used
not only as a flux density calibration but also to produce a bandpass
calibration, since the frequency structure of 3C147 is known
\citep{johnson02, pisano07}.

There are significant disadvantages of using a vector $T_{\rm cal}$
calibration for our data.  Any gain fluctuations (e.g., weather) are
not very easy to subtract since a baseline cannot be determined as
with the DCR data.  Also, if the $T_{\rm cal}$ values are applied to
the data they will add noise to the spectrum.  We therefore did not
pursue this calibration method.

\subsubsection{Flux line calibrator with the ACS}

If a spectral line of known flux density is within the RRL sub-band
then it can be used to calculate the $T_{\rm cal}$ value near this
frequency.  Since there are not many spectral lines with known flux
densities this method is typically not feasible.  Yet, any stable
spectral line with a high signal-to-noise ratio and well behaved
spectral baselines can provide a relative calibration.  For our
observations the RRL itself, if sufficiently bright, can be used as a
relative flux calibrator.

For example, \citet{quireza06b} used the \hii\ region W3 to measure a
relative intensity calibration scale for two adjacent RRLs.  Here we
use all target sources in our sample with a continuum intensity
greater than $5\,$K to determine an average relative calibration.
This calibration will average over variations in the spectral
baselines, weather, elevation gain, etc.

\begin{figure}
\includegraphics[angle=0,scale=0.55]{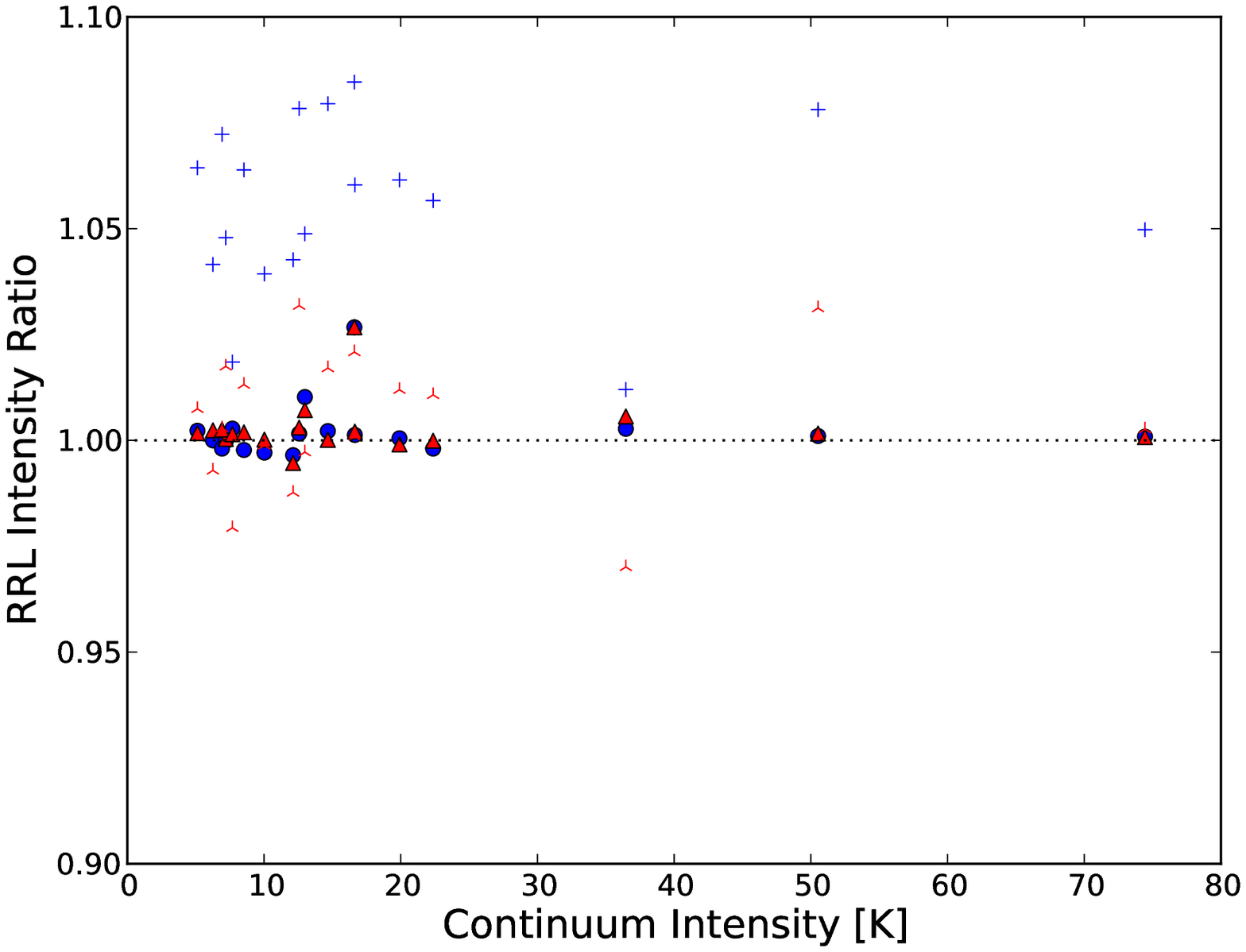}
\includegraphics[angle=0,scale=0.55]{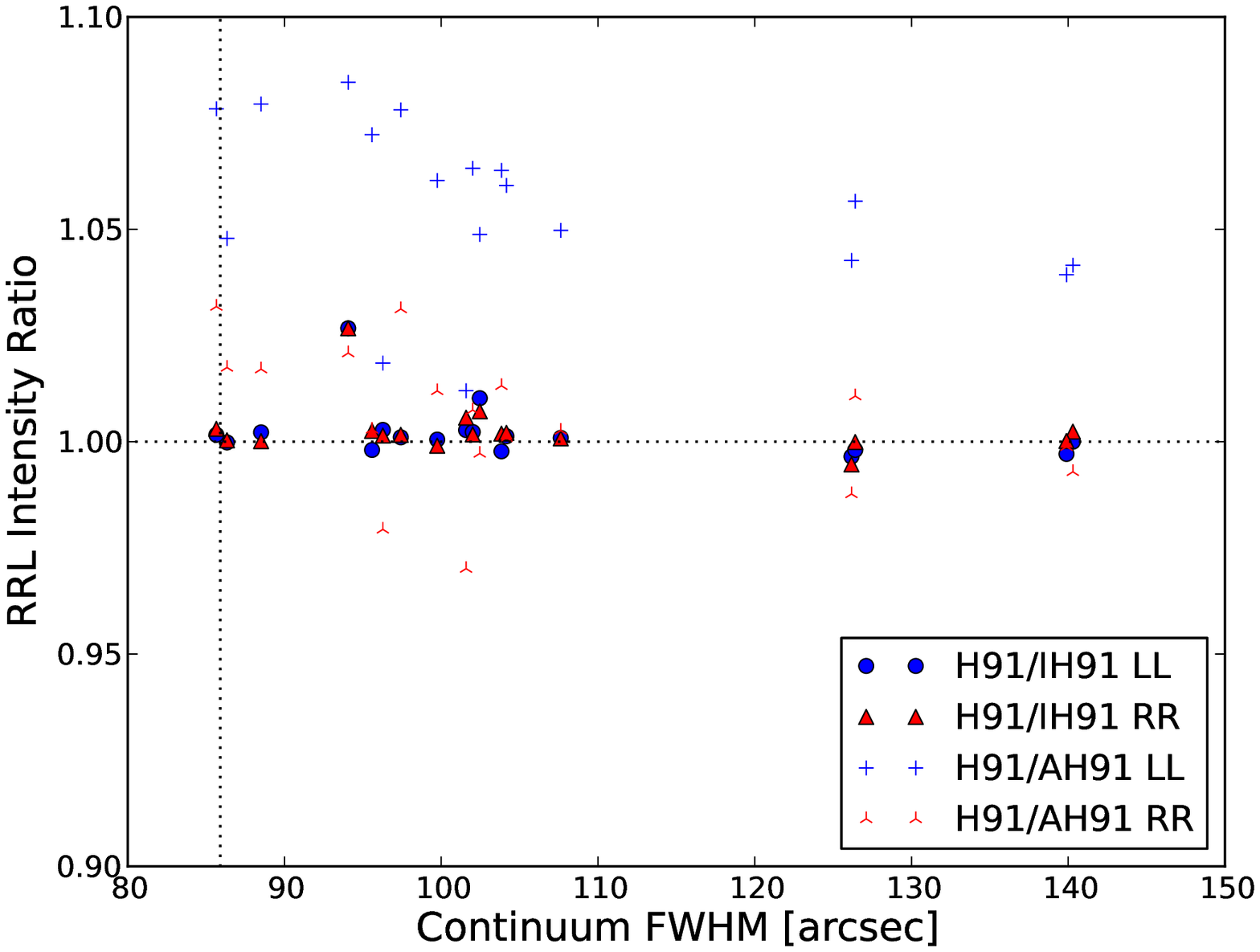}
\caption{RRL intensity ratios as a function of continuum intensity
  (top) and FWHM angular size (bottom).  Three different \hy91\
  intensities have been calculated for all sources with a continuum
  intensity above 5 K: H91 is the \hy91\ intensity; IH91 is the \hy91\
  intensity measured after the velocity scale has been interpolated;
  and AH91 is the \hy91\ intensity after the six adjacent RRLs
  have been interpolated and averaged.  Plotted are the ratios:
  H91/IH91 and H91/AH91 for each polarization.  The horizontal dashed
  line is a ratio of unity, whereas the vertical dashed line is the
  GBT's HPBW.  The H91/AH91 ratios are typically less than 5\% and are
  not correlated with either continuum intensity or angular size.}
\label{fig:ta}
\end{figure}

Let H91, IH91, and AH91 denote the \hy91\ intensity, the interpolated
\hy91\ intensity, and the averaged RRL intensity scaled to the \hy91\
frequency as discussed above.  Figure~\ref{fig:ta} plots RRL intensity
ratios for both circular polarizations for all sources with a
continuum intensity larger than 5\K.  Shown are the RRL intensity
ratios H91/IH91 and H91/AH91 for each polarization.  We excluded the
\hy90\ RRL sub-band since it was difficult to determine a good
baseline near the carbon RRL due to nearby higher order RRLs.  We
calculated ratios of $1.002 \pm 0.0067$ for H91/IH91 (LL); $1.003 \pm
0.0063$ for H91/IH91 (RR); $1.056 \pm 0.0197$ for H91/AH91 (LL); and
$1.005 \pm 0.0159$ for H91/AH91 (RR).  To within the uncertainty the
H91 and IH91 intensities are identical---this should be the case if
the interpolation is done properly.  There is a measurable systematic
offset when comparing the H91/AH91 ratios, however, that should stem
from errors in the $T_{\rm cal}$ values and any systematic errors in
our averaging technique.  These offsets are less than 5\%.  There is
no systematic effect of these ratios with respect to either the
continuum intensity or angular size.  The dispersion in H91/AH91 is a
result of fitting spectral baselines and Gaussian profiles, random
noise, weather fluctuations, elevation gain fluctuations, etc.  It is
therefore a good measure of the uncertainties in our observing and
data analysis procedures.

\begin{figure}
\includegraphics[angle=0,scale=0.55]{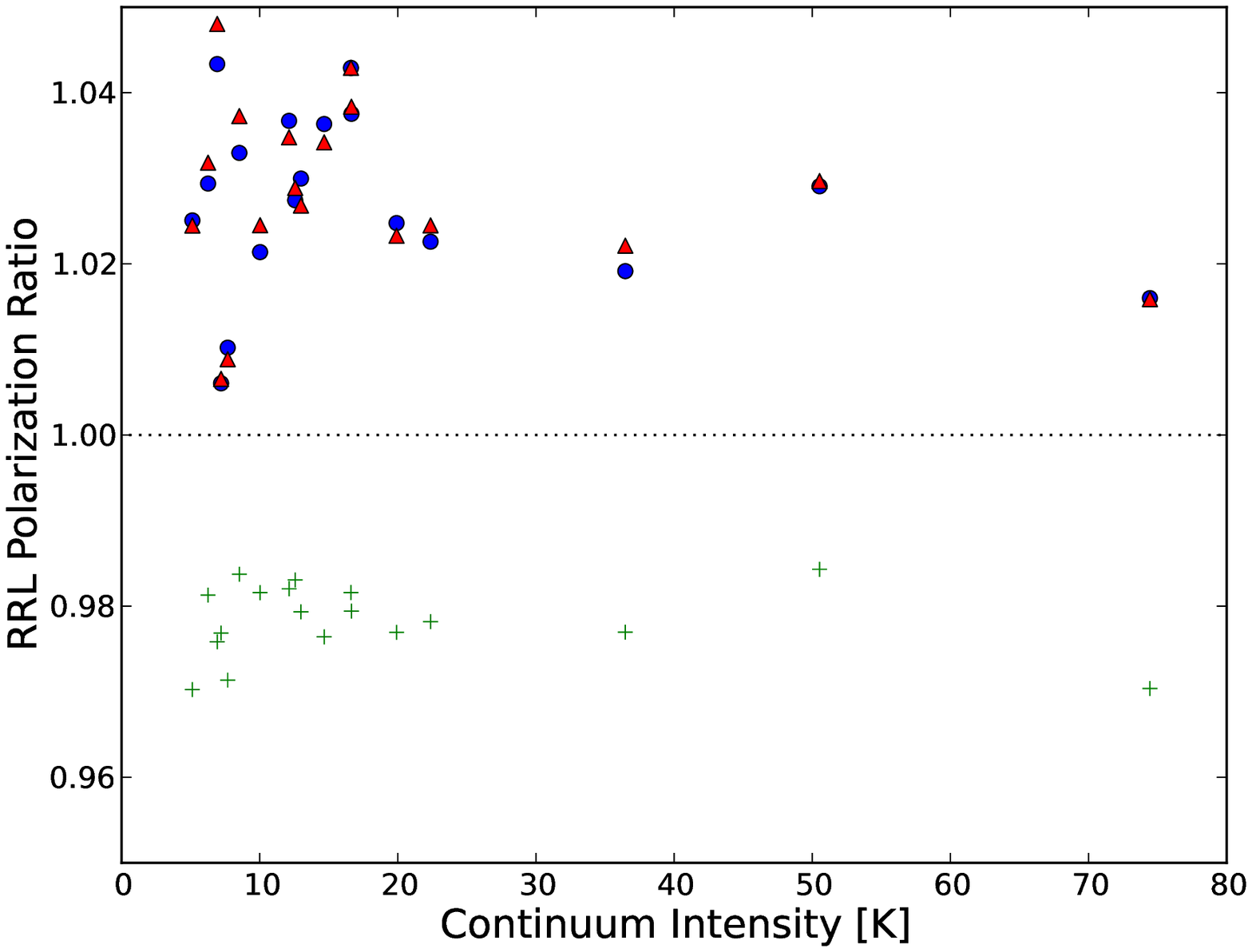}
\includegraphics[angle=0,scale=0.55]{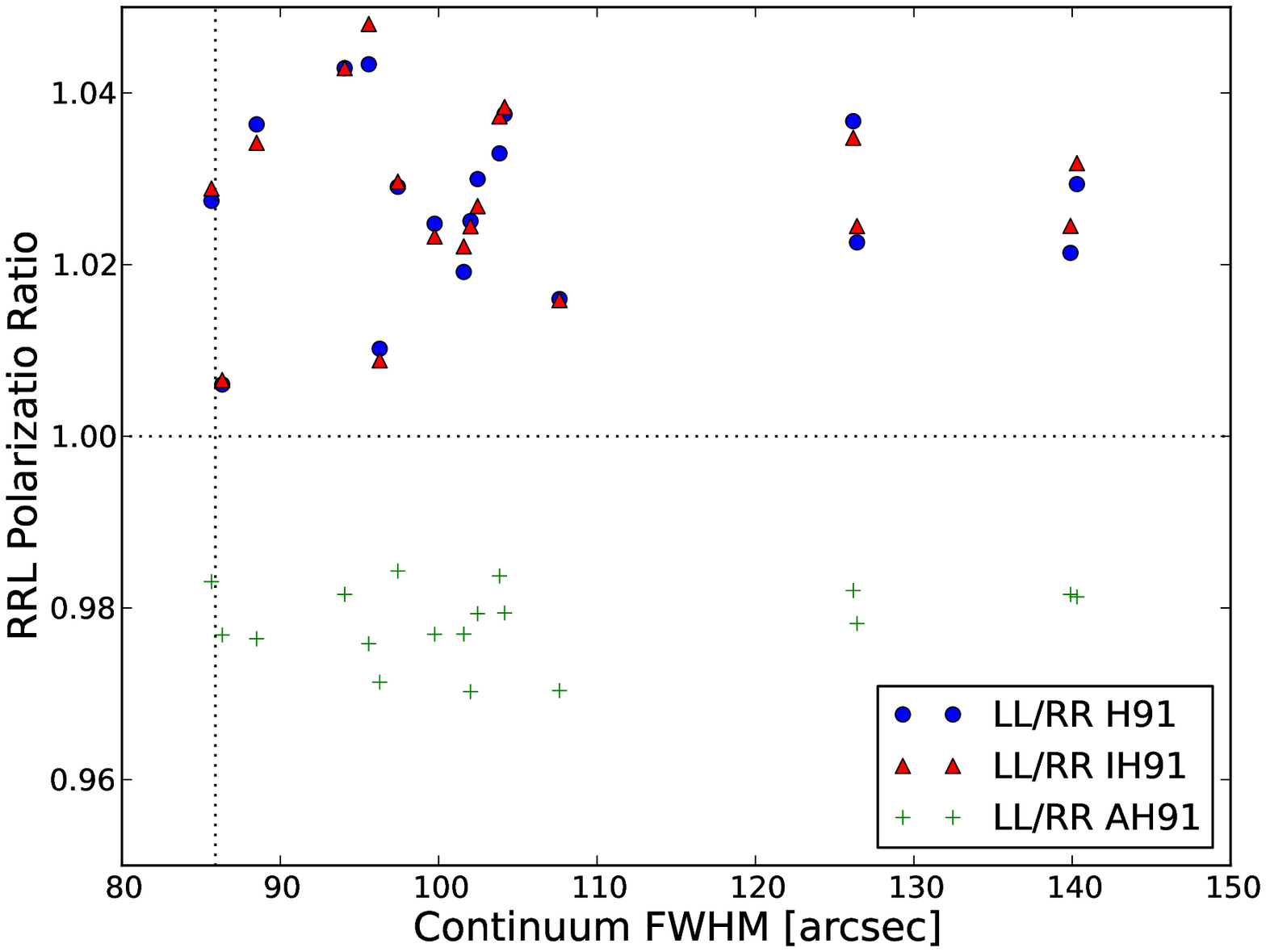}
\caption{RRL polarization ratios as a function of continuum intensity
  (top) and FWHM angular size (bottom).  Three different \hy91\
  intensities have been calculated for all sources with a continuum
  intensity above 5 K: H91 is the \hy91\ intensity; IH91 is the \hy91\
  intensity measured after the velocity scale has been interpolated;
  and AH91 is the \hy91\ intensity after the six adjacent RRLs
  have been interpolated and averaged.  Plotted are the polarization
  ratios (LL/RR) for the H91, IH91, and AH91 RRL intensities. The
  horizontal dashed line is a ratio of unity, whereas the vertical
  dashed line is the GBT's HPBW.  The AH91 polarization rations are
  typically less than 2\% and are consistent with random errors in
  $T_{\rm cal}$.}
\label{fig:pol}
\end{figure}

Figure~\ref{fig:pol} plots the RRL polarization ratios, LL/RR, for
H91, IH91, and AH91.  We calculate polarization ratios of $1.027 \pm
0.0101$ for H91; $1.028 \pm 0.0105$ for IH91; and $0.978 \pm 0.0043$
for AH91.  The polarization ratios are also not a function of either
the continuum intensity or angular size.  The H91 polarization ratios
are about 3\% from unity, well within the expected $T_{\rm cal}$
uncertainties of 10\%.  The polarization ratios for the other RRLs
show similar results.  To within the uncertainty the H91 and IH91
polarizations ratios are identical, again as expected.  The AH91
polarization ratios are within 2\% of unity, consistent with random
errors in $T_{\rm cal}$.  That is, if the errors in determining
$T_{\rm cal}$ are random they should average to unity given enough RRL
sub-bands.  Moreover, the dispersion in the distribution should
decrease as $1/\sqrt{N}$.  From Figure~\ref{fig:pol} we see a decrease
in the dispersion from 0.0101 to 0.0043 when we compare the single
sub-band H91 to the averaged AH91.  This is consistent with the six
RRL sub-bands used in the average.

Based on the H91/AH91 ratios for each polarization we derived and
applied corrections factors of CF = $1.056 \pm 0.0197$ for LL and CF
= $1.005 \pm 0.0159$ for RR.  Overall we deem that the relative
calibration scale is accurate to within 5\%.

\section{Accuracy of the Electron Temperature Derivation}

We estimate the accuracy of our derived electron temperatures by
calculating both measurement and calibration errors.  The measurement
errors are listed in Table~\ref{tab:prop} and were calculated by
propagating the uncertainty in the hydrogen and helium RRL line
intensity and width, together with the continuum intensity
uncertainty.  The average measurement error is $1.6\% \pm 1.1\%$ with
respect to the electron temperature.  The calibration errors are
estimated to be $< 5\%$ (see \S{\ref{sec:cal}}).  The accuracy of our
electron temperatures should therefore be $\sim 5$\% or better.  This
does not include systematic errors which are difficult to determine.
If we could estimate any systematic error we would remove it!  Our
sources, however, were selected to minimize potential systematic
errors.  For example, determining the baseline or zero level for the
continuum intensity is one potential source of systematic error.  This
is particularly difficult for sources located in complex regions where
extended emission often exists in the baseline region.  For such cases
our data analysis procedures will systematically underestimate the
continuum intensity and therefore the electron temperature.  We have
carefully chosen sources that are well isolated to minimize such
systematic effects.

{\it Facility:} \facility{GBT}

\clearpage

%
%

%
%

\clearpage

\begin{deluxetable}{lrrrrrrrrc}
\tabletypesize{\scriptsize}
\tablecaption{Properties of Galactic \hii\ Regions \label{tab:prop}}
\tablewidth{0pt}
\tablehead{
\colhead{} & \colhead{$\ell$} & \colhead{$b$} & \colhead{R.A.} & \colhead{Decl.} & \colhead{Az.} &
\colhead{$R_{\rm gal}$} & \colhead{$d_{\rm Sun}$} & \colhead{} & \colhead{$T_{\rm e}$} \\ 
\colhead{Name} & \colhead{(deg)} & \colhead{(deg)} & \colhead{(B1950)} & 
\colhead{(B1950)} & \colhead{(deg)} & \colhead{(kpc)} & \colhead{(kpc)} & 
\colhead{$y$} & \colhead{(K)} 
} 
\startdata 
      G7.47+0.1 & $7.472$ & $0.060$ & 17 59 11.6 & $-$22 27 55.0 & 169.635 & 21.9 & 30.3 & $0.083 \pm\ 0.007$ & $9929 \pm\ 87$ \\ 
            S64 & $28.788$ & $3.491$ & 18 28 50.4 & $-$02 07 27.4 & 123.218 &  8.7 & 15.2 & $0.059 \pm\ 0.006$ & $8399 \pm\ 73$ \\ 
            W43 & $30.782$ & $-0.028$ & 18 45 00.9 & $-$01 58 56.2 & 37.331 &  4.7 &  5.6 & $0.071 \pm\ 0.004$ & $6567 \pm\ 30$ \\ 
   G32.797+0.19 & $32.960$ & $0.276$ & 18 47 56.8 & $+$00 05 31.0 & 109.783 &  7.6 & 13.2 & $0.078 \pm\ 0.003$ & $8625 \pm\ 49$ \\ 
        NRAO584 & $34.254$ & $0.144$ & 18 50 47.8 & $+$01 10 46.0 & 19.242 &  6.0 &  3.5 & $0.053 \pm\ 0.004$ & $8084 \pm\ 55$ \\ 
            W48 & $35.196$ & $-1.746$ & 18 59 14.7 & $+$01 08 42.4 & 16.382 &  6.3 &  3.1 & $0.072 \pm\ 0.002$ & $8603 \pm\ 40$ \\ 
  G38.875+0.308 & $38.875$ & $0.308$ & 18 58 44.8 & $+$05 21 24.1 & 107.844 &  9.7 & 14.7 & $0.080 \pm\ 0.000$ & $8959 \pm\ 232$ \\ 
  G39.728-0.396 & $39.728$ & $-0.396$ & 19 02 50.5 & $+$05 47 13.4 & 76.086 &  6.0 &  9.2 & $0.080 \pm\ 0.000$ & $8503 \pm\ 255$ \\ 
            S76 & $40.502$ & $2.540$ & 18 53 47.2 & $+$07 49 40.6 & 7.369 &  7.4 &  1.5 & $0.045 \pm\ 0.004$ & $8223 \pm\ 55$ \\ 
            K47 & $45.454$ & $0.059$ & 19 12 00.1 & $+$11 03 56.3 & 61.592 &  6.3 &  7.8 & $0.076 \pm\ 0.003$ & $8026 \pm\ 63$ \\ 
   G46.495-0.25 & $46.499$ & $-0.251$ & 19 15 07.3 & $+$11 50 32.0 & 33.510 &  6.3 &  4.8 & $0.077 \pm\ 0.006$ & $5229 \pm\ 40$ \\ 
            W51 & $49.490$ & $-0.381$ & 19 21 24.9 & $+$14 24 52.8 & 40.497 &  6.5 &  5.5 & $0.084 \pm\ 0.001$ & $7166 \pm\ 25$ \\ 
     G52.75+0.3 & $52.757$ & $0.334$ & 19 25 18.9 & $+$17 37 32.0 & 70.451 &  8.1 &  9.6 & $0.065 \pm\ 0.014$ & $8970 \pm\ 186$ \\ 
     G55.11+2.4 & $55.114$ & $2.422$ & 19 22 19.8 & $+$20 41 36.0 & 97.723 & 15.3 & 18.4 & $0.087 \pm\ 0.009$ & $13126 \pm\ 144$ \\ 
     G59.80+0.2 & $59.796$ & $0.237$ & 19 40 26.6 & $+$23 42 44.0 & 63.685 &  8.8 &  9.1 & $0.057 \pm\ 0.009$ & $9068 \pm\ 120$ \\ 
            S87 & $60.883$ & $-0.133$ & 19 44 14.5 & $+$24 27 54.0 & 9.607 &  7.9 &  1.5 & $0.080 \pm\ 0.000$ & $7463 \pm\ 77$ \\ 
            S88 & $61.477$ & $0.094$ & 19 44 42.4 & $+$25 05 30.0 & 19.329 &  7.6 &  2.9 & $0.050 \pm\ 0.002$ & $8857 \pm\ 43$ \\ 
            S90 & $63.171$ & $0.448$ & 19 47 11.4 & $+$26 43 46.6 & 43.621 &  7.9 &  6.1 & $0.076 \pm\ 0.006$ & $7760 \pm\ 90$ \\ 
            S93 & $64.136$ & $-0.469$ & 19 52 56.9 & $+$27 05 00.0 & 20.353 &  7.7 &  3.0 & $0.050 \pm\ 0.005$ & $8452 \pm\ 58$ \\ 
            S98 & $68.144$ & $0.918$ & 19 57 10.5 & $+$31 13 19.0 & 73.879 & 12.8 & 13.3 & $0.065 \pm\ 0.013$ & $10834 \pm\ 207$ \\ 
     G69.94+1.5 & $69.922$ & $1.516$ & 19 59 14.0 & $+$33 02 49.0 & 71.327 & 12.8 & 12.9 & $0.085 \pm\ 0.004$ & $9703 \pm\ 50$ \\ 
          K3-50 & $70.292$ & $1.598$ & 19 59 50.5 & $+$33 24 13.2 & 55.730 &  9.9 &  8.7 & $0.041 \pm\ 0.007$ & $10297 \pm\ 121$ \\ 
     G75.77+0.3 & $75.767$ & $0.344$ & 20 19 49.0 & $+$37 16 18.0 & 38.157 &  9.0 &  5.7 & $0.076 \pm\ 0.003$ & $8590 \pm\ 47$ \\ 
   G75.834+0.40 & $75.834$ & $0.402$ & 20 19 46.3 & $+$37 21 35.0 & 35.173 &  8.8 &  5.2 & $0.077 \pm\ 0.002$ & $8363 \pm\ 32$ \\ 
     G76.15-0.3 & $76.152$ & $-0.281$ & 20 23 29.9 & $+$37 13 30.0 & 49.250 & 10.1 &  7.9 & $0.061 \pm\ 0.010$ & $9498 \pm\ 119$ \\ 
           S106 & $76.383$ & $-0.623$ & 20 25 34.2 & $+$37 12 46.0 & 13.513 &  8.5 &  2.0 & $0.028 \pm\ 0.006$ & $11245 \pm\ 92$ \\ 
     G77.98+0.0 & $77.972$ & $-0.012$ & 20 27 48.7 & $+$38 51 27.0 & 30.788 &  8.8 &  4.6 & $0.072 \pm\ 0.012$ & $7674 \pm\ 111$ \\ 
     G78.03+0.6 & $78.032$ & $0.607$ & 20 25 25.0 & $+$39 16 11.0 & 11.919 &  8.5 &  1.8 & $0.061 \pm\ 0.008$ & $8567 \pm\ 86$ \\ 
           S108 & $78.142$ & $1.814$ & 20 20 38.4 & $+$40 03 36.0 & 11.803 &  8.5 &  1.7 & $0.068 \pm\ 0.008$ & $8596 \pm\ 107$ \\ 
            DR7 & $79.293$ & $1.303$ & 20 26 20.0 & $+$40 41 57.0 & 48.709 & 10.6 &  8.1 & $0.079 \pm\ 0.006$ & $8693 \pm\ 86$ \\ 
     G79.42+2.4 & $79.417$ & $2.414$ & 20 21 55.7 & $+$41 26 48.0 & 10.567 &  8.4 &  1.6 & $0.080 \pm\ 0.000$ & $7977 \pm\ 222$ \\ 
     G79.96+0.9 & $79.957$ & $0.866$ & 20 30 16.5 & $+$40 58 31.0 & 34.420 &  9.2 &  5.3 & $0.080 \pm\ 0.000$ & $7921 \pm\ 294$ \\ 
     G80.35+0.7 & $80.352$ & $0.724$ & 20 32 08.0 & $+$41 12 26.0 & 57.567 & 12.5 & 10.7 & $0.057 \pm\ 0.011$ & $10250 \pm\ 155$ \\ 
     G80.88+0.4 & $80.880$ & $0.410$ & 20 35 09.9 & $+$41 26 16.0 & 9.076 &  8.5 &  1.3 & $0.089 \pm\ 0.010$ & $9032 \pm\ 109$ \\ 
     G80.94-0.1 & $80.938$ & $-0.130$ & 20 37 37.8 & $+$41 09 16.0 & 28.857 &  8.9 &  4.4 & $0.042 \pm\ 0.005$ & $8853 \pm\ 62$ \\ 
     G81.25+1.1 & $81.253$ & $1.123$ & 20 33 19.9 & $+$42 09 59.0 & 8.763 &  8.4 &  1.3 & $0.047 \pm\ 0.008$ & $8065 \pm\ 101$ \\ 
           DR21 & $81.681$ & $0.540$ & 20 37 14.0 & $+$42 09 06.0 & 22.947 &  8.7 &  3.4 & $0.063 \pm\ 0.003$ & $8829 \pm\ 36$ \\ 
     G82.57+0.4 & $82.582$ & $0.411$ & 20 40 47.7 & $+$42 46 55.0 & 31.587 &  9.2 &  4.9 & $0.064 \pm\ 0.011$ & $8030 \pm\ 128$ \\ 
           S112 & $83.781$ & $3.315$ & 20 32 02.1 & $+$45 30 09.0 & 6.214 &  8.4 &  0.9 & $0.093 \pm\ 0.016$ & $8643 \pm\ 184$ \\ 
     G85.24+0.0 & $85.246$ & $0.014$ & 20 51 49.6 & $+$44 35 33.0 & 39.751 & 10.3 &  6.6 & $0.081 \pm\ 0.015$ & $8824 \pm\ 177$ \\ 
          S127A & $96.291$ & $2.596$ & 21 27 06.0 & $+$54 24 06.0 & 52.465 & 16.3 & 13.0 & $0.060 \pm\ 0.018$ & $11428 \pm\ 305$ \\ 
          WB85A & $96.297$ & $2.597$ & 21 27 07.6 & $+$54 24 22.0 & 52.651 & 16.4 & 13.1 & $0.135 \pm\ 0.027$ & $11039 \pm\ 314$ \\ 
           S128 & $97.516$ & $3.172$ & 21 30 36.9 & $+$55 39 25.0 & 43.350 & 13.4 &  9.2 & $0.130 \pm\ 0.024$ & $10378 \pm\ 251$ \\ 
           S146 & $108.196$ & $0.577$ & 22 47 30.4 & $+$59 38 52.0 & 29.584 & 12.0 &  6.2 & $0.079 \pm\ 0.004$ & $9590 \pm\ 59$ \\ 
           S147 & $108.368$ & $-1.057$ & 22 54 15.3 & $+$58 15 12.0 & 28.707 & 11.8 &  6.0 & $0.080 \pm\ 0.000$ & $8992 \pm\ 131$ \\ 
           S152 & $108.759$ & $-0.951$ & 22 56 37.1 & $+$58 30 54.0 & 27.169 & 11.6 &  5.6 & $0.034 \pm\ 0.006$ & $9404 \pm\ 81$ \\ 
           S156 & $110.106$ & $0.044$ & 23 03 04.3 & $+$59 58 20.0 & 26.328 & 11.6 &  5.5 & $0.067 \pm\ 0.007$ & $9240 \pm\ 75$ \\ 
        NGC7538 & $111.525$ & $0.816$ & 23 11 21.8 & $+$61 13 38.0 & 29.300 & 12.5 &  6.6 & $0.082 \pm\ 0.003$ & $8483 \pm\ 51$ \\ 
           S159 & $111.612$ & $0.374$ & 23 13 21.3 & $+$60 50 49.0 & 30.947 & 13.0 &  7.2 & $0.031 \pm\ 0.006$ & $8428 \pm\ 68$ \\ 
           S162 & $112.223$ & $0.227$ & 23 18 29.2 & $+$60 55 24.0 & 22.635 & 11.1 &  4.6 & $0.063 \pm\ 0.009$ & $8641 \pm\ 118$ \\ 
           S168 & $115.789$ & $-1.579$ & 23 50 30.0 & $+$60 12 04.0 & 21.004 & 11.2 &  4.5 & $0.080 \pm\ 0.000$ & $8794 \pm\ 242$ \\ 
          WB380 & $124.644$ & $2.539$ & 01 04 36.3 & $+$65 05 22.0 & 28.891 & 15.7 &  9.2 & $0.080 \pm\ 0.000$ & $10758 \pm\ 288$ \\ 
           S186 & $124.897$ & $0.321$ & 01 05 38.8 & $+$62 51 35.3 & 17.133 & 11.3 &  4.1 & $0.080 \pm\ 0.000$ & $8975 \pm\ 460$ \\ 
         WB399B & $128.777$ & $2.012$ & 01 42 05.7 & $+$64 01 00.0 & 29.695 & 18.1 & 11.5 & $0.080 \pm\ 0.000$ & $10361 \pm\ 427$ \\ 
    G132.16-0.7 & $132.157$ & $-0.725$ & 02 04 29.1 & $+$60 31 46.0 & 19.811 & 13.4 &  6.1 & $0.076 \pm\ 0.010$ & $9785 \pm\ 123$ \\ 
             W3 & $133.720$ & $1.223$ & 02 21 56.9 & $+$61 52 40.0 & 14.703 & 11.7 &  4.1 & $0.082 \pm\ 0.002$ & $8977 \pm\ 38$ \\ 
   G133.790+1.4 & $133.785$ & $1.423$ & 02 23 04.0 & $+$62 02 32.1 & 17.348 & 12.7 &  5.2 & $0.088 \pm\ 0.006$ & $8752 \pm\ 74$ \\ 
    G136.91+1.0 & $136.900$ & $1.060$ & 02 45 41.8 & $+$60 28 46.0 & 14.085 & 12.0 &  4.3 & $0.080 \pm\ 0.000$ & $8204 \pm\ 257$ \\ 
           S201 & $138.494$ & $1.641$ & 02 59 18.9 & $+$60 16 15.0 & 12.261 & 11.5 &  3.7 & $0.080 \pm\ 0.000$ & $8302 \pm\ 131$ \\ 
           S206 & $150.593$ & $-0.951$ & 03 59 29.8 & $+$51 10 38.0 & 8.278 & 11.6 &  3.4 & $0.075 \pm\ 0.005$ & $10016 \pm\ 83$ \\ 
           S209 & $151.606$ & $-0.240$ & 04 07 19.7 & $+$51 01 58.0 & 14.918 & 17.3 &  9.4 & $0.070 \pm\ 0.006$ & $10795 \pm\ 98$ \\ 
           S211 & $154.649$ & $2.438$ & 04 32 59.6 & $+$50 46 31.0 & 10.334 & 14.0 &  5.9 & $0.080 \pm\ 0.000$ & $9734 \pm\ 175$ \\ 
           S212 & $155.357$ & $2.609$ & 04 36 46.8 & $+$50 21 58.0 & 12.620 & 17.0 &  8.9 & $0.088 \pm\ 0.024$ & $10253 \pm\ 309$ \\ 
           S228 & $169.191$ & $-0.903$ & 05 10 01.8 & $+$37 23 32.0 & 4.785 & 15.2 &  6.8 & $0.080 \pm\ 0.000$ & $9345 \pm\ 179$ \\ 
           S235 & $173.599$ & $2.798$ & 05 37 37.8 & $+$35 49 35.0 & 1.019 & 10.1 &  1.6 & $0.080 \pm\ 0.000$ & $8612 \pm\ 137$ \\ 
           S237 & $173.899$ & $0.288$ & 05 28 08.1 & $+$34 12 39.0 & 0.264 &  8.9 &  0.4 & $0.080 \pm\ 0.000$ & $8829 \pm\ 158$ \\ 
           S257 & $192.626$ & $-0.017$ & 06 10 08.7 & $+$17 59 46.0 & 358.435 &  9.7 &  1.2 & $0.080 \pm\ 0.000$ & $8833 \pm\ 107$ \\ 
           S269 & $196.456$ & $-1.670$ & 06 11 48.8 & $+$13 50 40.0 & 356.218 & 11.0 &  2.6 & $0.080 \pm\ 0.000$ & $9945 \pm\ 164$ \\ 
     G201.6+1.6 & $201.682$ & $1.652$ & 06 33 53.8 & $+$10 48 10.5 & 353.204 & 12.2 &  3.9 & $0.080 \pm\ 0.000$ & $10063 \pm\ 283$ \\ 
           OriA & $209.011$ & $-19.384$ & 05 32 49.0 & $-$05 25 16.0 & 358.714 &  8.9 &  0.4 & $0.082 \pm\ 0.001$ & $8322 \pm\ 55$ \\ 
          WB870 & $213.077$ & $-2.215$ & 06 41 16.4 & $-$01 05 14.0 & 343.352 & 16.4 &  8.6 & $0.080 \pm\ 0.000$ & $11343 \pm\ 162$ \\ 
          MONR2 & $213.704$ & $-12.606$ & 06 05 19.5 & $-$06 22 38.2 & 355.975 &  9.5 &  1.2 & $0.017 \pm\ 0.005$ & $8986 \pm\ 65$ \\ 
           S288 & $218.739$ & $1.848$ & 07 06 09.9 & $-$04 14 14.0 & 342.672 & 14.6 &  6.9 & $0.045 \pm\ 0.009$ & $14578 \pm\ 195$ \\ 
          WB952 & $218.740$ & $1.848$ & 07 06 10.0 & $-$04 14 17.0 & 342.657 & 14.6 &  6.9 & $0.080 \pm\ 0.000$ & $10671 \pm\ 143$ \\ 
           S291 & $220.521$ & $-2.767$ & 06 53 00.3 & $-$07 56 56.2 & 337.916 & 17.5 & 10.1 & $0.080 \pm\ 0.000$ & $12037 \pm\ 725$ \\ 
           S297 & $225.472$ & $-2.578$ & 07 02 55.7 & $-$12 15 12.0 & 358.434 &  8.7 &  0.3 & $0.080 \pm\ 0.000$ & $7537 \pm\ 158$ \\ 
           S298 & $227.796$ & $-0.135$ & 07 16 14.0 & $-$13 10 07.1 & 341.548 & 12.8 &  5.5 & $0.111 \pm\ 0.014$ & $12495 \pm\ 249$ \\ 
           RCW6 & $231.491$ & $-4.384$ & 07 07 46.8 & $-$18 25 00.7 & 339.802 & 12.8 &  5.7 & $0.080 \pm\ 0.000$ & $9098 \pm\ 286$ \\ 
           RCW8 & $233.760$ & $-0.203$ & 07 27 50.0 & $-$18 26 24.0 & 345.360 & 10.9 &  3.4 & $0.080 \pm\ 0.000$ & $9482 \pm\ 209$ \\ 
           S305 & $233.761$ & $-0.191$ & 07 27 52.8 & $-$18 26 08.0 & 345.310 & 10.9 &  3.4 & $0.080 \pm\ 0.000$ & $9024 \pm\ 206$ \\ 
           S311 & $243.166$ & $0.364$ & 07 50 17.6 & $-$26 18 28.2 & 336.623 & 11.9 &  5.3 & $0.079 \pm\ 0.012$ & $10477 \pm\ 214$ \\ 
\enddata 
\end{deluxetable}

\begin{deluxetable}{lcrrrrrrrrc}
\tabletypesize{\scriptsize}
\tablecaption{Radio Recombination Line Parameters of Galactic \hii\ Regions \label{tab:line}}
\tablewidth{0pt}
\tablehead{
\colhead{} & \colhead{} & \colhead{$T_{\rm L}$} & \colhead{$\sigma\,T_{\rm L}$} & 
\colhead{$\Delta{V}$} & \colhead{$\sigma\,\Delta{V}$} & 
\colhead{$V_{\rm LSR}$} & \colhead{$\sigma\,V_{\rm LSR}$} & 
\colhead{$t_{\rm intg}$} & \colhead{rms} & \colhead{} \\ 
\colhead{Name} & \colhead{Atom} & \colhead{(mK)} & \colhead{(mK)} & 
\colhead{(\kms)} & \colhead{(\kms)} & \colhead{(\kms)} & \colhead{(\kms)} & 
\colhead{(hr)} & \colhead{(mK)} & \colhead{QF} 
} 
\startdata 
      G7.47+0.1 & H &  192.51 &   0.90 &  32.98 &  0.18 & $ -16.93$ &   0.08 &  4.8 &  5.49 & B \\ 
                & He &   21.05 &   1.05 &  25.18 &  1.62 & $ -17.34$ &   0.63 & \nodata & \nodata & \nodata \\ 
                & C &   14.29 &   1.84 &   8.06 &  1.26 & $ -15.78$ &   0.52 & \nodata & \nodata & \nodata \\ 
            S64 & H &  391.96 &   1.82 &  21.00 &  0.11 & $  -1.07$ &   0.05 &  2.4 &  7.97 & B \\ 
                & He &   36.70 &   2.30 &  13.12 &  0.95 & $  -1.52$ &   0.40 & \nodata & \nodata & \nodata \\ 
                & C &   58.68 &   5.13 &   2.63 &  0.27 & $   8.88$ &   0.11 & \nodata & \nodata & \nodata \\ 
            W43 & H & 2032.18 &   3.68 &  33.51 &  0.07 & $  89.84$ &   0.03 &  2.4 & 12.11 & A \\ 
                & He &  158.98 &   4.04 &  30.47 &  1.31 & $  88.98$ &   0.51 & \nodata & \nodata & \nodata \\ 
   G32.797+0.19 & H &  539.84 &   1.34 &  29.49 &  0.08 & $  14.90$ &   0.04 &  4.8 &  5.85 & A \\ 
                & He &   52.52 &   1.51 &  23.76 &  0.81 & $  15.33$ &   0.33 & \nodata & \nodata & \nodata \\ 
                & C &   26.70 &   2.68 &   7.58 &  0.93 & $  12.99$ &   0.38 & \nodata & \nodata & \nodata \\ 
        NRAO584 & H & 1661.73 &   4.86 &  24.42 &  0.08 & $  53.11$ &   0.04 &  2.4 & 10.99 & A \\ 
                & He &  132.50 &   6.00 &  16.18 &  0.86 & $  52.62$ &   0.36 & \nodata & \nodata & \nodata \\ 
                & C &   40.11 &   9.04 &   7.07 &  1.88 & $  57.36$ &   0.79 & \nodata & \nodata & \nodata \\ 
            W48 & H & 1885.31 &   2.53 &  23.55 &  0.04 & $  46.00$ &   0.02 &  2.4 & 11.00 & A \\ 
                & He &  201.76 &   3.09 &  15.81 &  0.28 & $  46.23$ &   0.12 & \nodata & \nodata & \nodata \\ 
                & C &   84.68 &   5.22 &   5.53 &  0.39 & $  41.96$ &   0.17 & \nodata & \nodata & \nodata \\ 
  G38.875+0.308 & H &   48.24 &   0.90 &  28.79 &  0.63 & $ -14.98$ &   0.26 &  2.4 &  6.11 & C \\ 
  G39.728-0.396 & H &   46.92 &   1.01 &  25.67 &  0.67 & $  57.68$ &   0.27 &  2.4 &  5.78 & C \\ 
            S76 & H &  355.60 &   1.18 &  22.44 &  0.09 & $  21.74$ &   0.04 &  2.4 &  6.68 & A \\ 
                & He &   25.12 &   1.49 &  14.34 &  1.04 & $  19.84$ &   0.42 & \nodata & \nodata & \nodata \\ 
                & C &   35.86 &   3.43 &   2.69 &  0.30 & $  32.86$ &   0.13 & \nodata & \nodata & \nodata \\ 
            K47 & H &  742.94 &   1.64 &  27.62 &  0.07 & $  54.21$ &   0.03 &  2.4 &  7.95 & A \\ 
                & He &   76.45 &   1.98 &  20.40 &  0.74 & $  54.08$ &   0.28 & \nodata & \nodata & \nodata \\ 
   G46.495-0.25 & H &  257.77 &   1.10 &  18.32 &  0.09 & $  57.68$ &   0.04 &  4.4 &  4.91 & A \\ 
                & He &   26.45 &   1.27 &  13.81 &  0.78 & $  57.53$ &   0.32 & \nodata & \nodata & \nodata \\ 
            W51 & H & 6720.34 &   5.88 &  30.16 &  0.03 & $  55.95$ &   0.01 &  2.4 & 28.10 & A \\ 
                & He &  663.06 &   6.60 &  25.52 &  0.35 & $  56.00$ &   0.13 & \nodata & \nodata & \nodata \\ 
                & C &  110.95 &  10.31 &  10.28 &  1.23 & $  57.17$ &   0.50 & \nodata & \nodata & \nodata \\ 
     G52.75+0.3 & H &   59.22 &   0.74 &  25.45 &  0.37 & $  11.18$ &   0.16 &  4.8 &  4.18 & C \\ 
                & He &    6.81 &   0.98 &  14.29 &  2.40 & $   9.99$ &   1.01 & \nodata & \nodata & \nodata \\ 
     G55.11+2.4 & H &   49.39 &   0.29 &  32.01 &  0.22 & $ -76.22$ &   0.09 & 23.9 &  1.75 & B \\ 
                & He &    4.16 &   0.29 &  33.16 &  2.72 & $ -78.01$ &   1.12 & \nodata & \nodata & \nodata \\ 
     G59.80+0.2 & H &  106.76 &   0.82 &  21.83 &  0.19 & $  -3.73$ &   0.08 &  4.4 &  4.52 & B \\ 
                & He &    9.55 &   1.03 &  13.97 &  1.73 & $  -4.05$ &   0.74 & \nodata & \nodata & \nodata \\ 
            S87 & H &  114.01 &   0.86 &  21.21 &  0.18 & $  17.53$ &   0.08 &  5.0 &  4.65 & B \\ 
                & C &   38.96 &   1.90 &   4.34 &  0.25 & $  20.85$ &   0.10 & \nodata & \nodata & \nodata \\ 
            S88 & H &  810.84 &   1.61 &  26.03 &  0.06 & $  26.03$ &   0.03 &  2.4 &  7.96 & A \\ 
                & He &   61.00 &   1.98 &  17.34 &  0.65 & $  28.28$ &   0.28 & \nodata & \nodata & \nodata \\ 
                & C &  106.08 &   3.61 &   5.20 &  0.20 & $  20.41$ &   0.09 & \nodata & \nodata & \nodata \\ 
            S90 & H &  197.08 &   0.92 &  25.15 &  0.14 & $  16.77$ &   0.06 &  4.8 &  4.46 & B \\ 
                & He &   18.23 &   1.01 &  20.70 &  1.32 & $  16.45$ &   0.56 & \nodata & \nodata & \nodata \\ 
            S93 & H &  218.61 &   0.84 &  23.87 &  0.11 & $  23.32$ &   0.05 &  4.8 &  4.49 & B \\ 
                & He &   17.41 &   1.06 &  15.13 &  1.06 & $  23.56$ &   0.45 & \nodata & \nodata & \nodata \\ 
                & C &   13.61 &   1.60 &   6.61 &  0.91 & $  20.66$ &   0.38 & \nodata & \nodata & \nodata \\ 
            S98 & H &   47.82 &   0.54 &  24.70 &  0.32 & $ -64.69$ &   0.14 &  9.5 &  2.98 & C \\ 
                & He &    5.51 &   0.71 &  13.92 &  2.09 & $ -63.94$ &   0.88 & \nodata & \nodata & \nodata \\ 
     G69.94+1.5 & H &  364.80 &   0.90 &  26.98 &  0.08 & $ -64.77$ &   0.03 &  4.8 &  4.93 & A \\ 
                & He &   39.56 &   1.05 &  21.03 &  0.75 & $ -65.60$ &   0.29 & \nodata & \nodata & \nodata \\ 
          K3-50 & H &  656.17 &   3.49 &  36.96 &  0.23 & $ -25.57$ &   0.10 &  2.4 &  9.97 & A \\ 
                & He &   41.67 &   4.42 &  23.91 &  3.17 & $ -24.69$ &   1.26 & \nodata & \nodata & \nodata \\ 
                & C &   58.19 &   6.42 &  10.93 &  1.48 & $ -26.33$ &   0.62 & \nodata & \nodata & \nodata \\ 
     G75.77+0.3 & H &  547.20 &   1.11 &  28.17 &  0.07 & $  -8.81$ &   0.03 &  4.0 &  5.93 & A \\ 
                & He &   56.16 &   1.33 &  20.73 &  0.63 & $  -9.61$ &   0.25 & \nodata & \nodata & \nodata \\ 
                & C &   18.81 &   2.71 &   4.91 &  0.85 & $  -2.10$ &   0.35 & \nodata & \nodata & \nodata \\ 
   G75.834+0.40 & H & 1101.76 &   1.46 &  30.53 &  0.05 & $  -4.68$ &   0.02 &  2.4 &  8.75 & A \\ 
                & He &   97.48 &   1.60 &  26.50 &  0.58 & $  -5.81$ &   0.22 & \nodata & \nodata & \nodata \\ 
                & C &   46.21 &   4.00 &   4.31 &  0.45 & $   0.34$ &   0.18 & \nodata & \nodata & \nodata \\ 
     G76.15-0.3 & H &  117.59 &   0.79 &  30.88 &  0.24 & $ -30.48$ &   0.10 &  4.8 &  5.10 & B \\ 
                & He &    8.54 &   0.88 &  25.74 &  3.22 & $ -31.36$ &   1.30 & \nodata & \nodata & \nodata \\ 
                & C &   10.23 &   1.95 &   5.12 &  1.15 & $ -30.23$ &   0.48 & \nodata & \nodata & \nodata \\ 
           S106 & H &  675.66 &   3.22 &  42.02 &  0.23 & $   3.04$ &   0.10 &  2.4 & 14.43 & B \\ 
                & He &   29.79 &   4.10 &  26.50 &  4.37 & $   2.70$ &   1.79 & \nodata & \nodata & \nodata \\ 
                & C &   45.02 &   6.20 &  11.36 &  1.86 & $  -4.30$ &   0.78 & \nodata & \nodata & \nodata \\ 
     G77.98+0.0 & H &  169.85 &   1.22 &  28.69 &  0.24 & $  -3.59$ &   0.10 &  2.4 &  7.01 & B \\ 
                & He &   10.93 &   1.16 &  31.92 &  3.92 & $  -7.64$ &   1.66 & \nodata & \nodata & \nodata \\ 
     G78.03+0.6 & H &  213.28 &   1.22 &  27.24 &  0.18 & $   1.25$ &   0.08 &  2.4 &  6.82 & B \\ 
                & He &   17.39 &   1.41 &  20.22 &  1.89 & $   1.93$ &   0.80 & \nodata & \nodata & \nodata \\ 
           S108 & H &  180.08 &   1.23 &  24.60 &  0.19 & $   3.88$ &   0.08 &  2.4 &  7.23 & B \\ 
                & He &   19.73 &   1.55 &  15.37 &  1.39 & $   6.56$ &   0.59 & \nodata & \nodata & \nodata \\ 
            DR7 & H &  456.47 &   1.81 &  30.01 &  0.14 & $ -38.83$ &   0.06 &  4.8 &  5.56 & A \\ 
                & He &   50.09 &   2.16 &  21.73 &  1.17 & $ -37.82$ &   0.47 & \nodata & \nodata & \nodata \\ 
                & C &   14.91 &   2.92 &  12.43 &  3.69 & $ -39.60$ &   1.33 & \nodata & \nodata & \nodata \\ 
     G79.42+2.4 & H &   49.85 &   0.86 &  20.79 &  0.55 & $   5.17$ &   0.18 &  4.8 &  4.46 & C \\ 
     G79.96+0.9 & H &   39.27 &   1.05 &  23.01 &  0.73 & $ -12.76$ &   0.30 &  4.8 &  5.09 & C \\ 
     G80.35+0.7 & H &  107.68 &   0.93 &  26.46 &  0.26 & $ -65.13$ &   0.11 &  4.8 &  5.46 & C \\ 
                & He &    9.74 &   1.17 &  16.75 &  2.33 & $ -65.55$ &   0.98 & \nodata & \nodata & \nodata \\ 
     G80.88+0.4 & H &  133.09 &   0.90 &  23.18 &  0.18 & $   0.24$ &   0.08 &  4.8 &  4.99 & B \\ 
                & He &   13.82 &   0.98 &  19.76 &  1.62 & $   0.20$ &   0.68 & \nodata & \nodata & \nodata \\ 
     G80.94-0.1 & H &  595.13 &   1.82 &  28.66 &  0.10 & $  -6.89$ &   0.04 &  2.4 &  7.20 & A \\ 
                & He &   27.25 &   1.96 &  26.33 &  2.59 & $  -8.79$ &   0.99 & \nodata & \nodata & \nodata \\ 
                & C &   28.37 &   4.44 &   5.24 &  1.00 & $  -2.89$ &   0.39 & \nodata & \nodata & \nodata \\ 
     G81.25+1.1 & H &  122.62 &   0.93 &  22.90 &  0.20 & $  10.60$ &   0.09 &  4.8 &  4.94 & B \\ 
                & He &   11.15 &   1.30 &  11.82 &  1.59 & $   9.78$ &   0.68 & \nodata & \nodata & \nodata \\ 
           DR21 & H & 2215.95 &   3.86 &  35.91 &  0.07 & $  -1.57$ &   0.03 &  2.4 & 16.54 & A \\ 
                & He &  162.67 &   4.30 &  31.02 &  1.16 & $  -2.20$ &   0.44 & \nodata & \nodata & \nodata \\ 
                & C &   56.04 &   7.79 &  14.99 &  2.15 & $  -8.12$ &   1.09 & \nodata & \nodata & \nodata \\ 
     G82.57+0.4 & H &  165.36 &   1.55 &  23.43 &  0.25 & $ -13.91$ &   0.11 &  2.4 &  8.17 & B \\ 
                & He &   18.53 &   2.05 &  13.31 &  1.70 & $ -15.47$ &   0.72 & \nodata & \nodata & \nodata \\ 
           S112 & H &   66.11 &   0.84 &  23.08 &  0.34 & $   5.56$ &   0.14 &  4.8 &  4.52 & C \\ 
                & He &    9.13 &   1.02 &  15.56 &  2.00 & $   5.61$ &   0.85 & \nodata & \nodata & \nodata \\ 
     G85.24+0.0 & H &   84.41 &   0.89 &  26.88 &  0.33 & $ -35.09$ &   0.14 &  4.8 &  5.23 & C \\ 
                & He &    7.48 &   0.92 &  24.65 &  3.52 & $ -33.80$ &   1.49 & \nodata & \nodata & \nodata \\ 
          S127A & H &   41.61 &   0.67 &  28.72 &  0.54 & $ -99.92$ &   0.22 &  4.8 &  4.20 & C \\ 
                & He &    4.53 &   0.89 &  15.76 &  3.59 & $ -98.57$ &   1.52 & \nodata & \nodata & \nodata \\ 
          WB85A & H &   39.99 &   0.56 &  27.74 &  0.45 & $-100.57$ &   0.19 &  7.2 &  3.33 & C \\ 
                & He &    4.00 &   0.49 &  37.45 &  5.98 & $-107.36$ &   2.33 & \nodata & \nodata & \nodata \\ 
           S128 & H &   61.66 &   0.70 &  26.99 &  0.36 & $ -74.72$ &   0.15 &  4.8 &  4.20 & C \\ 
                & He &    5.14 &   0.57 &  42.20 &  6.11 & $ -82.96$ &   2.38 & \nodata & \nodata & \nodata \\ 
           S146 & H &  262.53 &   0.77 &  25.77 &  0.09 & $ -57.07$ &   0.04 &  4.8 &  4.29 & A \\ 
                & He &   29.43 &   0.94 &  18.05 &  0.76 & $ -58.03$ &   0.29 & \nodata & \nodata & \nodata \\ 
                & C &    9.28 &   1.50 &   6.90 &  1.38 & $ -49.75$ &   0.57 & \nodata & \nodata & \nodata \\ 
           S147 & H &   65.22 &   0.69 &  26.10 &  0.32 & $ -54.88$ &   0.14 &  4.8 &  3.92 & C \\ 
                & C &   12.92 &   1.52 &   5.43 &  0.74 & $ -54.18$ &   0.31 & \nodata & \nodata & \nodata \\ 
           S152 & H &  147.49 &   0.69 &  29.57 &  0.16 & $ -51.28$ &   0.07 &  4.8 &  4.08 & B \\ 
                & He &    7.38 &   0.83 &  20.23 &  2.67 & $ -52.95$ &   1.12 & \nodata & \nodata & \nodata \\ 
                & C &   23.69 &   1.86 &   4.04 &  0.37 & $ -50.77$ &   0.16 & \nodata & \nodata & \nodata \\ 
           S156 & H &  190.93 &   0.74 &  38.05 &  0.17 & $ -50.96$ &   0.07 &  4.8 &  4.52 & B \\ 
                & He &   12.47 &   0.75 &  38.91 &  3.46 & $ -53.92$ &   1.35 & \nodata & \nodata & \nodata \\ 
                & C &   22.07 &   2.26 &   4.49 &  0.57 & $ -50.69$ &   0.21 & \nodata & \nodata & \nodata \\ 
        NGC7538 & H & 1005.76 &   1.87 &  27.07 &  0.06 & $ -61.60$ &   0.02 &  2.4 &  9.37 & A \\ 
                & He &  106.75 &   2.19 &  20.92 &  0.56 & $ -62.37$ &   0.22 & \nodata & \nodata & \nodata \\ 
                & C &   13.05 &   3.61 &   7.52 &  2.57 & $ -56.75$ &   1.06 & \nodata & \nodata & \nodata \\ 
           S159 & H &  177.33 &   0.84 &  26.82 &  0.15 & $ -66.68$ &   0.06 &  4.8 &  4.75 & B \\ 
                & He &   10.77 &   1.20 &  13.75 &  1.92 & $ -70.46$ &   0.76 & \nodata & \nodata & \nodata \\ 
                & C &   31.13 &   2.14 &   4.22 &  0.35 & $ -57.67$ &   0.15 & \nodata & \nodata & \nodata \\ 
           S162 & H &  102.00 &   0.73 &  28.81 &  0.24 & $ -43.86$ &   0.10 &  4.8 &  4.34 & B \\ 
                & He &    9.80 &   0.90 &  18.96 &  2.02 & $ -47.93$ &   0.85 & \nodata & \nodata & \nodata \\ 
           S168 & H &   40.18 &   0.78 &  26.79 &  0.63 & $ -43.67$ &   0.25 &  4.8 &  4.28 & C \\ 
          WB380 & H &   32.67 &   0.62 &  29.73 &  0.70 & $ -79.13$ &   0.28 &  4.4 &  4.05 & C \\ 
           S186 & H &   19.48 &   0.73 &  26.98 &  1.19 & $ -41.63$ &   0.50 &  4.8 &  4.25 & D \\ 
         WB399B & H &    7.84 &   0.23 &  20.92 &  0.71 & $ -87.20$ &   0.30 & 49.0 &  1.18 & D \\ 
                & C &    7.71 &   0.63 &   2.73 &  0.26 & $ -83.07$ &   0.11 & \nodata & \nodata & \nodata \\ 
    G132.16-0.7 & H &   79.61 &   0.54 &  24.93 &  0.20 & $ -56.39$ &   0.08 &  9.5 &  3.09 & B \\ 
                & He &    6.76 &   0.58 &  22.37 &  2.21 & $ -54.41$ &   0.93 & \nodata & \nodata & \nodata \\ 
                & C &    8.36 &   1.26 &   4.68 &  0.82 & $ -56.49$ &   0.35 & \nodata & \nodata & \nodata \\ 
             W3 & H & 3837.85 &   4.75 &  27.69 &  0.04 & $ -40.68$ &   0.02 & 18.7 & 13.58 & A \\ 
                & He &  401.81 &   5.40 &  21.76 &  0.34 & $ -40.83$ &   0.14 & \nodata & \nodata & \nodata \\ 
                & C &  183.21 &  10.60 &   6.01 &  0.50 & $ -40.13$ &   0.19 & \nodata & \nodata & \nodata \\ 
   G133.790+1.4 & H &  787.00 &   2.97 &  27.46 &  0.12 & $ -49.42$ &   0.05 &  4.8 &  5.56 & A \\ 
                & He &   87.03 &   3.46 &  21.75 &  1.25 & $ -50.87$ &   0.48 & \nodata & \nodata & \nodata \\ 
    G136.91+1.0 & H &   26.84 &   0.58 &  23.49 &  0.60 & $ -40.65$ &   0.25 &  9.5 &  2.92 & C \\ 
           S201 & H &   79.63 &   0.78 &  23.80 &  0.27 & $ -35.47$ &   0.11 &  4.8 &  3.89 & B \\ 
                & C &   19.95 &   2.03 &   3.47 &  0.41 & $ -39.72$ &   0.17 & \nodata & \nodata & \nodata \\ 
           S206 & H &  219.24 &   0.80 &  27.83 &  0.12 & $ -26.61$ &   0.05 &  4.8 &  4.93 & B \\ 
                & He &   22.43 &   0.93 &  20.52 &  0.99 & $ -26.98$ &   0.42 & \nodata & \nodata & \nodata \\ 
           S209 & H &  160.45 &   0.71 &  31.66 &  0.16 & $ -51.18$ &   0.07 &  9.5 &  3.28 & B \\ 
                & He &   13.10 &   0.77 &  27.34 &  1.91 & $ -52.55$ &   0.79 & \nodata & \nodata & \nodata \\ 
                & C &   12.10 &   2.00 &   4.13 &  0.83 & $ -55.01$ &   0.34 & \nodata & \nodata & \nodata \\ 
           S211 & H &   50.04 &   0.66 &  28.55 &  0.44 & $ -35.22$ &   0.18 &  4.8 &  3.99 & C \\ 
                & C &    6.93 &   1.46 &   5.78 &  1.41 & $ -41.17$ &   0.60 & \nodata & \nodata & \nodata \\ 
           S212 & H &   45.43 &   0.76 &  25.57 &  0.50 & $ -43.95$ &   0.21 &  4.8 &  4.32 & C \\ 
                & He &    4.61 &   0.82 &  22.16 &  4.55 & $ -46.70$ &   1.93 & \nodata & \nodata & \nodata \\ 
           S228 & H &   60.90 &   0.82 &  23.09 &  0.36 & $ -17.31$ &   0.15 &  4.8 &  4.28 & C \\ 
                & C &   19.31 &   2.37 &   2.75 &  0.39 & $  -6.44$ &   0.17 & \nodata & \nodata & \nodata \\ 
           S235 & H &   81.78 &   0.94 &  20.86 &  0.28 & $ -25.61$ &   0.12 &  4.8 &  4.54 & C \\ 
           S237 & H &   62.54 &   0.76 &  22.98 &  0.32 & $  -0.64$ &   0.14 &  4.8 &  4.10 & C \\ 
                & C &   18.40 &   2.26 &   2.57 &  0.37 & $  -9.34$ &   0.15 & \nodata & \nodata & \nodata \\ 
           S257 & H &  105.83 &   0.88 &  22.12 &  0.22 & $   5.05$ &   0.09 &  4.8 &  4.31 & B \\ 
           S269 & H &   61.61 &   0.74 &  24.23 &  0.34 & $  12.87$ &   0.14 &  4.8 &  4.11 & C \\ 
                & C &   27.59 &   2.39 &   2.34 &  0.23 & $  18.10$ &   0.10 & \nodata & \nodata & \nodata \\ 
     G201.6+1.6 & H &   25.59 &   0.51 &  25.91 &  0.63 & $  23.14$ &   0.25 &  9.5 &  3.34 & C \\ 
           OriA & H & 9728.44 &   6.99 &  26.13 &  0.02 & $  -2.25$ &   0.01 &  2.4 & 35.84 & A \\ 
                & He & 1160.24 &   8.74 &  18.05 &  0.20 & $  -2.31$ &   0.07 & \nodata & \nodata & \nodata \\ 
                & C &  323.49 &  15.63 &   5.92 &  0.36 & $   8.91$ &   0.14 & \nodata & \nodata & \nodata \\ 
          WB870 & H &   68.31 &   0.70 &  28.56 &  0.34 & $  55.32$ &   0.14 &  4.8 &  4.46 & C \\ 
          MONR2 & H &  942.25 &   3.54 &  29.81 &  0.13 & $  10.82$ &   0.05 &  2.4 &  8.51 & A \\ 
                & He &   20.10 &   3.94 &  24.21 &  5.55 & $  11.65$ &   2.33 & \nodata & \nodata & \nodata \\ 
                & C &  144.47 &   8.44 &   5.38 &  0.38 & $   9.08$ &   0.15 & \nodata & \nodata & \nodata \\ 
           S288 & H &   70.43 &   0.55 &  24.52 &  0.22 & $  54.49$ &   0.09 &  9.5 &  3.04 & B \\ 
                & He &    5.85 &   0.76 &  13.14 &  1.96 & $  54.53$ &   0.83 & \nodata & \nodata & \nodata \\ 
                & C &    9.80 &   1.33 &   4.24 &  0.67 & $  56.05$ &   0.28 & \nodata & \nodata & \nodata \\ 
          WB952 & H &   73.83 &   0.74 &  24.62 &  0.29 & $  54.54$ &   0.12 &  4.8 &  4.32 & C \\ 
           S291 & H &    5.40 &   0.23 &  31.82 &  1.69 & $  70.33$ &   0.67 & 39.6 &  1.60 & D \\ 
           S297 & H &   40.34 &   0.58 &  22.59 &  0.37 & $   1.94$ &   0.16 &  9.1 &  3.01 & C \\ 
                & C &   23.10 &   1.98 &   1.90 &  0.19 & $  12.26$ &   0.08 & \nodata & \nodata & \nodata \\ 
           S298 & H &   15.09 &   0.17 &  28.89 &  0.37 & $  51.89$ &   0.16 & 108.8 &  0.97 & C \\ 
                & He &    2.31 &   0.19 &  20.90 &  2.04 & $  52.12$ &   0.86 & \nodata & \nodata & \nodata \\ 
           RCW6 & H &   42.21 &   0.98 &  20.54 &  0.55 & $  54.25$ &   0.23 &  4.8 &  5.06 & C \\ 
           RCW8 & H &   54.33 &   0.83 &  24.06 &  0.43 & $  35.31$ &   0.18 &  4.8 &  4.54 & C \\ 
           S305 & H &   51.88 &   0.79 &  26.05 &  0.47 & $  35.44$ &   0.20 &  4.8 &  4.57 & C \\ 
           S311 & H &   98.12 &   0.94 &  22.29 &  0.25 & $  51.68$ &   0.10 &  4.8 &  5.19 & C \\ 
                & He &   12.56 &   1.19 &  13.77 &  1.52 & $  53.05$ &   0.64 & \nodata & \nodata & \nodata \\ 
\enddata 
\end{deluxetable}

\begin{deluxetable}{lrrrrrrrrrrrrc}
\tabletypesize{\scriptsize}
\tablecaption{Radio Continuum Parameters of Galactic \hii\ Regions \label{tab:cont}}
\tablewidth{0pt}
\tablehead{
\colhead{} & \multicolumn{4}{c}{\underline{~~~~~~~~~~~~~~~~~R.A.~~~~~~~~~~~~~~~~~}} & 
\multicolumn{4}{c}{\underline{~~~~~~~~~~~~~~~~~Decl.~~~~~~~~~~~~~~~~~}} & 
\multicolumn{4}{c}{\underline{~~~~~~~~~~~~~~~Average~~~~~~~~~~~~~~~}} & \colhead{} \\ 
\colhead{} & \colhead{$T_{\rm C}$} & \colhead{$\sigma\,T_{\rm C}$} & 
\colhead{$\Theta$} & \colhead{$\sigma\,\Theta$} & 
\colhead{$T_{\rm C}$} & \colhead{$\sigma\,T_{\rm C}$} & 
\colhead{$\Theta$} & \colhead{$\sigma\,\Theta$} & 
\colhead{$T_{\rm C}$} & \colhead{$\sigma\,T_{\rm C}$} & 
\colhead{$\Theta$} & \colhead{$\sigma\,\Theta$} & \colhead{} \\ 
\colhead{Name} & \colhead{(K)} & \colhead{(K)} & \colhead{($\prime$)} & \colhead{($\prime$)} & 
\colhead{(K)} & \colhead{(K)} & \colhead{($\prime$)} & \colhead{($\prime$)} & 
\colhead{(K)} & \colhead{(K)} & \colhead{($\prime$)} & \colhead{($\prime$)} & \colhead{QF} 
} 
\startdata 
      G7.47+0.1 &   3.370 & 0.015 &   1.472 & 0.008 &   3.417 & 0.019 &   1.446 & 0.009 &   3.393 & 0.012 &   1.459 & 0.006 & A \\ 
            S64 &   3.583 & 0.029 &   5.609 & 0.099 &   3.507 & 0.017 &   4.960 & 0.039 &   3.545 & 0.017 &   5.274 & 0.051 & C \\ 
            W43 &  22.099 & 0.071 &   2.462 & 0.016 &  22.639 & 0.118 &   1.803 & 0.024 &  22.369 & 0.069 &   2.107 & 0.015 & C \\ 
   G32.797+0.19 &   7.250 & 0.043 &   1.439 & 0.010 &   7.155 & 0.046 &   1.439 & 0.011 &   7.202 & 0.031 &   1.439 & 0.007 & A \\ 
        NRAO584 &  16.404 & 0.144 &   1.824 & 0.019 &  16.877 & 0.116 &   1.652 & 0.013 &  16.641 & 0.092 &   1.736 & 0.011 & A \\ 
            W48 &  20.117 & 0.123 &   1.642 & 0.012 &  19.696 & 0.142 &   1.683 & 0.014 &  19.907 & 0.094 &   1.662 & 0.009 & A \\ 
  G38.875+0.308 &   0.638 & 0.008 &   1.367 & 0.020 &   0.676 & 0.005 &   1.428 & 0.013 &   0.657 & 0.005 &   1.397 & 0.012 & B \\ 
  G39.728-0.396 &   0.542 & 0.006 &   1.530 & 0.020 &   0.532 & 0.004 &   1.458 & 0.014 &   0.537 & 0.004 &   1.493 & 0.012 & B \\ 
            S76 &   3.322 & 0.026 &   1.922 & 0.018 &   3.303 & 0.011 &   1.791 & 0.007 &   3.312 & 0.014 &   1.855 & 0.009 & B \\ 
            K47 &   7.901 & 0.075 &   1.730 & 0.019 &   9.154 & 0.109 &   1.732 & 0.024 &   8.528 & 0.066 &   1.731 & 0.016 & C \\ 
   G46.495-0.25 &   1.156 & 0.006 &   2.638 & 0.023 &   1.245 & 0.003 &   3.967 & 0.012 &   1.201 & 0.003 &   3.235 & 0.015 & D \\ 
            W51 &  74.962 & 0.309 &   1.756 & 0.011 &  73.925 & 0.458 &   1.833 & 0.014 &  74.443 & 0.276 &   1.794 & 0.009 & B \\ 
     G52.75+0.3 &   0.711 & 0.005 &   1.529 & 0.013 &   0.697 & 0.005 &   1.538 & 0.012 &   0.704 & 0.004 &   1.534 & 0.009 & B \\ 
     G55.11+2.4 &   1.168 & 0.004 &   1.521 & 0.007 &   1.169 & 0.003 &   1.547 & 0.004 &   1.169 & 0.003 &   1.534 & 0.004 & A \\ 
     G59.80+0.2 &   1.081 & 0.007 &   1.833 & 0.014 &   1.109 & 0.005 &   1.841 & 0.009 &   1.095 & 0.004 &   1.837 & 0.008 & A \\ 
            S87 &   0.900 & 0.004 &   1.564 & 0.008 &   0.955 & 0.005 &   1.462 & 0.009 &   0.928 & 0.003 &   1.512 & 0.006 & B \\ 
            S88 &   9.476 & 0.069 &   1.487 & 0.013 &   9.700 & 0.037 &   1.460 & 0.007 &   9.588 & 0.039 &   1.474 & 0.007 & B \\ 
            S90 &   1.993 & 0.026 &   2.895 & 0.044 &   1.971 & 0.028 &   2.922 & 0.048 &   1.982 & 0.019 &   2.908 & 0.032 & B \\ 
            S93 &   2.270 & 0.009 &   1.655 & 0.008 &   2.224 & 0.009 &   1.616 & 0.008 &   2.247 & 0.006 &   1.635 & 0.005 & A \\ 
            S98 &   0.687 & 0.005 &   1.954 & 0.017 &   0.685 & 0.007 &   1.830 & 0.022 &   0.686 & 0.004 &   1.891 & 0.014 & B \\ 
     G69.94+1.5 &   5.115 & 0.029 &   1.731 & 0.011 &   5.139 & 0.012 &   1.670 & 0.005 &   5.127 & 0.016 &   1.700 & 0.006 & A \\ 
          K3-50 &  13.066 & 0.022 &   1.520 & 0.003 &  12.907 & 0.219 &   1.918 & 0.042 &  12.987 & 0.110 &   1.708 & 0.019 & B \\ 
     G75.77+0.3 &   6.924 & 0.051 &   1.595 & 0.014 &   6.923 & 0.045 &   1.591 & 0.012 &   6.923 & 0.034 &   1.593 & 0.009 & B \\ 
   G75.834+0.40 &  14.587 & 0.053 &   1.479 & 0.006 &  14.745 & 0.085 &   1.471 & 0.011 &  14.666 & 0.050 &   1.475 & 0.006 & B \\ 
     G76.15-0.3 &   1.776 & 0.011 &   1.555 & 0.011 &   1.834 & 0.012 &   1.552 & 0.011 &   1.805 & 0.008 &   1.553 & 0.008 & C \\ 
           S106 &  16.875 & 0.049 &   1.473 & 0.005 &  16.338 & 0.028 &   1.668 & 0.003 &  16.606 & 0.028 &   1.568 & 0.003 & B \\ 
     G77.98+0.0 &   1.932 & 0.020 &   2.855 & 0.035 &   1.900 & 0.014 &   2.150 & 0.024 &   1.916 & 0.012 &   2.478 & 0.020 & D \\ 
     G78.03+0.6 &   2.568 & 0.009 &   3.375 & 0.018 &   2.562 & 0.012 &   3.557 & 0.026 &   2.565 & 0.007 &   3.465 & 0.015 & B \\ 
           S108 &   2.036 & 0.014 &   3.597 & 0.036 &   1.920 & 0.019 &   2.944 & 0.041 &   1.978 & 0.012 &   3.254 & 0.028 & C \\ 
            DR7 &   6.129 & 0.060 &   2.169 & 0.035 &   6.393 & 0.083 &   2.521 & 0.039 &   6.261 & 0.051 &   2.338 & 0.026 & B \\ 
     G79.42+2.4 &   0.424 & 0.004 &   4.756 & 0.090 &   0.434 & 0.003 &   8.598 & 0.067 &   0.429 & 0.003 &   6.395 & 0.066 & C \\ 
     G79.96+0.9 &   0.388 & 0.005 &   4.299 & 0.100 &   0.354 & 0.005 &   5.050 & 0.093 &   0.371 & 0.004 &   4.659 & 0.069 & C \\ 
     G80.35+0.7 &   1.529 & 0.007 &   1.785 & 0.010 &   1.553 & 0.015 &   2.022 & 0.024 &   1.541 & 0.008 &   1.900 & 0.013 & C \\ 
     G80.88+0.4 &   1.530 & 0.004 &   2.871 & 0.026 &   1.441 & 0.009 &   2.627 & 0.033 &   1.486 & 0.005 &   2.747 & 0.021 & D \\ 
     G80.94-0.1 &   7.721 & 0.025 &   1.634 & 0.008 &   7.648 & 0.063 &   1.576 & 0.016 &   7.684 & 0.034 &   1.605 & 0.009 & B \\ 
     G81.25+1.1 &   1.121 & 0.004 &   6.672 & 0.056 &   1.163 & 0.007 &   6.392 & 0.105 &   1.142 & 0.004 &   6.530 & 0.060 & D \\ 
           DR21 &  35.969 & 0.157 &   1.981 & 0.010 &  36.966 & 0.140 &   1.447 & 0.006 &  36.467 & 0.105 &   1.693 & 0.006 & A \\ 
     G82.57+0.4 &   1.605 & 0.016 &   6.250 & 0.073 &   1.580 & 0.009 &   3.434 & 0.026 &   1.593 & 0.009 &   4.633 & 0.032 & C \\ 
           S112 &   0.713 & 0.004 &   4.226 & 0.039 &   0.690 & 0.003 &   5.001 & 0.036 &   0.701 & 0.003 &   4.597 & 0.027 & C \\ 
     G85.24+0.0 &   1.036 & 0.009 &   2.315 & 0.023 &   1.077 & 0.015 &   2.448 & 0.040 &   1.056 & 0.009 &   2.381 & 0.023 & C \\ 
          S127A &   0.739 & 0.005 &   1.552 & 0.013 &   0.730 & 0.009 &   1.833 & 0.026 &   0.734 & 0.005 &   1.687 & 0.014 & A \\ 
          WB85A &   0.696 & 0.005 &   1.576 & 0.013 &   0.707 & 0.006 &   1.815 & 0.019 &   0.701 & 0.004 &   1.691 & 0.011 & B \\ 
           S128 &   0.989 & 0.008 &   1.515 & 0.015 &   0.963 & 0.004 &   1.571 & 0.008 &   0.976 & 0.005 &   1.542 & 0.008 & A \\ 
           S146 &   3.493 & 0.015 &   1.526 & 0.008 &   3.423 & 0.024 &   1.548 & 0.012 &   3.458 & 0.014 &   1.537 & 0.007 & A \\ 
           S147 &   0.816 & 0.005 &   1.584 & 0.011 &   0.803 & 0.005 &   1.606 & 0.011 &   0.809 & 0.003 &   1.595 & 0.008 & A \\ 
           S152 &   2.073 & 0.006 &   1.501 & 0.005 &   2.107 & 0.015 &   1.569 & 0.013 &   2.090 & 0.008 &   1.535 & 0.007 & A \\ 
           S156 &   3.494 & 0.011 &   1.498 & 0.006 &   3.544 & 0.016 &   1.474 & 0.008 &   3.519 & 0.010 &   1.486 & 0.005 & A \\ 
        NGC7538 &  12.121 & 0.078 &   1.978 & 0.019 &  12.129 & 0.120 &   2.236 & 0.029 &  12.125 & 0.072 &   2.103 & 0.017 & C \\ 
           S159 &   1.967 & 0.004 &   1.503 & 0.004 &   2.040 & 0.007 &   1.528 & 0.007 &   2.004 & 0.004 &   1.516 & 0.004 & B \\ 
           S162 &   1.282 & 0.015 &   2.283 & 0.033 &   1.345 & 0.013 &   1.840 & 0.031 &   1.313 & 0.010 &   2.050 & 0.023 & B \\ 
           S168 &   0.497 & 0.007 &   2.923 & 0.061 &   0.500 & 0.005 &   2.185 & 0.040 &   0.499 & 0.004 &   2.527 & 0.035 & B \\ 
          WB380 &   0.557 & 0.006 &   1.396 & 0.018 &   0.578 & 0.004 &   1.422 & 0.012 &   0.567 & 0.004 &   1.409 & 0.011 & B \\ 
           S186 &   0.255 & 0.004 &   1.385 & 0.025 &   0.243 & 0.004 &   1.386 & 0.027 &   0.249 & 0.003 &   1.385 & 0.019 & B \\ 
         WB399B &   0.094 & 0.002 &   1.432 & 0.028 &   0.090 & 0.002 &   1.326 & 0.040 &   0.092 & 0.001 &   1.378 & 0.024 & B \\ 
    G132.16-0.7 &   1.004 & 0.007 &   2.889 & 0.022 &   1.068 & 0.005 &   1.474 & 0.008 &   1.036 & 0.004 &   2.064 & 0.010 & B \\ 
             W3 &  50.895 & 0.439 &   1.708 & 0.022 &  50.161 & 0.000 &   1.543 & 0.005 &  50.528 & 0.219 &   1.624 & 0.011 & B \\ 
   G133.790+1.4 &   9.927 & 0.097 &   2.376 & 0.028 &  10.127 & 0.051 &   2.288 & 0.017 &  10.027 & 0.055 &   2.332 & 0.016 & B \\ 
    G136.91+1.0 &   0.285 & 0.005 &   8.067 & 0.171 &   0.255 & 0.005 &   6.997 & 0.182 &   0.270 & 0.004 &   7.513 & 0.126 & C \\ 
           S201 &   0.869 & 0.016 &   2.091 & 0.044 &   0.774 & 0.007 &   1.771 & 0.017 &   0.822 & 0.008 &   1.924 & 0.022 & C \\ 
           S206 &   3.324 & 0.037 &   2.146 & 0.033 &   3.214 & 0.021 &   2.758 & 0.023 &   3.269 & 0.021 &   2.433 & 0.022 & B \\ 
           S209 &   2.890 & 0.024 &   2.956 & 0.047 &   3.015 & 0.019 &   2.628 & 0.026 &   2.952 & 0.015 &   2.787 & 0.026 & B \\ 
           S211 &   0.753 & 0.005 &   1.797 & 0.014 &   0.735 & 0.003 &   1.809 & 0.009 &   0.744 & 0.003 &   1.803 & 0.008 & A \\ 
           S212 &   0.654 & 0.007 &   2.985 & 0.040 &   0.639 & 0.005 &   2.929 & 0.029 &   0.647 & 0.005 &   2.957 & 0.025 & B \\ 
           S228 &   0.712 & 0.006 &   1.839 & 0.018 &   0.685 & 0.009 &   1.853 & 0.028 &   0.699 & 0.005 &   1.846 & 0.017 & B \\ 
           S235 &   0.769 & 0.005 &   4.244 & 0.032 &   0.774 & 0.006 &   3.603 & 0.034 &   0.772 & 0.004 &   3.910 & 0.024 & B \\ 
           S237 &   0.666 & 0.011 &   1.852 & 0.036 &   0.672 & 0.004 &   1.951 & 0.013 &   0.669 & 0.006 &   1.901 & 0.020 & A \\ 
           S257 &   1.072 & 0.010 &   2.401 & 0.027 &   1.109 & 0.008 &   2.516 & 0.021 &   1.090 & 0.006 &   2.458 & 0.017 & B \\ 
           S269 &   0.805 & 0.004 &   1.902 & 0.011 &   0.788 & 0.006 &   1.702 & 0.014 &   0.797 & 0.003 &   1.799 & 0.009 & A \\ 
     G201.6+1.6 &   0.277 & 0.003 &   2.233 & 0.055 &   0.441 & 0.004 &   3.795 & 0.069 &   0.359 & 0.003 &   2.911 & 0.044 & D \\ 
           OriA & 110.540 & 1.610 &   2.771 & 0.057 & 111.040 & 0.440 &   3.430 & 0.016 & 110.790 & 0.835 &   3.083 & 0.033 & B \\ 
          WB870 &   1.233 & 0.009 &   1.420 & 0.012 &   1.188 & 0.007 &   1.436 & 0.009 &   1.211 & 0.006 &   1.428 & 0.008 & B \\ 
          MONR2 &  12.545 & 0.063 &   1.416 & 0.008 &  12.590 & 0.057 &   1.439 & 0.008 &  12.568 & 0.043 &   1.428 & 0.006 & A \\ 
           S288 &   1.212 & 0.010 &   1.780 & 0.021 &   1.556 & 0.008 &   2.034 & 0.018 &   1.384 & 0.006 &   1.903 & 0.014 & D \\ 
          WB952 &   1.054 & 0.005 &   1.516 & 0.008 &   1.049 & 0.000 &   1.489 & 0.005 &   1.052 & 0.002 &   1.502 & 0.005 & B \\ 
           S291 &   0.121 & 0.002 &   3.550 & 0.110 &   0.108 & 0.001 &   3.757 & 0.063 &   0.114 & 0.001 &   3.652 & 0.064 & C \\ 
           S297 &   0.355 & 0.006 &   2.268 & 0.047 &   0.352 & 0.004 &   2.698 & 0.033 &   0.354 & 0.004 &   2.474 & 0.030 & B \\ 
           S298 &   0.297 & 0.003 &   3.341 & 0.045 &   0.325 & 0.004 &   2.770 & 0.060 &   0.311 & 0.003 &   3.042 & 0.039 & C \\ 
           RCW6 &   0.398 & 0.004 &   3.768 & 0.052 &   0.437 & 0.004 &   5.610 & 0.056 &   0.418 & 0.003 &   4.597 & 0.039 & C \\ 
           RCW8 &   0.659 & 0.008 &   2.554 & 0.045 &   0.663 & 0.010 &   2.795 & 0.053 &   0.661 & 0.006 &   2.672 & 0.035 & B \\ 
           S305 &   0.649 & 0.011 &   2.807 & 0.058 &   0.641 & 0.009 &   2.975 & 0.053 &   0.645 & 0.007 &   2.890 & 0.039 & B \\ 
           S311 &   1.235 & 0.016 &   4.861 & 0.096 &   1.241 & 0.033 &   4.683 & 0.171 &   1.238 & 0.019 &   4.771 & 0.099 & C \\ 
\enddata 
\end{deluxetable}

\begin{deluxetable}{rllccc}
\tablecaption{GBT Sample: LTE Electron Temperature Radial Gradient Fits \label{tab:GBTTeFits}}
\tablewidth{0pt}
\tablehead{
\colhead{Az Range} & \colhead{$a$} & \colhead{$b$} & \colhead{} & \colhead{} & \colhead{$\Delta{R_{\rm gal}}$} \\
\colhead{(deg)} & \colhead{(K)} & \colhead{(\Kkpc)} &    \colhead{r} & \colhead{N} & \colhead{(kpc)}
}
\startdata 

$0-360$   & $6404 \pm 442$   & $257 \pm 43$   & $0.689 \pm 0.069$ & 72 & $4.7-21.9$ \\
          & $6404 \pm 501$   & $257 \pm 50$   & \nodata\ & \nodata\ &   \nodata\  \\ \hline

$330-360$ & $4448 \pm 895$   & $455 \pm 87$   & $0.77 \pm 0.12$ & 12 & $8.5-16.4$ \\
          & $4419 \pm 1258$  & $458 \pm 120$  & \nodata\ & \nodata\ & \nodata\    \\ \hline

$0-30$    & $7104 \pm 425$   & $179 \pm 35$   & $0.62 \pm 0.19$ & 33 & $6.0-17.3$ \\
          & $7105 \pm 454$   & $178 \pm 38$   & \nodata\ & \nodata\ &  \nodata\    \\ \hline

$30-60$   & $4874 \pm 309$   & $389 \pm 31$   & $0.885 \pm 0.082$ & 16 & $4.7-16.4$ \\
          & $4876 \pm 343$   & $388 \pm 35$   & \nodata\ & \nodata\ &  \nodata\   \\

\enddata 
\tablecomments{Gradient fit defined by $T_{\rm e} = a + b\,R_{\rm
gal}$.  The second fit uses jackknife resampling.}
\end{deluxetable}

\begin{deluxetable}{rllccc}
\tablecaption{Green Bank Sample: LTE Electron Temperature Radial Gradient Fits \label{tab:GBT140TeFits}}
\tablewidth{0pt}
\tablehead{
\colhead{Az Range} & \colhead{$a$} & \colhead{$b$} & \colhead{} & \colhead{} & \colhead{$\Delta{R_{\rm gal}}$} \\
\colhead{(deg)} & \colhead{(K)} & \colhead{(\Kkpc)} &    \colhead{r} & \colhead{N} & \colhead{(kpc)}
}
\startdata 

$0-360$   & $5756 \pm 303$   & $299 \pm 31$   & $0.669 \pm 0.052$ & 133 & $0.1-21.9$ \\
          & $5756 \pm 315$   & $299 \pm 33$   & \nodata\ & \nodata\ &   \nodata\  \\ \hline

$330-360$ & $4977 \pm 718$   & $391 \pm 73$   & $0.778 \pm 0.066$ & 24 & $2.0-16.4$ \\
          & $4971 \pm 837$   & $392 \pm 84$   & \nodata\ & \nodata\ & \nodata\    \\ \hline

$0-30$    & $6511 \pm 303$   & $228 \pm 25$   & $0.74 \pm 0.104$ & 43 & $4.3-17.3$ \\
          & $6511 \pm 316$   & $228 \pm 26$   & \nodata\ & \nodata\ &  \nodata\    \\ \hline

$30-60$   & $4758 \pm 499$   & $404 \pm 40$   & $0.76 \pm 0.15$ & 28 & $3.7-16.4$ \\
          & $4758 \pm 516$   & $404 \pm 42$   & \nodata\ & \nodata\ &  \nodata\   \\

\enddata 
\tablecomments{Gradient fit defined by $T_{\rm e} = a + b\,R_{\rm
gal}$.  The second fit uses jackknife resampling.}
\end{deluxetable}

\begin{deluxetable}{rllll}
\tablecaption{O/H Radial Gradient Fits \label{tab:O2HFits}}
\tablewidth{0pt}
\tablehead{
\colhead{} & \multicolumn{2}{c}{\underline{~~~~~~~~~~~~~GBT Sample~~~~~~~~~~~~~}} & \multicolumn{2}{c}{\underline{~~~~~~~~~~~Green Bank Sample~~~~~~~~~~~}} \\
\colhead{Az Range} & \colhead{$a$} & \colhead{$b$} & \colhead{$a$} & \colhead{$b$} \\
\colhead{(deg)} & \colhead{(dex)} & \colhead{(\dexkpc)} & \colhead{(dex)} & \colhead{(\dexkpc)}
}
\startdata 

$0-360$   & $8.866 \pm 0.066$   & $-0.0383 \pm 0.0065$   & $8.962 \pm 0.045$   & $-0.0446 \pm 0.0046$ \\
          & $8.866 \pm 0.075$   & $-0.0383 \pm 0.0074$   & $8.962 \pm 0.047$   & $-0.0446 \pm 0.0049$ \\ \hline

$330-360$ & $9.16 \pm 0.13$  & $-0.068 \pm 0.013$   & $9.08 \pm 0.11$  & $-0.058 \pm 0.011$  \\
          & $9.16 \pm 0.19$  & $-0.068 \pm 0.018$   & $9.08 \pm 0.13$  & $-0.058 \pm 0.013$  \\ \hline

$0-30$    & $8.762 \pm 0.063$   & $-0.0266 \pm 0.0053$   & $8.850 \pm 0.045$   & $-0.0339 \pm 0.0037$ \\
          & $8.761 \pm 0.068$   & $-0.0266 \pm 0.0057$   & $8.850 \pm 0.047$   & $-0.0339 \pm 0.0039$ \\ \hline

$30-60$   & $9.094 \pm 0.046$   & $-0.0579 \pm 0.0046$   & $9.111 \pm 0.074$   & $-0.0602 \pm 0.0060$ \\
          & $9.093 \pm 0.051$   & $-0.0579 \pm 0.0053$   & $9.111 \pm 0.077$   & $-0.0602 \pm 0.0063$ \\ \hline

\enddata 
\tablecomments{Gradient fit defined by $12 + {\rm Log(O/H)} = a + b\,R_{\rm
gal}$.  The second fit uses jackknife resampling.}
\end{deluxetable}

\begin{deluxetable}{lccccc}
\tablecaption{3C147 DCR Continuum Calibration \label{tab:cal}}
\tablewidth{0pt}
\tablehead{
\colhead{Frequency} & \colhead{Bandwidth} & \colhead{} & \colhead{Intensity} & \colhead{FWHM} & \colhead{} \\
\colhead{(MHz)} & \colhead{(MHz)} & \colhead{Pol.} & \colhead{(K)} & \colhead{(arcmin)} & \colhead{CF}
} 
\startdata 

8665.0000 & 320.0000  & LL &    $9.1427 \pm 0.1240$  & $83.7267 \pm 1.1384$   & 1.018 \\
8665.0000 & 320.0000  & RR &    $8.9200 \pm 0.1303$  & $83.2861 \pm 0.6417$   & 1.043 \\
8045.6050 &  80.0000  & LL &    $9.4322 \pm 0.0053$  & $91.0121 \pm 0.0165$   & 1.053 \\
8045.6050 &  80.0000  & RR &    $9.2940 \pm 0.0940$  & $91.2244 \pm 0.1271$   & 1.069 \\
8300.0000 &  80.0000  & LL &    $8.7954 \pm 0.0412$  & $86.2657 \pm 1.3826$   & 1.099 \\
8300.0000 &  80.0000  & RR &    $8.9534 \pm 0.0404$  & $86.3678 \pm 1.0142$   & 1.079 \\
8584.8232 &  80.0000  & LL &    $8.9331 \pm 0.1130$  & $85.7964 \pm 0.9476$   & 1.050 \\
8584.8232 &  80.0000  & RR &    $8.7353 \pm 0.1283$  & $86.0558 \pm 1.1671$   & 1.074 \\
8665.3000 &  80.0000  & LL &    $9.1793 \pm 0.0078$  & $84.2725 \pm 0.2749$   & 1.014 \\
8665.3000 &  80.0000  & RR &    $9.0220 \pm 0.0277$  & $84.1787 \pm 0.9257$   & 1.031 \\
8877.0000 &  80.0000  & LL &    $9.6489 \pm 0.0450$  & $80.6957 \pm 1.6937$   & 0.944 \\
8877.0000 &  80.0000  & RR &    $8.5722 \pm 0.0327$  & $80.8070 \pm 1.2275$   & 1.063 \\
9183.0000 &  80.0000  & LL &    $8.2036 \pm 0.0726$  & $79.6762 \pm 1.4276$   & 1.078 \\
9183.0000 &  80.0000  & RR &    $8.5481 \pm 0.1100$  & $79.1310 \pm 1.1645$   & 1.034 \\
9505.0000 &  80.0000  & LL &    $8.2952 \pm 0.2131$  & $78.0025 \pm 1.5026$   & 1.034 \\
9505.0000 &  80.0000  & RR &    $7.8862 \pm 0.1243$  & $77.3615 \pm 1.7872$   & 1.088 \\
9812.0000 &  80.0000  & LL &    $7.5754 \pm 0.1698$  & $73.7876 \pm 0.9622$   & 1.101 \\
9812.0000 &  80.0000  & RR &    $8.2367 \pm 0.1512$  & $74.0209 \pm 0.6705$   & 1.013 \\
\enddata 
\tablecomments{Data taken on 6 January 2008 at Hour Angle = $-1.8\,$hr and elevation =  67\degree.}
\end{deluxetable}


\begin{thebibliography} {}

\bibitem[Afflerbach et al.(1996)]{afflerbach96} 
Afflerbach, A., Churchwell, E., Acord, J. M., Hofner, P., Kurtz, S., \& De Pree,
C. G. 1996, \apjs, 106, 423

\bibitem[Afflerbach et al.(1997)]{afflerbach97}
Afflerbach, A., Churchwell, E., Werner, M. W. 1997, \apj, 478, 190

\bibitem[Anderson \& Bania(2009)]{anderson09}
Anderson, L. D., \& Bania, T. M. 2009,
\apj, 690, 706

\bibitem[Anderson et al.(2011)]{anderson11}
Anderson, L. D., Bania, T. M., Balser, D. S., \& Rood, R. T. 2011,
\apj, in press

\bibitem[Andreuzzi et al.(2010)]{andreuzzi10}
Andreuzzi, G. Bragaglia, A., Tos, M., \& Marconi, G. 2010, \mnras, 412, 1265

\bibitem[Andrievsky et al.(2002)]{andrievsky02}
Andrievsky, S. M., Kovtyukh, V. V., Luck, R. E., L\'{e}pine, J. R. D.,
Maciel, W. J., \& Beletsky, Y. V. 2002, \aap, 392, 491

\bibitem[Andrievsky et al.(2004)]{andrievsky04}
Andrievsky, S. M., Luck, R. E., Martin, P., \& L\'{e}pine, J. R. D.,
2004, \aap, 413, 159

\bibitem[Audouze \& Tinsley(1976)]{audouze76}
Audouze, J. \& Tinsley, B. M. 1976, \araa, 14, 43

\bibitem[Baldwin et al.(1991)]{baldwin91} 
Baldwin, J. A., Ferland, G.  J., Martin, P. G., Corbin, M. R., Cota,
S. A., Peterson, B. M., \& Slettebak, A. 1991, \apj, 374, 580

\bibitem[Balser(1995)]{balser95}
Balser, D. S. 1995, Ph.D. Thesis, Boston Univ.
(see http://find.nrao.edu/theses.)

\bibitem[Balser(2006)]{balser06}
Balser, D. S. 2006, \aj, 132, 2326

\bibitem[Bania et al.(2010)]{bania10}
Bania, T. M., Anderson, L. D., Balser, D. S., \& Rood, R. T. 2010,
\apj, 718, L106

\bibitem[Bania et al.(1997)]{bania97}
Bania, T. M., Balser, D. S., Rood, R. T., Wilson, T. L., \&
Wilson, T. A. 1997, \apjs, 113, 353

\bibitem[Blitz et al.(1982)]{blitz82}
Blitz, L., Fich, M., \& Stark, A. A. 1982, \apjs, 49, 183

\bibitem[Boesgaard \& Steigman(1985)]{boesgaard85}
Boesgaard, A. M., \& Steigman, G., 1985, \araa, 23, 319

\bibitem[Boissier \& Prantzos(1999)]{boissier99}
Boissier, S., \& Prantzos, N. 1999, \mnras, 307, 857

\bibitem[Bragaglia \& Tosi(2006)]{bragaglia06}
Bragaglia, A., \& Tosi, M. 2006, \aj, 131, 1544

\bibitem[Brand(1986)]{brand86} 
Brand, J. 1986, PhD Thesis, Leiden Univ. (Netherlands)

\bibitem[Brand \& Wouterloot(2007)]{brand07}
Brand, J., \& Wouterloot, J. G. A. 2007, \aap, 464, 909

\bibitem[Bresolin(2011)]{bresolin11}
Bresolin, F. 2011, \apj, 730, 129

\bibitem[Bresolin et al.(2009)]{bresolin09} Bresolin, F., Gieren, W.,
  Kudritzki, R.-P., Pietrzy\'{n}ski, G., Urbaneja, M. A., \& Carraro,
  G. 2009, \apj, 700, 309

\bibitem[Caputo et al.(2001)]{caputo01}
Caputo, F., Marconi, M., Musella, I., Pont, F. 2001, \aap, 372, 544

\bibitem[Carigi et al.(2005)]{carigi05}
Carigi, L., Peimbert, M., Esteban, C., \& Garc\'{i}a-Rojas 2005, \apj,
623, 213

\bibitem[Carraro et al.(2007)]{carraro07}
Carraro, G., Geisler, D., Villanova, S., Frinchaboy, P. M., \&
Majewski, S. R. 2007, \aap, 476, 217

\bibitem[Chen et al.(2003)]{chen03}
Chen, L., Hou, J. L., \& Wang, J. J. 2003, \aj, 125, 1397

\bibitem[Chiappini et al.(1997)]{chiappini97}
Chiappini, C., Matteucci, F., \& Gratton, R. 1997, \apj, 477, 765

\bibitem[Churchwell et al.(1978)] {churchwell78} 
Churchwell, E., Smith, L. F., Mathis, J., Mezger, P. G., \&
Huchtmeier, W. 1978, \aap, 70, 719

\bibitem[Churchwell \& Walmsley(1975)] {churchwell75}
Churchwell, E., \& Walmsley, C. M. 1975, \aap, 38, 451

\bibitem[Colavitti et al.(2008)]{colavitti08}
Colavitti, E., Matteucci. F., \& Murante, G. 2008, \aap, 483, 401

\bibitem[Daflon \& Cunha(2004)]{daflon04}
Daflon, S., \& Cunha, K. 2004, \apj, 617, 1115

\bibitem[Daflon et al.(2004)]{daflon04b}
Daflon, S., Cunha, K., \& Butler, K. 2004, \apj, 606, 519

\bibitem[Deharveng et al.(2000)]{deharveng00}
Deharveng, L., Pe$\tilde{\rm n}$a, M., Caplan, J., \& Costero,
R. 2000, \mnras, 311, 329

\bibitem[Digel et al.(1994)]{digel94}
Digel, S., De Geus, E., \& Thaddeus, P. 1994, \apj, 422, 92

\bibitem[Efron et al.(1979)]{efron79}
Efron, B. 1979, SIAM, 21, 460

\bibitem[Eisenstein et al.(2011)]{eisenstein11}
Eisenstein, D. J., et al. 2011, arXiv:1101.1529

\bibitem[Esteban et al.(2005)]{esteban05}
Esteban, C., Garc\'{i}a-Rojas, J., Peimbert, M. Peimbert, A., Ruiz,
M. T., Rodr\'{i}guez, \& Carigi, L. 2005, \apj, 618, L95

\bibitem[Feigelson \& Babu(1992)]{feigelson92}
Feigelson, E. D., \& Babu, G. J. 1992, \apj, 397, 55

\bibitem[Felli \& Churchwell(1972)]{felli72}
Felli, M., \& Churchwell, E. 1972, \aaps, 5, 369

\bibitem[Felli et al.(1978)]{felli78}
Felli, M., Harten, R. H., Habing, H. J., \& Israel, F. P. 1978, \aaps,
32, 423

\bibitem[Fich et al.(1989)]{fich89}
Fich, M., Blitz, L., \& Stark, A. A. 1989, \apj, 342, 272

\bibitem[Fich \& Silkey(1991)]{fich91}
Fich, M., \& Silkey, M. 1991, \apj, 366, 107

\bibitem[Fish et al.(2003)]{fish03}
Fish, V. L., Reid, M. J., Wilner, D. J., \& Churchwell, E. 2003, \apj, 587, 701

\bibitem[Fitzsimmons et al.(1990)]{fitzsimmons90}
Fitzsimmons, A., Brown, P. J. F., Dufton, P. L., \& Lennon,
D. J. 1990, \aap, 232, 437

\bibitem[Friel(1995)]{friel95}
Friel, E. D. 1995, \araa, 33, 381

\bibitem[Friel et al.(2002)]{friel02}
Friel, E. D., Janes, K. A., Tavarez, M., Scott, J., Katsanis, R.,
Lotz, J., Hong, L., \& Miller, N. 2002, \aj, 124, 2693

\bibitem[Fu et al.(2009)]{fu09}
Fu, J., Hou, J. L., Yin, J., Chang, R. X. 2009, \apj, 696, 668

\bibitem[Garay \& Rodr\'{\i}guez(1983)]{garay83}
Garay, G., \& Rodr\'{\i}guez, L. F. 1983, \apj, 266, 263

\bibitem[Garc\'{i}a-Rojas et al.(2007a)]{garcia-rojas07a}
  Garc\'{i}a-Rojas, J., \& Esteban, C. 2007, \apj, 670, 457

\bibitem[Garc\'{i}a-Rojas et al.(2007b)]{garcia-rojas07b}
  Garc\'{i}a-Rojas, J., Esteban, C., Peimbert, A., Rodr\'{i}guez, M.,
  Peimbert, M., \& Ruiz, M. T. 2007, Rev. Mex. AA, 43, 3

\bibitem[Ghigo et al.(2001)]{ghigo01}
Ghigo, F., Maddalena, R., Balser, D., \& Langston, G. 2001, GBT
Commissioning Memo 10

\bibitem[Gordon(1976)]{gordon76} Gordon, M. A., 1976, in Methods of
  Experimental Physics. Vol. 12. Astrophysics. Part C: Radio
  observations, 277

\bibitem[Gordon \& Wixom(1978)]{gordon78}
Gordon, W. J., \& Wixom, J. A. 1978, Mathematics of Computation, 32, 253

\bibitem[Gummersbach et al.(1998)]{gummersbach98}
Gummersbach, C. A., Kaufer, A., Sch\"{a}fer, D. R., Szeifert, T., \&
Wolf, B., 1998, \aap, 338, 881

\bibitem[Hawley (1978)]{hawley78}
Hawley, S. A. 1978, \apj, 224, 417

\bibitem[Henry et al.(2004)]{henry04}
Henry, R. B. C., Kwitter, K. B., \& Balick, B. 2004, \apj, 127, 2284

\bibitem[Henry et al.(2010)]{henry10}
Henry, R. B. C., Kwitter, K. B., Jaskot, A. E., Balick, B., Morrison,
M. A., \& Milingo, J. B. 2010, \apj, 724, 748

\bibitem[Henry \& Worthey(1999)]{henry99}
Henry, R. B. C., \& Worthey, G. 1999, \pasp, 111, 919

\bibitem[Hou et al.(2000)]{hou00}
Hou, J. L., Prantzos, N., \& Boissier, S. 2000, \aap, 362, 921

\bibitem[Isobe et al.(1990)]{isobe90}
Isobe, T., Feigelson, E. D., Akritas, M. G., \& Babu, G. J. 1990,
\apj, 364, 104

\bibitem[Janes(1979)]{janes79}
Janes, K. A. 1979, \apjs, 39, 135 

\bibitem[Johnson et al.(2002)] {johnson02}
Johnson, C., Maddalena, R., Ghigo, F., \& Balser, D. 2002, GBT
Commissioning Memo 22

\bibitem[Kaufer et al.(1994)]{kaufer94}
Kaufer, A., Szeifert, Th., Krenzin, R., Baschek, B., \& Wolf, B. 1994,
\aap, 289, 740

\bibitem[Kennicutt \& Garnett(1996)]{kennicutt96}
Kennicutt, R. C., \& Garnett, D. R. 1996, \apj, 456, 504

\bibitem[Kilian-Montenbruck et al.(1994)]{kilian-montenbruck94}
Kilian-Montenbruck, J., Gehren, T., \& Nissen, P. E. 1994, \aap, 291, 757

\bibitem[King(1971)]{king71}
King, I. R. 1971, \pasp, 83, 377

\bibitem[Kuchar \& Bania(1994)]{kuchar94} 
Kuchar, T. A. \& Bania, T. M. 1994, \apj, 436, 117

\bibitem[Kolpak et al.(2003)]{kolpak03}
Kolpak, M. A., Jackson, J. M., Bania, T. M., \& Clemens, D. P. 2003, \apj, 582, 756

\bibitem[Lemasle et al.(2008)]{lemasle08}
Lemasle, B., Fran\c{c}ois, P., Piersimoni, A., Pedicelli, S., Bono,
G., Laney, D., Primas, F., \& Romaniello, M. 2008, \aap, 490, 613

\bibitem[Lichten et al.(1979)]{lichten79}
Lichten, S. M., Rodriguez, L. F., \& Chaisson, E. J. 1979, \apj, 229, 524

\bibitem[Liu et al.(2000)]{liu00} Liu, X.-W., Storey, P. J., Barlow,
  M. J., Danziger, I. J., Cohen, M., \& Bryce, M.  2000, MNRAS, 312,
  585

\bibitem[Lockman(1989)]{lockman89}
Lockman, F. J. 1989, \apjs, 71, 469

\bibitem[Lockman et al.(1996)]{lockman96}
Lockman, F. J, Pisano, D. J., \& Howard, G. J., 1996, \apj, 472, 173

\bibitem[Luck et al.(2003)]{luck03}
Luck, R. E., Gieren, W. P., Andrievsky, S. M., Kovtyukh, V. V.,
Fouqu\'{e}, P., Pont, F., \& Kienzle, F. 2003, \aap, 401, 939 

\bibitem[Luck et al.(2006)]{luck06}
Luck, R. E., Kovtyukh, V. V., \& Andrievsky, S. M. 2006, \aj, 132, 902

\bibitem[Mathis(1986)]{mathis86} 
Mathis, J. S. 1986, \pasp, 98, 995

\bibitem[Maciel et al.(2003)]{maciel03}
Maciel, W. J., Costa, R. D. D., \& Uchida, M. M. M. 2003, \aap, 397, 667

\bibitem[Maciel \& Qurieza(1999)]{maciel99}
Maciel, W. J., \& Quireza, C. 1999, \aap, 345, 629

\bibitem[Magrini et al.(2009)]{magrini09}
Magrini, L., Sestito, P., Randich, S., \& Galli, D. 2009,
\aap, 494, 95

\bibitem[Markwardt(2009)]{markwardt09}
Markwardt, C. B. 2009, Astronomical Data Analysis Software and Systems XVIII (ASP Conf. Ser. 411), 
ed. D. A. Bohlender, D. Durand, \& P. Dowler (San Francisco, CA: ASP), 251

\bibitem[Matteucci \& Fran\c{c}ois(1989)]{matteucci89}
Matteucci, F., \& , Fran\c{c}oi, P. 1989, \mnras, 239, 885

\bibitem[Menten et al.(2007)]{menten07} 
Menten, K.~M., Reid, M.~J., Forbrich, J., \& Brunthaler, A. 2007, \aap, 474, 515 

\bibitem[Menzel(1968)]{menzel68}
Menzel, D. H. 1968, Nature, 218, 756

\bibitem[Mezger et al.(1979)]{mezger79}
Mezger, P. G., Pankonin, V., Schmid-Burgk, J., Thum, C., \& Wink,
J. 1979, \aap, 80, L3

\bibitem[Minchev \& Famaey(2010)]{minchev10}
Minchev, I., \& Famaey, B. 2010, \apj, 722, 112

\bibitem[Moll\`{a} et al.(1997)]{molla97}
Moll\`{a}, M., Ferrini, F., \& Diaz, A. I. 1997, \apj, 475, 519

\bibitem[Nahar et al.(2010)]{nahar10}
Nahar, S. N., Montenegro, M., Eissner, W., \& Pradhan, A., K. 2010,
Physical Review A, 82, 5401

\bibitem[Oliveira \& Maciel(1986)]{oliveira86}
Oliveira, S, \& Maciel, W. J. 1986, Ap\&SS, 128, 421

\bibitem[Pagel et al.(1979)]{pagel79}
Pagel, B. E. J., Edmunds, M. G., Blackwell, D. E., Chun, M. S., \&
  Smith, G. 1979, \mnras, 189, 95

\bibitem[Pagel(1997)]{pagel97}
Pagel, B. E. J. 1997, Nucleosynthesis and Chemical Evolution of
Galaxies (Cambridge: Cambridge University Press)

\bibitem[Pagel \& Edmunds(1981)]{pagel81}
Pagel, B. E. J., \& Edmunds, M. G. 1981, \araa, 19, 77

\bibitem[Pedicelli et al.(2009)]{pedicelli09}
Pedicelli, S., Bono, G., Lemasle, B., Fran\c{c}ois, P., Groenewegen,
M., Lub, J., Pel, J. W., Laney, D., Piersimoni, A., Romaniello, M.,
Buonanno, R., Caputo, F., Cassisi, S., Castelli, F., Leurini, S.,
Pietrinferni, A., Primas, F., \& Pritchard, J. 2009, \aap, 504, 81

\bibitem[Peimbert(1975)]{peimbert75}
Peimbert, M. 1975, \araa, 13, 113

\bibitem[Peimbert(1978)]{peimbert78}
Peimbert M., 1978, in IAU Symp. 76, Planetary Nebulae, ed. Y. Terzian
(Reidel: Dordrecht), 215

\bibitem[Peimbert \& Peimbert(2010)]{peimbert10}
Peimbert, A., \& Peimbert, M. 2010, \apj, 724, 791

\bibitem[Peng et al.(2000)] {peng00}
Peng, B., Kraus, A., Krichbaum, T. P., \& Witzel, A. 2000, \aaps, 145, 1

\bibitem[Pilyugin et al.(2003)]{pilyugin03}
Pilyugin, L. S., Ferrini, F., \& Shkvarun, R. V. 2003, \aap, 401, 557

\bibitem[Pisano et al.(2007)] {pisano07} Pisano, D.J., Maddalena, R.,
Figura, C., \& Wagg, J. 2007, GBT Memo 246

\bibitem[Perinotto \& Morbidelli(2006)]{perinotto06}
Perinotto, M., \& Morbidelli, L. 2006, \mnras, 372, 45

\bibitem[Pottasch \& Bernard-Salas(2006)]{pottasch06}
Pottasch, S. R., \& Bernard-Salas, J., 2006, \aap, 457, 189

\bibitem[Prantzos(2003)]{prantzos03}
Prantzos, N. 2003, Rev. Mex. AA, 17, 121

\bibitem[Quireza et al.(2006a)]{quireza06a}
Quireza. C., Rood, R. T., Balser, D. S., \& Bania, T. M. 2006a, \apjs,
165, 338

\bibitem[Quireza et al.(2006b)]{quireza06b} 
Quireza. C., Rood, R. T., Bania, T. M., Balser, D. S., \& Maciel, W.
J.  2006b, \apj, 653, 1126

\bibitem[Reid et al.(2009)]{reid09}
Reid, M. J., Menten, K. M., Brunthaler, A., Zheng, X. W., Moscadelli, L., 
\& Xu, Y. 2009, \apj, 693, 397

\bibitem[Roelfsema et al.(1992)]{roelfsema92}
Roelfsema, P. R., Goss, W. M., \& Mallik, D. C., V. 1992, \apj, 394, 188

\bibitem[Rolleston et al.(2000)]{rolleston00}
Rolleston, W. R. J., Smartt, S. J., Dufton, P. L., \& Ryans,
R. S. I. 2000, \aap, 363, 537
 
\bibitem[Rosolowsky \& Simon(2008)]{rosolowsky08}
Rosolowsky, E., \& Simon, J. D. 2008, \apj, 675, 1213

\bibitem[Ro\v{s}kar et al.(2008a)]{roskar08a}
Ro\v{s}kar, R., Debattista, Stinson, G. S., V. P., Quinn, T. R., 
Kaufmann, T., \& Wadsley, J. 2008, \apjl, 675, L65

\bibitem[Ro\v{s}kar et al.(2008b)]{roskar08b}
Ro\v{s}kar, R., Debattista, V. P., Quinn, T. R., Stinson, G. S., \& Wadsley, J.
\apjl, 684, L79

\bibitem[Roy \& Kunth(1995)]{roy95}
Roy, J.-R., \& Kunth, D. 1995, \aap, 294, 432

\bibitem[Rubin(1985)]{rubin85}
Rubin, R. H. 1985, \apjs, 57, 349

\bibitem[Rubin et al.(1972)] {rubin72}
Rubin, V. C., Kumar, C. K., \& Ford, W. K., Jr. 1972, \apj, 177, 31

\bibitem[Rudolph et al.(1996)]{rudolph96}
Rudolph, A. L., Brand, J., De Geus, E. J., \& Wouterloot,
J. G. A. 1996, \apj, 458

\bibitem[Rudolph et al.(2006)]{rudolph06} 
Rudolph, A. L., Fich, M., Bell, G. R., Norsen, T., Simspon, J. P.,
Haas, M. R., \& Erickson, E. F. 2006, \apjs, 162, 346

\bibitem[Rudolph et al.(1997)]{rudolph97}
Rudolph, A. L., Simspon, J. P., Haas, M. R., Erickson, E. F., \& Fich,
M. 1997, \apj, 489, 94

\bibitem[Samland \& Gerhard(2003)]{samland03}
Samland, M., \& Gerhard, O. E. 2003, \aap, 399, 961

\bibitem[Samland et al.(1997)]{samland97}
Samland, M., Hensler, G., \& Theis, Ch. 1997, \apj, 476, 544

\bibitem[Scalo \& Elemgreen(2004)]{scalo04}
Scalo, J., \& Elemgreen, B. G. 2004, \araa, 42, 275

\bibitem[Sch\"{o}nrich \& Binney(2009)]{schonrich09}
Sch\"{o}nrich, R., \& Binney, J. 2009, \mnras, 396, 203

\bibitem[Searle(1971)]{searle71}
Searle, L. 1971, \apj, 168, 327

\bibitem[Sellwood \& Binney(2002)]{sellwood02}
Sellwood, J. A., \& Binney, J. J. 2002, \mnras, 336, 785

\bibitem[Sestito et al.(2008)]{sestito08}
Sestito, P., Bragaglia, A., Randich, S., Pallavicini, R., Andrievsky,
S. M., \& Korotin, S. A. 2008, \aap, 488, 943

\bibitem[Shaver et al.(1983)]{shaver83}
Shaver, P. A., McGee, R. X., Newton, L. M., Danks, A. C., \& Pottasch,
S. R. 1983, \mnras, 204, 53

\bibitem[Shepard(1968)]{shepard68}
Shepard, D. 1968, Proc. 1968 ACM Nat. Conf., 517

\bibitem[Shields \& Kennicutt(1995)]{shields95} 
Shields, J. C., \& Kennicutt, Jr. R. C., 1995, \apj, 454, 807

\bibitem[Simpson et al.(1995)]{simpson95}
Simpson, J. P., Colgan, S. W. J., Rubin, R. H., Erickson, E. F., \&
Haas, M. R. 1995, \apj, 444, 721

\bibitem[Smartt \& Rolleston(1997)]{smartt97}
Smartt, S. J., \& Rolleston, W. R. 1997, \apj, 481, L47

\bibitem[Smith(1975)]{smith75}
Smith, H. E. 1975, \apj, 199, 591

\bibitem[Sofia \& Meyer(2001)]{sofia01}
Sofia, U. J., \& Meyer, D. M. 2001, \apj, 554, L221

\bibitem[Stanghellini et al.(2006)]{stanghellini06}
Stanghellini, L., Guerrero, M. A., Cunha, K., Manchado, A., \&
Villaver, E. 2006, \apj, 651, 898

\bibitem[Stanghellini \& Haywood(2010)]{stanghellini10} 
Stanghellini, L., \& Haywood, M. 2010

\bibitem[Steigman(2007)]{steigman07}
Steigman, G. 2007, Annu. Rev. Nucl. Part. Sci., 57, 463

\bibitem[Stil et al.(2006)]{stil06} 
Stil, J. M., Taylor, A. R., Dickey, J. M., Kavars, D. W., 
Martin, P. G., Rothwell, T. A., Boothroyd, A. I., Lockman, F. J., 
\& McClure-Griffiths, N. M., \aj, 132, 1158

\bibitem[Talent \& Dufour(1979)]{talent79}
Talent, D. L., \& Dufour, R. J. 1979, \apj, 233, 888

\bibitem[Tosi(1988)]{tosi88}
Tosi M. 1988, \aap, 197, 33

\bibitem[Tosi(1998)]{tosi98}
Tosi M. 1998, in Nuclei and their Galactic Evolution, eds.
N. Prantzos, M. Tosi, \& R. von Steiger (Dordrecht: Kluwer), 207

\bibitem[Tosi(2000)]{tosi00}
Tosi, M. 2000, in The Evolution of the Milky Way, ed. F. Matteucci \& F.
Giovannelli ( Dordrecht: Kluwer), 505

\bibitem[Trimble(1975)]{trimble75}
Trimble, V. 1975, Rev. Mod. Physics, 47, 877

\bibitem[Twarog et al.(1997)]{twarog97}
Twarog, B. A., Ashman, K. M., \& Anthony-Twarog, B. J. 1997, \aj, 114,
2556

\bibitem[van den Bergh(1975)]{vandenbergh75}
van den Bergh, S. 1974, \araa, 13, 217

\bibitem[van Zee et al.(1998)]{vanzee98} 
van Zee, L., Salzer, J. J., Haynes, M. P., O'Donoghue, A. A., \&
  Balonek, T. J. 1998, \aj, 116, 2805

\bibitem[Vila-Costas \& Edmunds(1992)]{vila-costas92}
Vila-Costas, M. B., \& Edmunds, M. G. 1992, \mnras, 259, 121

\bibitem[V\'{i}lchez \& Esteban(1996)]{vilchez96}
V\'{i}lchez, J. M., \& Esteban, C. 1996, \mnras, 280, 720

\bibitem[von Proch\'{a}zka et al.(2010)]{vonprochazka10}
von Proch\'{a}zka, A. A., Remijan, A. J., Balser, D. S., Ryans, R. S. I.,
Marshall, A. H., Schwab, F. R., Hollis, J. M., Jewell, P. R., \&
Lovas, F. J. 2010, \pasp, 122, 354

\bibitem[Wilson et al.(1979)] {wilson79}
Wilson, T. L., Bieging, J., \& Wilson, W. E. 1979, \aap, 71, 205

\bibitem[Wilson \& Rood(1994)]{wilson94}
Wilson, T. L., \& Rood, R. T. 1994, \araa, 32, 191

\bibitem[Wink et al.(1983)]{wink83}
Wink, J. E., Wilson, T. L., \& Bieging, J. H. 1983, \aap, 127, 211

\bibitem[Yong et al.(2005)]{yong05}
Yong, D., Carney, B., \& Teixera de Almeida, M. L. 2005, \aj, 130, 597

\bibitem[Yong et al.(2006)]{yong06}
Yong, D., Carney, B., Teixera de Almeida, M. L., \& Pohl, B. L. 2006, \aj, 131, 2256

\end{thebibliography}
\end{document}